\documentclass[11pt]{article}
\usepackage{amsmath,amsfonts,amssymb,graphicx,theorem}
\usepackage[usenames]{color}
\usepackage{pstricks}

\usepackage[backref=false]{hyperref}
\hypersetup{
colorlinks=true,
citecolor=red,
linkcolor=darkblue,
urlcolor=darkblue
}

\long\def\ignore#1{}
\definecolor{darkblue}{rgb}{0,0,.8}
\definecolor{red}{rgb}{1,0,0}
\definecolor{purple}{rgb}{1,0,1}
\definecolor{coloroflink}{rgb}{0.7,0,1}
\definecolor{coloroflink}{rgb}{0.180392, 0.545098, 0.341176}
\definecolor{darkpurple}{rgb}{1,.2,1}
\definecolor{pink}{rgb}{1,.7,.7}
\definecolor{lightblue}{rgb}{.61,.61,1}
\definecolor{midblue}{rgb}{.7,.7,1}
\definecolor{lightlightblue}{rgb}{.9,.9,1}
\definecolor{lightestblue}{rgb}{.96,.96,1}
\definecolor{lightpurple}{rgb}{1,.65,1}
\definecolor{darkgreen}{rgb}{0.180392, 0.545098, 0.341176}

\newcommand{\mince}{0.5pt}
\newcommand{\elegant}{1.0pt}

\newcommand{\boxfilling}{lightlightblue}

\theorembodyfont{\itshape} 
\theoremheaderfont{\scshape}
\theoremstyle{plain}  
\newtheorem{Lemme}{Lemma}
\newtheorem{Theoreme}[Lemme]{Theorem}
\newtheorem{Proposition}[Lemme]{Proposition}

\numberwithin{equation}{section}

\newcommand{\tl}[1]{\mathsf{TL}_{#1}} 
\newcommand{\Stan}[1]{\mathsf{V}_{#1}} 
\newcommand{\Irre}[1]{\mathsf{I}_{#1}} 
\newcommand{\Proj}[1]{\mathsf{P}_{#1}} 

\newcommand{\ctwotimes}[1]{(\mathbb C^2)^{\otimes{\hspace{0.025cm}#1}}}
\newcommand{\eigenSz}[2]{({\mathbb C^2})^{\otimes\,{#1}}_{#2}}
\newcommand{\theArcs}[1]{\mathcal A({#1})}
\newcommand{\theArcsb}[1]{{\mathcal S}({#1})}
\newcommand{\hxxz}{H_{\textrm{XXZ}}}
\newcommand{\hloop}{{\mathcal H}}
\newcommand{\hspin}{{\mathbb H}}
\newcommand{\qn}[1]{\{{#1}\}}
\newcommand{\qqn}[1]{\big\{{#1}\big\}}    

\DeclareMathOperator{\id}{id}

\newcommand{\nc}{\newcommand}
\nc{\bib}{\bibitem}
\nc{\be}{\begin{equation}}
\nc{\ee}{\end{equation}}
\nc{\hs}{\hspace{0.025cm}}
\nc{\sm}{\sigma^-}
\nc{\sz}{\sigma^z}
\nc{\ket}[1]{|{#1}\rangle}
\nc{\bket}[1]{\big|{#1}\big\rangle}
\nc{\chit}{\raisebox{0.25ex}{$\chi$}}
\nc{\chih}{\raisebox{0.25ex}{$\hat\chi$}}
\nc{\ir}{\mathrm{i}}

\def\arXiv#1#2{\href{http://arxiv.org/abs/#1}{arXiv:#1 #2}}

\renewcommand{\ge}{\geqslant}
\renewcommand{\le}{\leqslant}

\begin{document}

\topmargin -15mm
\oddsidemargin 05mm

%
%

\title{\mbox{}\vspace{-.2in}
\bf 
On the reality of spectra\\[0.6cm] of $\boldsymbol{U_q(sl_2)}$-invariant XXZ Hamiltonians}

\date{\vspace{-.5truecm}}  

\maketitle

\begin{center}
{\vspace{-11mm}
{\large Alexi Morin-Duchesne$^\ast$, J{\o}rgen Rasmussen$^\S$, Philippe Ruelle$^\ast$,  Yvan Saint-Aubin$^\ddagger$}}
\\[.8cm]
{\em {}$^\ast$Institut de Recherche en Math\'ematique et Physique\\ Universit\'e Catholique de Louvain, Louvain-la-Neuve, B-1348, Belgium}
\\[.2cm]
{\em {}$^\S$School of Mathematics and Physics, University of Queensland}\\
{\em St Lucia, Brisbane, Queensland 4072, Australia}
\\[.2cm]
{\em {}$^\ddagger$D\'epartement de math\'ematiques et de statistique, Universit\'e de Montr\'eal}\\
{\em Montr\'eal, Qu\'ebec, Canada, H3C 3J7}
\\[.7cm] 
\begin{tabular}{cc}
{\tt alexi.morin-duchesne\,@\,uclouvain.be}
 & \qquad
{\tt j.rasmussen\,@\,uq.edu.au}
\\
{\tt philippe.ruelle\,@\,uclouvain.be}
 & \qquad
{\tt yvan.saint-aubin\,@\,umontreal.ca}
\end{tabular}
\end{center}

%
%

\vspace{.6cm}

\begin{abstract}

A new inner product is constructed on each standard module over the Temperley-Lieb algebra $\tl n(\beta)$
for $\beta\in \mathbb R$ and $n \ge 2$.
On these modules, the Hamiltonian $h = -\sum_i e_i$ is shown to be self-adjoint with respect to this inner product. 
This implies that its action on these modules is
diagonalisable with real eigenvalues. A representation theoretic argument shows that the reality of spectra of the 
Hamiltonian extends to all other Temperley-Lieb representations. In particular, this result applies to the celebrated 
$U_q(sl_2)$-invariant XXZ Hamiltonian, for all $q+q^{-1}\in \mathbb R$.

\bigskip

\noindent Keywords: Temperley-Lieb algebra, XXZ Hamiltonian, 
loop Hamiltonian, densely packed loops, self-adjoint Hamiltonian, reality of spectra. 

\end{abstract}

\clearpage

%
%

\tableofcontents

\clearpage

%
\section{Introduction}
\label{sec:Introduction}
%

Many important lattice models are described by an evolution operator in the form of a matrix realisation of the Hamiltonian element 
\begin{equation}
h=-\sum_{i=1}^{n-1}e_i
\label{eq:leh}
\end{equation} 
of the Temperley-Lieb algebra $\tl n(\beta)$~\cite{TL}. 
An important example is the XXZ Hamiltonian introduced by Alcaraz {\it et al\/} in~\cite{ABBBQ87}. Indeed,
Pasquier and Saleur~\cite{PS90} made the far-reaching observation that the choice of boundary terms made in~\cite{ABBBQ87} 
endows the spin chain with both an invariance under the quantum group $U_q(sl_2)$ and a natural representation of the Temperley-Lieb algebra.
In that model, the generators $e_i, 1\le i \le n-1$, are represented by endomorphisms of $\ctwotimes n$,
\begin{equation}\label{eq:lesEi}
\chit(e_i)=
\frac12\Big(\sigma_i^x\sigma_{i+1}^x+\sigma_i^y\sigma_{i+1}^y
-\frac12(q+q^{-1})(\sigma_i^z\sigma_{i+1}^z-I)
 +\frac12(q-q^{-1})(\sigma_i^z-\sigma_{i+1}^z)\Big),
\end{equation}
where $\sigma_i^\alpha$, $\alpha=x,y,z$, are the usual Pauli matrices.
The parameter $\beta$ labeling the algebra $\tl n(\beta)$ is given by
\be
 \beta=q+q^{-1},\qquad q\in\mathbb C^\times,
\label{beta}
\ee 
while the XXZ spin-chain Hamiltonian is 
\be
 \hxxz=\chit(h)=-\sum_{i=1}^{n-1}\chit(e_i).
\label{eq:leH}
\ee 
Another example of a lattice system whose evolution is prescribed by $h$ is the densely packed loop model, in which several Temperley-Lieb representations are involved depending on the choice of boundary conditions~\cite{PRZ06}. In particular, these lattice models realise the so-called standard modules $\Stan{n,d}$ addressed below in Theorem~\ref{thm:thm0}.

Although the Hamiltonian $\hxxz$ is not hermitian, the authors of~\cite{ABBBQ87} remark that the spectrum of $\hxxz$ is real for $|q|=1$. Their argument is short 
and essentially amounts to observing that the Hamiltonian is invariant under the simultaneous application 
of complex conjugation and left-right reflection. 
However, the Hamiltonian $\hat H=\frac12(q-q^{-1})\sum_{i=1}^{n-1}(\sigma_i^z-\sigma_{i+1}^z)$ is likewise invariant under these two operations, but its spectrum is purely imaginary for generic $q$ on the unit circle. This reasoning for the reality of the eigenvalues is therefore incomplete, and it is not clear how to fix it.

The reality of the spectrum of the $U_q(sl_2)$-invariant $\hxxz$ is also addressed by Korff and Weston~\cite{KW07}, albeit in a roundabout way. They recall that $PT$-invariance, the invariance under a simultaneous change of parity $P$ and time reversal $T$ (complex conjugation), is a sufficient criterion to 
identify non-hermitian Hamiltonians with real spectra 
if the invariance is realised in the strong sense, meaning that every eigenstate is separately invariant (up to a constant). They subsequently use the invariance under $U_q(sl_2)$ to identify a quotient of $\ctwotimes n$ on which the Hamiltonian $\hxxz$ is $PT$-symmetric and diagonalisable with real eigenvalues. 
However, since the Hamiltonian $\hxxz$ defined by \eqref{eq:lesEi} and \eqref{eq:leH} is known to have nontrivial Jordan blocks when $q$ is a root of unity, the set of representations for which their result holds is rather limited. Still, their paper contains several results of interest to our studies.
First, they show that the full Hamiltonian $\hxxz$
for the spin chain with $n$ spins is diagonalisable with real spectrum if $q=e^{i\theta}$ with $|\theta|<\pi/n$. 
Second, for $q^{2\ell}=1$ with $\ell\ge 2$ integer, they 
demonstrate the reality of the spectrum of $\hxxz$ for irreducible representations lying to the left of the first critical line. 
(These concepts are defined in Section~\ref{sec:Reps}.) They even display orthonormal bases for the inner product on these 
representations. 
However, these two constraints limit the number of representations for which the reality of the spectrum has been established, and this 
limitation is especially unfortunate for the study of the XXZ spin models in their large $n$ limit.

A straightforward way to prove the reality of the spectrum of a linear map $H\in\text{End}(V)$, where $V$ is a finite-dimensional vector 
space over $\mathbb C$, is to find an inner product $(\ |\ ):V\times V\rightarrow \mathbb C$ such that $(v|Hw)=(Hv|w)$ for all $v,w\in V$. 
In any basis $\{ v_i\}$ of $V$, the matrix $S_{ij}=(v_i|v_j)$ is a hermitian positive-definite matrix and the condition $(v|Hw)=(Hv|w)$ reads
\begin{equation}\label{eq:laCondition}
SH=H^\dagger S.
\end{equation}
Here we are using the same symbol for both the linear map $H$ and its matrix representation in the chosen basis. 
Since $S$ is positive-definite, a hermitian square root $S^{\frac12}$ can be constructed and it follows from \eqref{eq:laCondition} that 
the matrix $S^{\frac12}HS^{-\frac12}$ is hermitian. Thus, $S^{\frac12}HS^{-\frac12}$ and $H$ are similar, diagonalisable and have real spectra. 
A goal of this paper is therefore to construct such an inner product on the family of standard modules over the Temperley-Lieb algebra $\tl n(\beta)$. 

In fact, any diagonalisable matrix $H$ with a real spectrum is similar to a hermitian matrix. This alone implies that $H$ and its adjoint are conjugate by a hermitian positive-definite matrix $S$ which can be used to define an inner product. The construction of $S$ is therefore not only a straightforward way to prove the diagonalisability and the reality of the spectrum; if $H$ has these properties, a matrix $S$ always exists. Operators with this generalised hermiticity property are sometimes referred to as {\it pseudo-hermitian} or {\it quasi-hermitian}. 

Our main results are summarised as follows.
\begin{Theoreme}\label{thm:thm0} 
Let $0\le  d\le  n$, $d\equiv n\: \mathrm{mod}\: 2$, $n\ge 2$ and $\beta \in \mathbb R$.
On each standard module $\Stan{n,d}$ over $\tl n(\beta)$, there exists an inner product 
$(\ |\ )_{n,d}:\Stan{n,d}\times\Stan{n,d}\rightarrow\mathbb C$ 
with respect to which the Hamiltonian $h$ is self-adjoint:
\be 
(v|hw)_{n,d}=(hv|w)_{n,d}, \qquad v,w \in \Stan{n,d}.
\ee
\end{Theoreme}
Following the previous discussion, the Hamiltonian $h$
on any standard module is therefore diagonalisable with real spectrum, a result 
previously assumed in~\cite{PRZ06}. As we shall see, a simple 
representation theoretic argument then shows that the reality of the 
eigenvalues of $h$ extends to all Temperley-Lieb representations.
\begin{Theoreme}
For $n \ge 2$ and $\beta \in \mathbb R$, the action of $h $ on any $\tl n(\beta)$-module has real eigenvalues.
\label{thm:Theorem2}
\end{Theoreme}
As an immediate corollary, we have the following result.

\begin{Theoreme}
\label{cor:cor1}
The $U_q(sl_2)$-invariant XXZ Hamiltonian $\hxxz$ has a real spectrum for all $n\ge 2$ and $q+q^{-1} \in \mathbb R$.
\end{Theoreme}

The paper is structured as follows. In Section~\ref{sec:background}, we review the definition of the algebra $\tl n(\beta)$, describe its 
standard, irreducible and projective representations, and write down the module decomposition of its spin-chain modules into 
indecomposable ones. Section~\ref{sec:strategy} introduces key tools used in the proof of the three theorems above, 
namely the hermitian spin-chain Hamiltonian $\hspin_{n-1}$ and the linear map 
$f_{n,d}$ intertwining $\hspin_{n-1}$ and the Temperley-Lieb Hamiltonian $h$ acting on 
the standard module $\Stan{n,d}$. The inner product $S_{n,d}$ on $\Stan{n,d}$ is then constructed in terms of $f_{n,d}$ 
whose intertwining property is established in Appendix~\ref{sec:proof}.
Section~\ref{sec:diagandreal} provides the proofs of the theorems and discusses some consequences for $h$ on general
indecomposable modules. Section~\ref{sec:conclusion} contains some concluding remarks.

\subsection*{Glossary of terms and symbols}

\begin{tabular}{ll}  
$\tl{n} = \tl{n}(\beta)$: Temperley-Lieb algebra on $n$ sites
\qquad\qquad\qquad\qquad\qquad\qquad
& Section \ref{sec:background} \\[.1cm]
$\Stan{n,d}$: standard module over $\tl n$ &  Section \ref{sec:standardModules}\\[0.1cm]
$\Irre{n,d}$: irreducible quotient of $\Stan{n,d}$ & Section \ref{sec:Reps}\\[0.1cm]
$\Proj{n,d}$: projective cover of $\Irre{n,d}$  & Section \ref{sec:Reps}\\[0.1cm]
$\mathcal B_{n,d}$: $(n,d)$-link basis of $\Stan{n,d}$ & Section \ref{sec:standardModules}\\[0.1cm]
&\\  
$h= -\sum_{i=1}^{n-1} e_i\in\tl n$:
Temperley-Lieb Hamiltonian & Equation \eqref{eq:leh}\\[0.1cm]
$\hxxz$: $U_q(sl_2)$-invariant XXZ Hamiltonian & Equation \eqref{eq:leH}\\[0.1cm]
$\hloop=\hloop_{n,d}$: 
matrix representative of $h$ on the standard module $\Stan{n,d}$ & Section \ref{sec:loopHamiltonians}\\[0.1cm]
$\hspin = \hspin_{n-1}$: hermitian spin-chain Hamiltonian on $\ctwotimes {n-1}$ & Equation \eqref{eq:hspin}\\[0.1cm]
&\\
$S^z$: total magnetisation & Equation (\ref{Sz}) \\[.1cm] 
$\eigenSz{n}{s}$: $S^z$-eigensubspace of $\ctwotimes n$ with eigenvalue $s$ & Section \ref{sub:decomposition}\\[0.1cm]
&\\
$(\ |\ )_{n,d}: \Stan{n,d} \times \Stan{n,d}\rightarrow \mathbb C$: inner product on $\Stan{n,d}$ & Section \ref{sec:strategy}\\[.1cm]
$S_{n,d}$: matrix realisation of the inner product $(\ |\ )_{n,d}$  & Section \ref{sec:strategy}\\[.1cm]
$f_{n,d}:\Stan{n,d}\rightarrow\ctwotimes{n-1}_{(d-1)/2}$ : injective map & Section \ref{sec:fnd} 
\end{tabular}

%
\section{The Temperley-Lieb algebra and its representations}
\label{sec:background}
%

\subsection{The algebra $\tl n(\beta)$} 

The Temperley-Lieb algebra $\tl n(\beta)$ \cite{TL} is the associative unital algebra generated by the identity $\id$ and the $n-1$ Temperley-Lieb generators
$\{e_i, 1\le i\le n-1\}$ satisfying 
\be
e_j^2=\beta e_j, \qquad e_j e_{j\pm1} e_j = e_j,  \qquad e_i e_j = e_j e_i, \qquad (|i-j|>1).
\ee
The algebra is labeled by an integer $n\ge 2$ and a complex parameter $\beta$, often written as in (\ref{beta}). For convenience, we occasionally suppress the $\beta$ dependence and denote the algebra by $\tl n$. The structure of the Temperley-Lieb algebras was first described by Martin~\cite{M91} and Goodman and Wenzl~\cite{GW93}.

The algebra $\tl n$ is isomorphic to another one defined diagrammatically, as described in~\cite{W95,RSA14} and reviewed in the following.
An $n$-diagram is a box with $n$ marked nodes on its top edge and as many on its bottom one, where all $2n$ nodes are connected pairwise by non-intersecting curves drawn within the box. For example,
\begin{equation}
\psset{unit=0.8} 
a_1 =\ \begin{pspicture}[shift=-0.4](-0.0,0)(2.4,1) 
\psline[linewidth=\mince,fillstyle=solid,fillcolor=\boxfilling]{-}(0,0)(2.4,0)(2.4,1)(0,1)(0,0)
\psline[linecolor=blue, linewidth=\elegant]{-}(0.2,0)(0.2,1)
\psline[linecolor=blue, linewidth=\elegant]{-}(2.2,0)(2.2,1)
\psarc[linecolor=blue,linewidth=\elegant]{-}(1.2,1){0.2}{180}{360}
\psbezier[linecolor=blue,linewidth=\elegant]{-}(0.6,1)(0.6,0.5)(1.8,0.5)(1.8,1)
\psarc[linecolor=blue,linewidth=\elegant]{-}(0.8,0){0.2}{0}{180}
\psarc[linecolor=blue,linewidth=\elegant]{-}(1.6,0){0.2}{0}{180}
\end{pspicture}
\qquad\textrm{and}\qquad
a_2 = \ \begin{pspicture}[shift=-0.4](-0.0,0)(2.4,0.5)
\psline[linewidth=\mince,fillstyle=solid,fillcolor=\boxfilling]{-}(0,0)(2.4,0)(2.4,1)(0,1)(0,0)
\psarc[linecolor=blue,linewidth=\elegant]{-}(0.4,1){0.2}{180}{360}
\psarc[linecolor=blue,linewidth=\elegant]{-}(1.2,1){0.2}{180}{360}
\psarc[linecolor=blue,linewidth=\elegant]{-}(1.2,0){0.2}{0}{180}
\psbezier[linecolor=blue,linewidth=\elegant]{-}(0.6,0)(0.6,0.5)(1.8,0.5)(1.8,0)
\psbezier[linecolor=blue,linewidth=\elegant]{-}(0.2,0)(0.2,0.5)(1.8,0.5)(1.8,1)
\psline[linecolor=blue,linewidth=\elegant]{-}(2.2,0)(2.2,1)
\end{pspicture}
\end{equation}
are two $6$-diagrams. Two $n$-diagrams that differ only by continuous deformations of the non-intersecting curves are identified. 
The dimension of $\tl n$ is then the number of distinct $n$-diagrams, 
\be
 \dim\tl n=\frac1{n+1}\begin{pmatrix}2n\\n\end{pmatrix}.
\ee
The diagrammatic algebra isomorphic to $\tl n$ is the vector space over $\mathbb C$ of formal linear combinations of $n$-diagrams with 
multiplication defined by vertical concatenation of diagrams. More precisely, if $a_1$ and $a_2$ are $n$-diagrams, their product 
$a_1a_2$ is obtained by drawing $a_2$ on top of $a_1$, removing their common edge, and reading off the new connections between the nodes on the top and bottom edges. The product is then the resulting $n$-diagram weighted by a factor 
$\beta^\#$ where $\#$ is the number of loops closed in the concatenation process. For example,
\begin{equation}
\psset{unit=0.8}
a_1 a_2 =\ \begin{pspicture}[shift=-0.9](-0.0,0)(2.4,2)
\psline[linewidth=\mince,fillstyle=solid,fillcolor=\boxfilling]{-}(0,0)(2.4,0)(2.4,1)(0,1)(0,0)
\psline[linecolor=blue, linewidth=\elegant]{-}(0.2,0)(0.2,1)
\psline[linecolor=blue, linewidth=\elegant]{-}(2.2,0)(2.2,1)
\psarc[linecolor=blue,linewidth=\elegant]{-}(1.2,1){0.2}{180}{360}
\psbezier[linecolor=blue,linewidth=\elegant]{-}(0.6,1)(0.6,0.5)(1.8,0.5)(1.8,1)
\psarc[linecolor=blue,linewidth=\elegant]{-}(0.8,0){0.2}{0}{180}
\psarc[linecolor=blue,linewidth=\elegant]{-}(1.6,0){0.2}{0}{180}
\psline[linewidth=\mince,fillstyle=solid,fillcolor=\boxfilling]{-}(0,2)(2.4,2)(2.4,1)(0,1)(0,2)
\rput(0,1){
\psarc[linecolor=blue,linewidth=\elegant]{-}(0.4,1){0.2}{180}{360}
\psarc[linecolor=blue,linewidth=\elegant]{-}(1.2,1){0.2}{180}{360}
\psarc[linecolor=blue,linewidth=\elegant]{-}(1.2,0){0.2}{0}{180}
\psbezier[linecolor=blue,linewidth=\elegant]{-}(0.6,0)(0.6,0.5)(1.8,0.5)(1.8,0)
\psbezier[linecolor=blue,linewidth=\elegant]{-}(0.2,0)(0.2,0.5)(1.8,0.5)(1.8,1)
\psline[linecolor=blue,linewidth=\elegant]{-}(2.2,0)(2.2,1)
}
\end{pspicture}
\ = \beta^2\ 
\begin{pspicture}[shift=-0.4](-0.0,0)(2.4,1)
\psline[linewidth=\mince,fillstyle=solid,fillcolor=\boxfilling]{-}(0,0)(2.4,0)(2.4,1)(0,1)(0,0)
\psarc[linecolor=blue,linewidth=\elegant]{-}(0.4,1){0.2}{180}{360}
\psarc[linecolor=blue,linewidth=\elegant]{-}(1.2,1){0.2}{180}{360}
\psbezier[linecolor=blue,linewidth=\elegant]{-}(0.2,0)(0.2,0.5)(1.8,0.5)(1.8,1)
\psline[linecolor=blue,linewidth=\elegant]{-}(2.2,0)(2.2,1)
\psarc[linecolor=blue,linewidth=\elegant]{-}(0.8,0){0.2}{0}{180}
\psarc[linecolor=blue,linewidth=\elegant]{-}(1.6,0){0.2}{0}{180}
\end{pspicture}
\ = \beta^2 a_3
,\qquad
a_3=\
\begin{pspicture}[shift=-0.4](-0.0,0)(2.4,1)
\psline[linewidth=\mince,fillstyle=solid,fillcolor=\boxfilling]{-}(0,0)(2.4,0)(2.4,1)(0,1)(0,0)
\psarc[linecolor=blue,linewidth=\elegant]{-}(0.4,1){0.2}{180}{360}
\psarc[linecolor=blue,linewidth=\elegant]{-}(1.2,1){0.2}{180}{360}
\psbezier[linecolor=blue,linewidth=\elegant]{-}(0.2,0)(0.2,0.5)(1.8,0.5)(1.8,1)
\psline[linecolor=blue,linewidth=\elegant]{-}(2.2,0)(2.2,1)
\psarc[linecolor=blue,linewidth=\elegant]{-}(0.8,0){0.2}{0}{180}
\psarc[linecolor=blue,linewidth=\elegant]{-}(1.6,0){0.2}{0}{180}
\end{pspicture}
\ .
\end{equation}
The diagrammatic algebra thus defined is clearly associative with unit
\begin{equation}
\psset{unit=0.8}
\id =\ \begin{pspicture}[shift=-0.4](-0.0,0)(2.4,1)
\psline[linewidth=\mince,fillstyle=solid,fillcolor=\boxfilling]{-}(0,0)(2.4,0)(2.4,1)(0,1)(0,0)
\psline[linecolor=blue, linewidth=\elegant]{-}(0.2,0)(0.2,1)
\psline[linecolor=blue, linewidth=\elegant]{-}(0.6,0)(0.6,1)
\psline[linecolor=blue, linewidth=\elegant]{-}(1,0)(1,1)
\rput(1.6,0.5){$\dots$}
\psline[linecolor=blue, linewidth=\elegant]{-}(2.2,0)(2.2,1)
\end{pspicture}\ .
\end{equation}
The
 isomorphism between the algebraic and diagrammatic formulations is given by
\begin{equation}
\psset{unit=0.8}
e_i = \ 
\begin{pspicture}[shift=-0.7](-0.0,-0.3)(3.8,1.0)
\psline[linewidth=\mince,fillstyle=solid,fillcolor=\boxfilling]{-}(0,0)(3.8,0)(3.8,1)(0,1)(0,0)
\psline[linecolor=blue, linewidth=\elegant]{-}(0.2,0)(0.2,1)
\psline[linecolor=blue, linewidth=\elegant]{-}(0.6,0)(0.6,1)
\rput(1.1,0.5){$\dots$}
\rput(1.9,-0.25){$\scriptstyle\phantom{+}i\phantom{+}$}
\rput(2.4,-0.25){$\scriptstyle{i+1}$}
\psline[linecolor=blue, linewidth=\elegant]{-}(1.5,0)(1.5,1)
\psarc[linecolor=blue,linewidth=\elegant]{-}(2.1,0){0.2}{0}{180}
\psarc[linecolor=blue,linewidth=\elegant]{-}(2.1,1){0.2}{180}{360}
\psline[linecolor=blue, linewidth=\elegant]{-}(2.7,0)(2.7,1)
\rput(3.2,0.5){$\dots$}
\psline[linecolor=blue, linewidth=\elegant]{-}(3.6,0)(3.6,1)
\end{pspicture}\ ,\qquad i=1,2,\ldots,n-1.
\end{equation}

\subsection{Standard modules}\label{sec:standardModules}

The {\em standard modules} $\Stan{n,d}$ over $\tl n(\beta)$ are defined for $0\le d\le n$ with $d\equiv n\textrm{\:mod\:}2$ and are 
constructed using a natural diagrammatic action of connectivities on links.
An $n$-link is the diagram obtained by erasing, from an $n$-diagram, its sides and top edge, 
as well as all curves joining two nodes on this top edge. Curves that were connecting the bottom edge with the top one become unattached 
at their upper extremities and are called {\em defects}. An $n$-link with $d$ defects is referred to as an $(n,d)$-link. As for $n$-diagrams, 
two $n$-links differing only by an isotopy transformation are identified. We denote by $\mathcal B_{n,d}$ the set of $(n,d)$-links and by 
$\Stan{n,d}$ the set of formal linear combinations over $\mathbb C$ of these links. Examples are 
\begin{alignat}{2}
\label{eq:V64basis}
&\mathcal B_{6,4} = \Big\{\,
\psset{unit=0.8}
\begin{pspicture}[shift=-0.15](-0.0,0)(2.55,0.5)
\psline{-}(0,0)(2.4,0)
\psline[linecolor=blue, linewidth=\elegant]{-}(0.2,0)(0.2,0.5)
\psline[linecolor=blue, linewidth=\elegant]{-}(0.6,0)(0.6,0.5)
\psline[linecolor=blue, linewidth=\elegant]{-}(1,0)(1,0.5)
\psline[linecolor=blue, linewidth=\elegant]{-}(1.4,0)(1.4,0.5)
\psarc[linecolor=blue,linewidth=\elegant]{-}(2.0,0){0.2}{0}{180}
\rput(2.55,0){,}
\end{pspicture}
\quad
\begin{pspicture}[shift=-0.15](-0.0,0)(2.55,0.5)
\psline{-}(0,0)(2.4,0)
\psline[linecolor=blue, linewidth=\elegant]{-}(0.2,0)(0.2,0.5)
\psline[linecolor=blue, linewidth=\elegant]{-}(0.6,0)(0.6,0.5)
\psline[linecolor=blue, linewidth=\elegant]{-}(1,0)(1,0.5)
\psline[linecolor=blue, linewidth=\elegant]{-}(2.2,0)(2.2,0.5)
\psarc[linecolor=blue,linewidth=\elegant]{-}(1.6,0){0.2}{0}{180}
\rput(2.55,0){,}
\end{pspicture}
\quad
\begin{pspicture}[shift=-0.15](-0.0,0)(2.55,0.5)
\psline{-}(0,0)(2.4,0)
\psline[linecolor=blue, linewidth=\elegant]{-}(0.2,0)(0.2,0.5)
\psline[linecolor=blue, linewidth=\elegant]{-}(0.6,0)(0.6,0.5)
\psline[linecolor=blue, linewidth=\elegant]{-}(2.2,0)(2.2,0.5)
\psline[linecolor=blue, linewidth=\elegant]{-}(1.8,0)(1.8,0.5)
\psarc[linecolor=blue,linewidth=\elegant]{-}(1.2,0){0.2}{0}{180}
\rput(2.55,0){,}
\end{pspicture}
\quad
\begin{pspicture}[shift=-0.15](-0.0,0)(2.55,0.5)
\psline{-}(0,0)(2.4,0)
\psline[linecolor=blue, linewidth=\elegant]{-}(0.2,0)(0.2,0.5)
\psline[linecolor=blue, linewidth=\elegant]{-}(1.4,0)(1.4,0.5)
\psline[linecolor=blue, linewidth=\elegant]{-}(2.2,0)(2.2,0.5)
\psline[linecolor=blue, linewidth=\elegant]{-}(1.8,0)(1.8,0.5)
\psarc[linecolor=blue,linewidth=\elegant]{-}(0.8,0){0.2}{0}{180}
\rput(2.55,0){,}
\end{pspicture}
\quad
\begin{pspicture}[shift=-0.15](-0.0,0)(2.4,0.5)
\psline{-}(0,0)(2.4,0)
\psline[linecolor=blue, linewidth=\elegant]{-}(2.2,0)(2.2,0.5)
\psline[linecolor=blue, linewidth=\elegant]{-}(1.8,0)(1.8,0.5)
\psline[linecolor=blue, linewidth=\elegant]{-}(1,0)(1,0.5)
\psline[linecolor=blue, linewidth=\elegant]{-}(1.4,0)(1.4,0.5)
\psarc[linecolor=blue,linewidth=\elegant]{-}(0.4,0){0.2}{0}{180}
\end{pspicture}
\,\,\Big\}, \\[0.1cm]
&
\psset{unit=0.8}
\mathcal B_{6,0} = \Big\{\,
\begin{pspicture}[shift=-0.15](-0.0,0)(2.55,0.5)
\psline{-}(0,0)(2.4,0)
\psarc[linecolor=blue,linewidth=\elegant]{-}(0.4,0){0.2}{0}{180}
\psarc[linecolor=blue,linewidth=\elegant]{-}(1.2,0){0.2}{0}{180}
\psarc[linecolor=blue,linewidth=\elegant]{-}(2.0,0){0.2}{0}{180}
\rput(2.55,0){,}
\end{pspicture}
\quad
\begin{pspicture}[shift=-0.15](-0.0,0)(2.55,0.5)
\psline{-}(0,0)(2.4,0)
\psarc[linecolor=blue,linewidth=\elegant]{-}(0.4,0){0.2}{0}{180}
\psarc[linecolor=blue,linewidth=\elegant]{-}(1.6,0){0.2}{0}{180}
\psbezier[linecolor=blue,linewidth=\elegant]{-}(1.0,0)(1.0,0.5)(2.2,0.5)(2.2,0)
\rput(2.55,0){,}
\end{pspicture}
\quad
\begin{pspicture}[shift=-0.15](-0.0,0)(2.55,0.5)
\psline{-}(0,0)(2.4,0)
\psarc[linecolor=blue,linewidth=\elegant]{-}(0.8,0){0.2}{0}{180}
\psarc[linecolor=blue,linewidth=\elegant]{-}(2.0,0){0.2}{0}{180}
\psbezier[linecolor=blue,linewidth=\elegant]{-}(0.2,0)(0.2,0.5)(1.4,0.5)(1.4,0)
\rput(2.55,0){,}
\end{pspicture}
\quad
\begin{pspicture}[shift=-0.15](-0.0,0)(2.55,0.5)
\psline{-}(0,0)(2.4,0)
\psarc[linecolor=blue,linewidth=\elegant]{-}(0.8,0){0.2}{0}{180}
\psarc[linecolor=blue,linewidth=\elegant]{-}(1.6,0){0.2}{0}{180}
\psbezier[linecolor=blue,linewidth=\elegant]{-}(0.2,0)(0.2,0.7)(2.2,0.7)(2.2,0)
\rput(2.55,0){,}
\end{pspicture}
\quad
\begin{pspicture}[shift=-0.15](-0.0,0)(2.4,0.5)
\psline{-}(0,0)(2.4,0)
\psarc[linecolor=blue,linewidth=\elegant]{-}(1.2,0){0.2}{0}{180}
\psbezier[linecolor=blue,linewidth=\elegant]{-}(0.6,0)(0.6,0.5)(1.8,0.5)(1.8,0)
\psbezier[linecolor=blue,linewidth=\elegant]{-}(0.2,0)(0.2,0.7)(2.2,0.7)(2.2,0)
\end{pspicture}\,\, \Big\},
\label{eq:V60basis}
\end{alignat}
and the dimension of $\Stan{n,d}$ is 
\be
 \dim\Stan{n,d}=\begin{pmatrix}n\\ \frac{n-d}{2}\end{pmatrix}-\begin{pmatrix}n\\ \frac{n-d}{2}-1\end{pmatrix}.
\ee

The algebra $\tl n(\beta)$ acts naturally on $\Stan{n,d}$ by concatenation, thereby giving rise to the representation $\rho_{n,d}$. 
For an $n$-diagram $a\in\tl n$ and an $(n,d)$-link $v\in\mathcal B_{n,d}$, the state $a v\in\Stan{n,d}$ is obtained by drawing the link 
atop the diagram and removing the top and lateral edges of $a$. The ensuing diagram has the form of a link.
If there are less than $d$ defects in this link, $a v$ is zero. If the number of defects in the link is $d$, 
then $a v$ is this $(n,d)$-link weighted by $\beta^\#$, 
where $\#$ is the number of loops closed in the concatenation process. For example, with $v=\psset{unit=0.5}
\begin{pspicture}[shift=-0.0](-0.0,0)(2.4,0.5)
\psline{-}(0,0)(2.4,0)
\psarc[linecolor=blue,linewidth=\elegant]{-}(2.0,0){0.2}{0}{180}
\psarc[linecolor=blue,linewidth=\elegant]{-}(1.2,0){0.2}{0}{180}
\psline[linecolor=blue, linewidth=\elegant]{-}(0.2,0)(0.2,0.5)
\psline[linecolor=blue, linewidth=\elegant]{-}(0.6,0)(0.6,0.5)
\end{pspicture} \in\mathcal B_{6,2}$,
\begin{equation}
\psset{unit=0.8}
a_1 v =\ \begin{pspicture}[shift=-0.5](-0.0,0)(2.4,1.5)
\rput(0,1){
\psarc[linecolor=blue,linewidth=\elegant]{-}(2.0,0){0.2}{0}{180}
\psarc[linecolor=blue,linewidth=\elegant]{-}(1.2,0){0.2}{0}{180}
\psline[linecolor=blue, linewidth=\elegant]{-}(0.2,0)(0.2,0.5)
\psline[linecolor=blue, linewidth=\elegant]{-}(0.6,0)(0.6,0.5)}
\psline[linewidth=\mince,fillstyle=solid,fillcolor=\boxfilling]{-}(0,0)(2.4,0)(2.4,1)(0,1)(0,0)
\psline[linecolor=blue, linewidth=\elegant]{-}(0.2,0)(0.2,1)
\psline[linecolor=blue, linewidth=\elegant]{-}(2.2,0)(2.2,1)
\psarc[linecolor=blue,linewidth=\elegant]{-}(1.2,1){0.2}{180}{360}
\psbezier[linecolor=blue,linewidth=\elegant]{-}(0.6,1)(0.6,0.5)(1.8,0.5)(1.8,1)
\psarc[linecolor=blue,linewidth=\elegant]{-}(0.8,0){0.2}{0}{180}
\psarc[linecolor=blue,linewidth=\elegant]{-}(1.6,0){0.2}{0}{180}
\end{pspicture}\ = \beta\ 
\begin{pspicture}[shift=-0.1](-0.0,0)(2.4,0.5)
\psline[linewidth=\mince]{-}(0,0)(2.4,0)
\psline[linecolor=blue, linewidth=\elegant]{-}(0.2,0)(0.2,0.5)
\psline[linecolor=blue, linewidth=\elegant]{-}(2.2,0)(2.2,0.5)
\psarc[linecolor=blue,linewidth=\elegant]{-}(0.8,0){0.2}{0}{180}
\psarc[linecolor=blue,linewidth=\elegant]{-}(1.6,0){0.2}{0}{180}
\end{pspicture}
\qquad\textrm{but}\qquad
a_2 v = \ \begin{pspicture}[shift=-0.5](-0.0,0)(2.4,1.5)
\psline[linewidth=\mince,fillstyle=solid,fillcolor=\boxfilling]{-}(0,0)(2.4,0)(2.4,1)(0,1)(0,0)
\psarc[linecolor=blue,linewidth=\elegant]{-}(0.4,1){0.2}{180}{360}
\psarc[linecolor=blue,linewidth=\elegant]{-}(1.2,1){0.2}{180}{360}
\psarc[linecolor=blue,linewidth=\elegant]{-}(1.2,0){0.2}{0}{180}
\psbezier[linecolor=blue,linewidth=\elegant]{-}(0.6,0)(0.6,0.5)(1.8,0.5)(1.8,0)
\psbezier[linecolor=blue,linewidth=\elegant]{-}(0.2,0)(0.2,0.5)(1.8,0.5)(1.8,1)
\psline[linecolor=blue,linewidth=\elegant]{-}(2.2,0)(2.2,1)
\rput(0,1){
\psarc[linecolor=blue,linewidth=\elegant]{-}(2.0,0){0.2}{0}{180}
\psarc[linecolor=blue,linewidth=\elegant]{-}(1.2,0){0.2}{0}{180}
\psline[linecolor=blue, linewidth=\elegant]{-}(0.2,0)(0.2,0.5)
\psline[linecolor=blue, linewidth=\elegant]{-}(0.6,0)(0.6,0.5)}
\end{pspicture}\ = 0.
\end{equation}
This action, defined for $n$-diagrams and $(n,d)$-links, is extended linearly on both factors. It is easily verified, for example 
diagrammatically, that $\Stan{n,d}$ is a $\tl n$-module under this action. 

\subsection{Loop Hamiltonians}\label{sec:loopHamiltonians}

As discussed in the introduction, the Hamiltonian $h=-\sum_{i=1}^{n-1}e_i\in\tl n$ 
plays a prominent role in applications of the Temperley-Lieb algebra.
Its matrix representative acting on the standard module $\Stan{n,d}$ is denoted by 
\be
 \hloop_{n,d}=\rho_{n,d}(h). 
\ee
For $n=6$, $d=0$ and in the ordered basis \eqref{eq:V60basis}, for example, it is given by
\be
\hloop_{6,0} = -
\left(
\begin{array}{ccccc}
 3 \beta  & 2 & 2 & 0 & 2 \\
 1 & 2 \beta  & 0 & 1 & 0 \\
 1 & 0 & 2 \beta  & 1 & 0 \\
 0 & 1 & 1 & 2 \beta  & 2 \\
 0 & 0 & 0 & 1 & \beta  \\
\end{array}
\right).
\ee
This matrix is obviously non-hermitian, immediately sparking the question of its diagonalisability and of the reality of its spectrum. In this 
example, one can verify by an explicit computation that both the diagonalisability and the reality of the spectrum hold for $\beta \in \mathbb R$ (while the reality of spectra breaks down for $\beta\in\mathbb{C}\setminus\mathbb{R}$ since $-2\beta$ is an eigenvalue of $\hloop_{6,0}$). However, this conclusion can also be reached by showing that the matrix 
\be
S_{6,0} = 
\left(
\begin{array}{ccccc}
 1 & 0 & 0 & 0 & 0 \\
 0 & 2 & 0 & 0 & 1 \\
 0 & 0 & 2 & 0 & 1 \\
 0 & 0 & 0 & 3 & -\beta  \\
 0 & 1 & 1 & -\beta  & \beta ^2+4 \\
\end{array}
\right)
\label{eq:S60mat}
\ee
satisfies $S_{6,0} \hloop_{6,0} = \hloop_{6,0}^\dagger S_{6,0}$. For $\beta \in \mathbb R$, 
$S_{6,0}$ is real and symmetric.
Its eigenvalues can be explicitly computed and are all positive, so $S_{6,0}$ defines an inner product on $\Stan{6,0}$. 
The matrix $\hloop_{6,0}$ is thus self-adjoint with respect to this inner product and is therefore diagonalisable with real eigenvalues. 

Our main objective is to extend this analysis to all standard modules $\Stan{n,d}$. 
Section~\ref{sec:strategy} will thus describe the construction of $S_{n,d}$ for all $n$ and $d$, with 
the crucial property $S_{n,d} \hloop_{n,d} = \hloop_{n,d}^\dagger S_{n,d}$ proven in Section~\ref{sec:diagandreal} 
and Appendix~\ref{sec:proof}.

\subsection{Module structures}\label{sec:Reps}

A {\it composition series} of a module $\mathsf M$ over a finite associative algebra is a filtration of $\mathsf M$ by submodules,
\begin{equation}
0 = \mathsf M_0 \subset \mathsf M_1 \subset \mathsf M_2 \subset \dots \subset \mathsf M_k = \mathsf M,
\end{equation}
such that each {\it composition factor} $\mathsf M_{i+1}/\mathsf M_i$ is irreducible. 
Although a module $\mathsf M$ often can be described by more than one composition series, the Jordan-H\"older theorem asserts
that the set of composition factors is unique. 

The structure of $\mathsf M$ is alternatively encoded in its {\it Loewy diagram}. 
This diagram is an oriented graph whose vertices are given by the composition factors of $\mathsf M$ connected by arrows, where
an arrow pointing from the factor $\mathsf A$ to the factor $\mathsf B$ means that vectors in $\mathsf B$ can be reached from vectors 
in $\mathsf A$ by the action of the algebra. 
A composition factor with no outwards pointing arrow is thus an (irreducible) submodule of $\mathsf M$.
The {\it socle} of the module $\mathsf M$ is the direct sum of its irreducible submodules and appears 
at the bottom of the associated Loewy diagram. Likewise, the {\it head} of $\mathsf M$ 
is the quotient of $\mathsf M$ by the intersection of all its maximal submodules 
(known as the {\it radical}) and appears at the top of the Loewy diagram. The Loewy diagram of $\mathsf M$ is a connected graph if and only if the module is indecomposable.
Explicit examples of Loewy diagrams are given below in \eqref{eq:Loewystandard}, \eqref{eq:Loewyprojective} and \eqref{proj}.

Recal that the {\it regular representation} of $\tl n$ is afforded by the action of $\tl n$ on itself.
The {\it principal indecomposable modules} over $\tl n$ are the modules appearing as the indecomposable summands in the 
decomposition of this representation. They are also precisely the {\it projective covers} of the irreducible modules.
For semi-simple algebras, the principal indecomposable modules are irreducible and yield a complete set of such modules. 
The non-semi-simple case is more complicated as it involves modules that are reducible yet indecomposable.

With $\beta$ parameterised as in \eqref{beta}, the structure of the standard and projective modules over $\tl n(\beta)$ 
depends on whether $q$ is a root of unity.
For $q$ not a root of unity, the algebra $\tl n(\beta=q+q^{-1})$ is semi-simple and the corresponding $q$ is termed {\it generic}. 
In this case, the standard modules $\Stan{n,d}$, $0\le d\le n$ with $d\equiv n\textrm{\:mod\:}2$, form a complete set of non-isomorphic irreducible 
modules, also denoted by $\Irre{n,d}$. 

For $q$ a root of unity, let $\ell$ be the smallest positive integer such that $q^{2\ell}=1$. Semi-simplicity extends to some of these cases, namely $\ell = 1$, $\ell =2$ with $n$ odd, and $\ell \ge 3$ with $n< \ell$. 
For $\ell=2$ with $n$ even or $\ell \ge 3$ with $n\ge \ell$, the algebra $\tl n$ is non-semi-simple and has reducible yet indecomposable representations. 
Let us write the integers $\{d\,|\, 0\le d\le n;\; d\equiv n\textrm{\:mod\:}2\}$ in increasing order along a horizontal line,
with the goal to introduce $\ell$-dependent partitions of this set. Vertical (dashed) lines, called {\it critical lines},
are drawn through the integer positions $i\equiv \ell-1\textrm{\:mod\:}\ell$ on the horizontal line.
If such a line goes through $d$, the latter is also called a {\it critical integer}
and $\{d\}$ is then one of the subsets of the partition.
All other elements of $\{d\,|\, 0\le d\le n;\; d\equiv n\textrm{\:mod\:}2\}$ are called {\it non-critical} and are organised into (non-critical) {\it orbits} under
reflection through the critical lines. An example for $n=20$ and $\ell=5$ is given in Figure~\ref{fig:example5,20}.
\begin{figure}[h!]
\begin{center}
\psset{unit=0.75}
\begin{pspicture}[shift=0](-0.5,-1)(20.5,0.5)
\psline[linewidth=\elegant, linestyle=dotted]{-}(4,-1)(4,-0.25)
\psline[linewidth=\elegant, linestyle=dotted]{-}(9,-1)(9,0.25)
\psline[linewidth=\elegant, linestyle=dotted]{-}(14,-1)(14,-0.25)
\psline[linewidth=\elegant, linestyle=dotted]{-}(19,-1)(19,0.25)
\psarc[linecolor=blue,linewidth=\elegant]{-}(0.25,-0.25){0.25}{180}{270}
\psarc[linecolor=blue,linewidth=\elegant]{-}(7.75,-0.25){0.25}{270}{360}
\psline[linecolor=blue,linewidth=\elegant]{-}(0.25,-0.5)(7.75,-0.5)
\psarc[linecolor=blue,linewidth=\elegant]{-}(8.25,-0.25){0.25}{180}{270}
\psarc[linecolor=blue,linewidth=\elegant]{-}(9.75,-0.25){0.25}{270}{360}
\psline[linecolor=blue,linewidth=\elegant]{-}(8.25,-0.5)(9.75,-0.5)
\psarc[linecolor=blue,linewidth=\elegant]{-}(10.25,-0.25){0.25}{180}{270}
\psarc[linecolor=blue,linewidth=\elegant]{-}(17.75,-0.25){0.25}{270}{360}
\psline[linecolor=blue,linewidth=\elegant]{-}(10.25,-0.5)(17.75,-0.5)
\psarc[linecolor=blue,linewidth=\elegant]{-}(18.25,-0.25){0.25}{180}{270}
\psarc[linecolor=blue,linewidth=\elegant]{-}(19.75,-0.25){0.25}{270}{360}
\psline[linecolor=blue,linewidth=\elegant]{-}(18.25,-0.5)(19.75,-0.5)
\psline[linecolor=red,linewidth=\elegant]{-}(2,-0.5)(2,-0.25)
\psline[linecolor=red,linewidth=\elegant]{-}(6,-0.5)(6,-0.25)
\psline[linecolor=red,linewidth=\elegant]{-}(12,-0.5)(12,-0.25)
\psline[linecolor=red,linewidth=\elegant]{-}(16,-0.5)(16,-0.25)
\psarc[linecolor=red,linewidth=\elegant]{-}(2.25,-0.5){0.25}{180}{270}
\psarc[linecolor=red,linewidth=\elegant]{-}(5.75,-0.5){0.25}{270}{360}
\psline[linecolor=red,linewidth=\elegant]{-}(2.25,-0.75)(5.75,-0.75)
\psarc[linecolor=red,linewidth=\elegant]{-}(6.25,-0.5){0.25}{180}{270}
\psarc[linecolor=red,linewidth=\elegant]{-}(11.75,-0.5){0.25}{270}{360}
\psline[linecolor=red,linewidth=\elegant]{-}(6.25,-0.75)(11.75,-0.75)
\psarc[linecolor=red,linewidth=\elegant]{-}(12.25,-0.5){0.25}{180}{270}
\psarc[linecolor=red,linewidth=\elegant]{-}(15.75,-0.5){0.25}{270}{360}
\psline[linecolor=red,linewidth=\elegant]{-}(12.25,-0.75)(15.75,-0.75)
\rput(0,0.25){$0$}\rput(2,0.25){$2$}
\rput(4,0.25){$4$}\rput(6,0.25){$6$}
\rput(8,0.25){$8$}\rput(10,0.25){$10$}
\rput(12,0.25){$12$}\rput(14,0.25){$14$}
\rput(16,0.25){$16$}\rput(18,0.25){$18$}
\rput(20,0.25){$20$}
\end{pspicture}
\caption{For $\ell = 5$, the set $\{0, 2, \dots, 20\}$ partitions into two orbits, $\{0,8,10,18, 20\}$ and $\{2,6,12,16\}$, 
and two critical integers, $\{4\}$ and $\{14\}$.}
\label{fig:example5,20}
\end{center}
\end{figure}
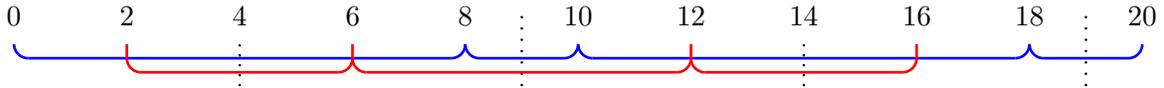
We shall write the elements of an orbit $\{d_l,\dots,d_r\}$ in increasing order, implying that $d_l$ and $d_r$ are the smallest and largest 
elements, respectively. If they exist, the elements immediately to the left and right of $d$ in the orbit 
$\{d_l,\dots,d_r\}$ are respectively denoted by $d_-$ and $d_+$. For $\ell=2$ and $n$ even, the integers $d=0,2,\ldots,n$ form a single orbit.

The standard modules $\Stan{n,d}$ for $d$ critical are both irreducible and 
projective. 
Let $d_r$ be the rightmost element of an orbit. The module $\Stan{n,d_r}$ is then irreducible but not projective. 
(Its projective cover is described below.) If $d$ is in an orbit and $d<d_r$, the module $\Stan{n,d}$ is not irreducible. 
Instead, it is a reducible yet indecomposable module whose structure is described by the Loewy diagram
\be
\begin{pspicture}[shift=-0.9](-0.2,-0.2)(1.4,1.4)
\rput(0,1.2){$\Irre{n,d}$}
\rput(1.2,0){$\Irre{n,d_+}$}
\psline[linewidth=\elegant]{->}(0.3,0.9)(0.9,0.3)
\end{pspicture}
\label{eq:Loewystandard}
\ee
where $\Irre{n,d}$ and $\Irre{n,d_+}$ are irreducible modules.
Equivalently, the structure of $\Stan{n,d}$ is given by the non-split short exact sequence 
\be
 0\rightarrow \Irre{n,d_+}\rightarrow \Stan{n,d}\rightarrow\Irre{n,d}\rightarrow 0 
\ee
implying that $\Irre{n,d_+}$ is a submodule of $\Stan{n,d}$ and $\Irre{n,d}\simeq\Stan{n,d}/\Irre{n,d_+}$. 

The irreducible quotients $\Irre{n,d}$ of the standard modules $\Stan{n,d}$ form a complete set of irreducible modules over $\tl n$. 
These irreducible modules are non-isomorphic except for $\beta=0$. In that case, for $n$ even, the standard module $\Stan{n,0}$ is 
isomorphic to the irreducible module $\Irre{n,2}$, and a complete set of irreducible modules is given by $\{\Irre {n,d}| d = 2, 4, \dots, n\}$. 

For $d_l<d<d_r$, the (indecomposable) projective cover $\Proj{n,d}$ of $\Irre{n,d}$ has four composition factors: $\Irre{n,d_-}$, $\Irre{n,d_+}$ and two copies of $\Irre{n,d}$. 
Its structure is described by the non-split short exact sequence 
$0\rightarrow\Stan{n,d_-}\rightarrow\Proj{n,d}\rightarrow\Stan{n,d}\rightarrow 0$ and its Loewy diagram is
\be
\begin{pspicture}[shift=-1.0](-0.2,-0.2)(2.6,2.5)
\rput(0.2,1.2){$\Irre{n,d_-}$}
\rput(1.2,0){$\Irre{n,d}$}
\rput(2.65,1.2){$\Irre{n,d_+}\ \ \ .$}
\rput(1.2,2.4){$\Irre{n,d}$}
\psline[linewidth=\elegant]{->}(0.3,0.9)(0.9,0.3)
\psline[linewidth=\elegant]{->}(2.1,0.9)(1.5,0.3)
\psline[linewidth=\elegant]{->}(0.9,2.1)(0.3,1.5)
\psline[linewidth=\elegant]{->}(1.5,2.1)(2.1,1.5)
\end{pspicture}
\label{eq:Loewyprojective}
\ee
The boundary cases $d=d_l$ and $d=d_r$ are special as the projective covers have two or three composition factors, and their Loewy 
diagrams are given by 
\be
\begin{array}{ccccccc}
\begin{pspicture}[shift=-0.7](-0.2,-0.2)(1.4,1.4)
\rput(0,1.2){$\Irre{n,d_l}$}
\rput(1.2,0){$\Irre{n,(d_l)_+}$}
\psline[linewidth=\elegant]{->}(0.3,0.9)(0.9,0.3)
\end{pspicture} &\qquad &
\begin{pspicture}[shift=-1.3](1.0,-0.2)(2.4,2.6)
\rput(1.2,0){$\Irre{n,2}$}
\rput(2.4,1.2){$\Irre{n,4}$}
\rput(1.2,2.4){$\Irre{n,2}$}
\psline[linewidth=\elegant]{->}(1.5,2.1)(2.1,1.5)
\psline[linewidth=\elegant]{->}(2.1,0.9)(1.5,0.3)
\end{pspicture}
&\qquad& 
\begin{pspicture}[shift=-1.3](-0.2,-0.2)(1.4,2.6)
\rput(0,1.2){$\Irre{n,(d_r)_-}$}
\rput(1.2,0){$\Irre{n,d_r}$}
\rput(1.2,2.4){$\Irre{n,d_r}$}
\psline[linewidth=\elegant]{->}(0.3,0.9)(0.9,0.3)
\psline[linewidth=\elegant]{->}(0.9,2.1)(0.3,1.5)
\end{pspicture}
&\quad {\rm and}  & 
\begin{pspicture}[shift=-0.9](1,-1)(2,1)
\rput(1.5,0.75){$\Irre{2,2}$}
\rput(1.5,-0.75){$\Irre{2,2}$}
\psline[linewidth=.8pt,arrowsize=3pt 2]{->}(1.5,0.35)(1.5,-0.35)
\end{pspicture}
.\!\!
\\ \\ 
(\beta\neq0)& &(\beta=0, d_l=2)& &\textrm{(all roots of unity)}
& & (\beta = 0, n=2) 
\end{array}
\label{proj}
\ee
Standard and projective modules over $\tl n$ appear in the decomposition of the XXZ spin-chain modules. 
This is described in the next subsection.

\subsection[Decomposition of the XXZ spin-chain representations]{Decomposition of the XXZ spin-chain representations}
\label{sub:decomposition}

The map $\chit : \tl n(\beta) \rightarrow \textrm{End}\big(\ctwotimes n\big)$ given in \eqref{eq:lesEi} defines a representation of 
$\tl n(\beta)$. Goodman and Wenzl~\cite{GW93}, using an equivalent representation, showed that $\chit$ is faithful, 
while Martin~\cite{M92} determined its decomposition in terms of indecomposable representations. It is readily verified that $\chit(e_i)$ commutes with the total magnetisation 
\be
 S^z=\tfrac12\sum_{i=1}^n\sigma^z_i.
\label{Sz}
\ee 
As a module, the space $\ctwotimes n$ therefore splits into the direct sum of eigenspaces 
$\eigenSz{n}{s}$ of $S^z$ labeled by the corresponding eigenvalues $s\in\{-\frac n2,-\frac{n}2+1,\dots, \frac n2-1,\frac n2\}$. The submodules $\eigenSz{n}{s}$ and $\eigenSz{n}{-s}$ are isomorphic. The full decomposition is described in the following. 

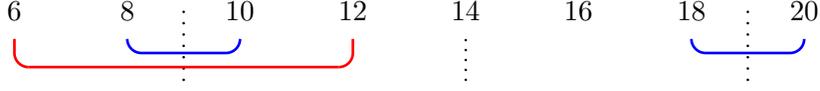
\begin{figure}[h!]
\begin{center}
\psset{unit=0.75}
\begin{pspicture}[shift=-1.0](5.5,-1)(20.5,0.5)
\psline[linewidth=\elegant, linestyle=dotted]{-}(9,-1)(9,0.25)
\psline[linewidth=\elegant, linestyle=dotted]{-}(14,-1)(14,-0.25)
\psline[linewidth=\elegant, linestyle=dotted]{-}(19,-1)(19,0.25)
\psarc[linecolor=blue,linewidth=\elegant]{-}(8.25,-0.25){0.25}{180}{270}
\psarc[linecolor=blue,linewidth=\elegant]{-}(9.75,-0.25){0.25}{270}{360}
\psline[linecolor=blue,linewidth=\elegant]{-}(8.25,-0.5)(9.75,-0.5)
\psarc[linecolor=blue,linewidth=\elegant]{-}(18.25,-0.25){0.25}{180}{270}
\psarc[linecolor=blue,linewidth=\elegant]{-}(19.75,-0.25){0.25}{270}{360}
\psline[linecolor=blue,linewidth=\elegant]{-}(18.25,-0.5)(19.75,-0.5)
\psline[linecolor=red,linewidth=\elegant]{-}(6,-0.5)(6,-0.25)
\psline[linecolor=red,linewidth=\elegant]{-}(12,-0.5)(12,-0.25)
\psarc[linecolor=red,linewidth=\elegant]{-}(6.25,-0.5){0.25}{180}{270}
\psarc[linecolor=red,linewidth=\elegant]{-}(11.75,-0.5){0.25}{270}{360}
\psline[linecolor=red,linewidth=\elegant]{-}(6.25,-0.75)(11.75,-0.75)
\rput(6,0.25){$6$}
\rput(8,0.25){$8$}\rput(10,0.25){$10$}
\rput(12,0.25){$12$}\rput(14,0.25){$14$}
\rput(16,0.25){$16$}\rput(18,0.25){$18$}
\rput(20,0.25){$20$}
\end{pspicture}
\caption{Diagrammatic decomposition of $\ctwotimes {20}_3$ for $\ell = 5$. The pairings are $(8,10), (6,12)$ and $(18,20)$ and 
they contribute the projective modules $\Proj{20,10}$, $\Proj{20,12}$ and $\Proj{20,20}$. The integer $d=14$ is critical and contributes a 
factor of $\Stan{20,14}$, whereas $d=16$, being unpaired and the rightmost element of the orbit $\{6,12,16\}$, contributes 
$\Stan{20,16}$. 
The final result is $\eigenSz{20}{s=3} \simeq \Proj{20,10}\oplus\Proj{20,12}\oplus\Stan{20,14}\oplus\Stan{20,16}\oplus\Proj{20,20}.$}
\label{fig:exampleofdec}
\end{center}
\end{figure}

For $q$ generic, the $\tl n$-submodule $\ctwotimes n_s$ for $s\ge 0$ decomposes into a direct sum of irreducible modules,
\be 
\ctwotimes n_s \simeq\!\!\! \bigoplus_{d = 2s, 2s+2, \dots, n}\!\!\! \Irre{n,d} \qquad (q\ \textrm{generic}).
\ee
For $q$ a root of unity, the decomposition takes the form of a direct sum of projective and standard modules, 
\be
\ctwotimes n_s \simeq \Big(\bigoplus_{d}\, m^{\mathsf P}_d\ \Proj{n,d}\Big) \oplus \Big(\bigoplus_{d}\, m^{\mathsf V}_{d}\  \Stan{n,d}\Big) \qquad 
 (q\ \textrm{a root of unity}),
\label{eq:Csrootofunity}
\ee
where $m^{\mathsf P}_d$ and $m^{\mathsf V}_d$ (both in $\{0,1\}$)
are the multiplicities of the corresponding modules. Gai\-nutdinov and Vasseur~\cite{GV13} wrote down an explicit formula for these multiplicities, and Provencher and Saint-Aubin~\cite{PSA13} gave an easy diagrammatic rule to obtain the overall decomposition. We review the latter and illustrate it with a concrete example in Figure~\ref{fig:exampleofdec}. 

Starting from the partition of integers described above with $q^{2\ell} = 1$, 
we build a new partition by keeping only the integers $d$ larger 
than or equal to $2s$. Each critical integer $d$ in this new partition
contributes a standard module $\Stan{n,d}$ to the direct sum decomposition \eqref{eq:Csrootofunity}. 
The non-critical integers are again grouped into orbits with respect to the critical lines. Each orbit is subsequently split
into pairs $(d_-, d)$, starting from the left, with a given non-critical integer $d$ appearing in at most one pair. If the orbit has odd cardinality, then the rightmost element $d_r$ is left unpaired.
Each pair $(d_-,d)$ then contributes a projective module $\Proj{n,d}$ to the decomposition \eqref{eq:Csrootofunity}, whereas the 
rightmost element $d_r$ of each orbit, if left unpaired, contributes a standard module $\Stan{n,d_r}$.
This rule gives the multiplicities $m^{\mathsf P}_d$ and $m^{\mathsf V}_d$ in the direct sum \eqref{eq:Csrootofunity}, and thus reveals 
the decomposition of $\eigenSz{n}{s}$ in terms of indecomposable modules. 

Of course, the decomposition of $\ctwotimes n$ is the direct sum of $\eigenSz{n}{s}$ over the possible eigenvalues $s$ of $S^z$ and is thus a direct sum of standard and projective modules.

%
\section{Intertwiners and inner products}
\label{sec:strategy} 
%

This section introduces a new positive-definite hermitian matrix $S_{n,d}$ satisfying 
\be
 S_{n,d}\hloop_{n,d}=\hloop_{n,d}^\dagger S_{n,d}.
\label{SHHS}
\ee 
There are two key hurdles to overcome. First, $S_{n,d}$ must be positive-definite, and second, the relation \eqref{SHHS} must be satisfied.

There is a straightforward way to obtain a positive-definite hermitian matrix $S$ of size $m\times m$.
Any $p\times m$ matrix $f$ of rank $m$ 
(and thus with $p\ge m$) defines such an $m\times m$ matrix through $S=f^\dagger f$. Its hermiticity is clear. Its positive-definiteness 
requires that $v^\dagger Sv$ be positive for all nonzero $v$ in $\mathbb C^m$ and follows because $v^\dagger Sv=||fv||^2$, where 
$||\cdot||$ denotes the usual norm on $\mathbb C^p$. Since $f$ has full rank $m$, its kernel is trivial and $fv\neq 0$, 
and hence $v^\dagger Sv >0$ for 
all $v\neq 0$. Thus, $S=f^\dagger f$ is positive-definite and hermitian. 
With an appropriately defined $f$, see Section~\ref{sec:fnd}, this observation will resolve the first hurdle, with the injectivity of this $f$ established in Section~\ref{sec:injectivity}.
To overcome the second hurdle and establish that $\hloop_{n,d}$ is self-adjoint with respect to $S_{n,d}$,
we subsequently show that the linear map $f$ intertwines $\hloop_{n,d}$ and a particular {\em hermitian} spin-chain Hamiltonian
defined in Section~\ref{sec:newH}. 

\subsection{Another spin-chain Hamiltonian}\label{sec:newH}

In a recent study~\cite{MDRR14} of dimers on the square lattice, a spin-chain representation of $\tl n(\beta=0)$ on the space 
$\ctwotimes{n-1}$ was constructed. It is given by
\be 
\tau(e_i)=\sigma_{i-1}^-\sigma_i^+ + \sigma_i^+\sigma_{i+1}^-,
\label{eq:tau}
\ee
where the convention $\sigma_0^\pm = \sigma_n^\pm = 0$ is used. This construction is special because, first, the module over 
$\tl n(0)$ is $\ctwotimes{n-1}$ (and not $\ctwotimes n$, like that underlying the 
$n$-spin Hamiltonian $\hxxz$ in \eqref{eq:leH}), and second, the action of $h$ is the {\em hermitian} Hamiltonian 
\be
\hspin_{n-1}\big|_{\beta=0}=-\sum_{i=1}^{n-1}\tau(e_i)=-\sum_{j=1}^{n-2}(\sigma_j^+\sigma_{j+1}^- + \sigma_j^-\sigma_{j+1}^+).
\label{eq:dimerH}
\ee
In~\cite{MDRR14}, the decomposition of $\ctwotimes{n-1}$ as a module over $\tl n$ was unravelled using intertwiners between $\tau$ and representations with known decompositions.
In this case, the composition factors of $\ctwotimes{n-1}$ are irreducible modules $\Irre{n,d}$ over $\tl n(0)$, and the multiplicity of each $\Irre{n,d}$ is greater or equal to $1$ in $\ctwotimes{n-1}$.
As will be argued in Section~\ref{sub:consequences}, this implies that the eigenvalues of $\hspin_{n-1}|_{\beta=0}, \hloop_{n,d}|_{\beta=0}$ and $\hxxz|_{q=i}$ are identical (up to degeneracies) and real.

Consider the hermitian spin-chain Hamiltonian
\begin{equation}\label{eq:hspin}
\hspin_{n-1} = -\Big(\sum_{j=1}^{n-2} \big(\sigma^-_j \sigma^+_{j+1} + \sigma^-_{j+1} \sigma^+_{j} -\beta\, \sigma^-_j\sigma^+_j \sigma^-_{j+1}\sigma^+_{j+1}\big) +\beta \sum_{j=1}^{n-1}\sigma^-_j \sigma^+_j \Big) \in \textrm{\,End\,}\big(\ctwotimes {n-1}\big).
\end{equation}
It is obviously real and symmetric for $\beta \in \mathbb R$ and coincides with \eqref{eq:dimerH} for $\beta=0$. 
Just like $\hxxz$, $\hspin_{n-1}$ 
commutes with the total magnetisation $S^z$ and therefore splits into sectors. We denote by $\hspin_{n-1,m}$ the restriction of 
$\hspin_{n-1}$ to $\ctwotimes {n-1}_{s = (m-1)/2}$, with $m = -(n\!-\!2), -(n\!-\!4), \dots, n$. The notation is manifestly asymmetric, but this is 
justified by what follows.

We shall define a linear map $f_{n,d}$ in Section~\ref{sec:fnd} and show in Proposition~\ref{sec:int} 
that the following intertwining condition
\begin{equation}\label{eq:entrelacement}
f_{n,d}\,\hloop_{n,d}=\hspin_{n-1,d}\, f_{n,d}
\end{equation} 
holds. The symbol $f_{n,d}$ is used both for the linear map and its matrix realisation. In Proposition~\ref{sec:f.injective}, 
we furthermore show that the map is injective. Together, these two properties of $f_{n,d}$ imply that the spectrum of $\hloop_{n,d}$ 
is included in that of $\hspin_{n-1,d}$. Because $\hspin_{n-1}$ is hermitian, the properties also imply the desired condition 
(\ref{SHHS}), where the inner product is defined by
\be
 S_{n,d}=f_{n,d}^\dagger\,f^{}_{n,d}\,.
\label{Sff}
\ee
Indeed, with the indices omitted, the argument reads
\be
 S\hloop=f^\dagger f\hloop=f^\dagger\hspin f=(f^\dagger \hspin f)^\dagger = (S\hloop)^\dagger=\hloop^\dagger S.
\ee

That $\hxxz$, $\hloop_{n,d}$ and $\hspin_{n-1}$ have overlapping spectra may seem puzzling at first. 
The existence of the injective map $f_{n,d}$ implies that an eigenvalue of $\hloop_{n,d}$ is an eigenvalue of $\hspin_{n-1}$,
that is $\cup_d {\rm \,spec\,} \hloop_{n,d} \subset {\rm \,spec\,} \hspin_{n-1}$,
where ${\rm spec\,}A$ denotes the set of distinct eigenvalues of the matrix $A$.
To understand how the spectrum of $\hxxz$ fits in, we turn to 
the Bethe ansatz equations for the more general XXZ Hamiltonian 
\begin{equation}\label{eq:alcaraz}
\bar H=-\frac12\Big(\sum_{i=1}^{L-1}(\sigma_i^x\sigma_{i+1}^x+\sigma_i^y\sigma_{i+1}^y+\Delta\sigma_i^z\sigma_{i+1}^z)
 +p\sigma_1^z+p'\sigma_{L}^z\Big) - \alpha\, I 
\end{equation}
which are known for all $L, \Delta, \alpha, p$ and $p'$~\cite{ABBBQ87}.
This family includes the Hamiltonian $\hxxz$ defined by \eqref{eq:lesEi} and (\ref{eq:leH}), for which the parameters must be chosen as
\begin{equation}
L=n, \qquad \Delta=
-{\textstyle \frac12}(q+q^{-1}),\qquad 
p=-p'={\textstyle\frac12}(q-q^{-1}), 
\qquad \alpha = 
\tfrac{n-1}4(q+q^{-1}).
\label{eq:criteriahspin}
\end{equation}
The Hamiltonian $\hspin_{n-1}$ can also be written as $\bar H$,
with parameters
\be\label{eq:paraHspin}
 L=n-1, \qquad \Delta=p=p'= -\tfrac\beta2= - \tfrac12(q+q^{-1}),
\qquad \alpha = \tfrac {n\beta} 4= \tfrac  n4 (q+q^{-1}).
\ee
Strikingly, the Bethe ansatz equations for the two Hamiltonians corresponding to the
sets of parameters \eqref{eq:criteriahspin} and \eqref{eq:paraHspin}
turn out to be the same, thus implying that $\hxxz$ and $\hspin_{n-1}$ have overlapping spectra.

Because each irreducible module appears at least once as a composition factor in $\hxxz$, and in at least one of the standard modules, 
${\rm spec\,}\hxxz$ is the same as $\cup_d {\rm \,spec\,} \hloop_{n,d}$.
We thus have
\be 
 {\rm \,spec\,} \hxxz =\bigcup_d {\rm \,spec\,} \hloop_{n,d} \subset {\rm \,spec\,} \hspin_{n-1}.
\label{eq:overlap}
\ee
It is noted that a common eigenvalue of the Hamiltonians need not appear with the same multiplicity in the corresponding spectra. 

For small system sizes, we observe that the last inclusion in \eqref{eq:overlap}
is in general not an equality. This means that, for $\beta \neq 0$, 
$\hspin_{n-1}$ typically has eigenvalues not appearing in the spectra of $\hloop_{n,d}$ or $\hxxz$. This also implies that, in general, 
$\hspin_{n-1}$ does not belong to a representation of $\tl n(\beta)$ on $\ctwotimes{n-1}$, as it does for $\beta = 0$.

\subsection[Construction of the intertwiner $f_{n,d}$]{Construction of the intertwiner $\boldsymbol {f_{n,d}}$}
\label{sec:fnd}

This subsection constructs the linear map 
\be 
 f_{n,d}:\Stan{n,d}\rightarrow \eigenSz{n-1}{(d-1)/2}
\ee 
and discusses its intertwining property (\ref{eq:entrelacement}). Its injectivity is discussed in Section~\ref{sec:injectivity}. The $(n,d)$-link 
basis is used for the standard module $\Stan{n,d}$, while the usual spin basis $|s\rangle=|s_1s_2\dots s_{n-1}\rangle$ with 
$s_i\in\{+,-\}\simeq\{+1,-1\}\simeq\{\uparrow,\downarrow\}$ 
and $\tfrac12\sum_i s_i = \frac{d-1}2$ is used for $\eigenSz{n-1}{(d-1)/2}$. 

A link
$w \in \mathcal B_{n,d}$ is characterised by the set 
\be
 \theArcs w=\big\{(i_k,j_k)\,|\,i_k<j_k,\, 1\le k\le \tfrac{n-d}{2}\big\}
\ee 
of pairs of end nodes of its $\frac{n-d}2$ arcs, where the $n$ nodes are labeled from left to right by $1,\ldots,n$. For example, 
\be
\theArcs{\psset{unit=0.5}\,
\begin{pspicture}[shift=-0.0](-0.0,0)(2.4,0.5)
\psline{-}(0,0)(2.4,0)
\psarc[linecolor=blue,linewidth=\elegant]{-}(2.0,0){0.2}{0}{180}
\psline[linecolor=blue, linewidth=\elegant]{-}(0.2,0)(0.2,0.5)
\psline[linecolor=blue, linewidth=\elegant]{-}(0.6,0)(0.6,0.5)
\psline[linecolor=blue, linewidth=\elegant]{-}(1.0,0)(1.0,0.5)
\psline[linecolor=blue, linewidth=\elegant]{-}(1.4,0)(1.4,0.5)
\end{pspicture}\,}=\{(5,6)\}, \qquad
\theArcs{\psset{unit=0.5}\,
\begin{pspicture}[shift=-0.0](-0.0,0)(2.4,0.5)
\psline{-}(0,0)(2.4,0)
\psarc[linecolor=blue,linewidth=\elegant]{-}(0.8,0){0.2}{0}{180}
\psbezier[linecolor=blue,linewidth=\elegant]{-}(0.2,0)(0.2,0.7)(1.4,0.7)(1.4,0)
\psarc[linecolor=blue,linewidth=\elegant]{-}(2.0,0){0.2}{0}{180}
\end{pspicture}\,}=\{(1,4),(2,3),(5,6)\}.
\ee

The linear map $f_{n,d}$ is now defined by its action on the elements $w$ of the basis $\mathcal B_{n,d}$:
\be
f_{n,d}(w)=\sum_{|s\rangle}c_{s}|s\rangle,
\label{eq:fnddecomp}
\ee
where the coefficients $c_s$ depend on the link $w$, while the sum is over the elements $|s\rangle$ of the spin basis of $\eigenSz{n-1}{(d-1)/2}$.
For each half-arc $(i,j) \in \theArcs w$, define the $|s \rangle$-dependent integer 
\be 
 m_s^{i,j}=\frac12\Big(1-\sum_{k=i}^{j-1}s_k\Big). 
\label{eq:msij}
\ee 
The coefficient $c_{s}$ in \eqref{eq:fnddecomp} is then defined as
\begin{equation}\label{eq:Alexi}
c_{s}=\hspace{-0.1cm}\prod_{(i,j)\in\theArcs w}\hspace{-0.2cm}\qn{m_s^{i,j}},\qquad 
 \qn{m} = (-1)^{m-1}\Big(\frac{q^m-q^{-m}}{q-q^{-1}}\Big) = [m]_{-q},
\end{equation}
where $[m]_q$ is the $m$-th $q$-number. Note that $\beta = q + q^{-1} = -\qn{2}$.

This algebraic definition has the following diagrammatic interpretation. Let us draw the link $w$ above the state 
$|s\rangle=|s_1\ldots s_{n-1}\rangle$ 
in such a way that $s_i$ is located between the node positions $i$ and $i+1$ of $w$. For example, for\!
$\psset{unit=0.5}\,
w=
\begin{pspicture}[shift=-0.0](-0.0,0)(2.4,0.5)
\psline{-}(0,0)(2.4,0)
\psarc[linecolor=blue,linewidth=\elegant]{-}(0.8,0){0.2}{0}{180}
\psbezier[linecolor=blue,linewidth=\elegant]{-}(0.2,0)(0.2,0.7)(1.4,0.7)(1.4,0)
\psarc[linecolor=blue,linewidth=\elegant]{-}(2.0,0){0.2}{0}{180}
\end{pspicture}\, \in \mathcal B_{6,0}$
 and $| s \rangle = | + - - +\, - \rangle\in\eigenSz{5}{-1/2}$, this diagram is 
\begin{equation}\psset{unit=1.2}
\begin{pspicture}[shift=-0.15](-0.0,-0.45)(2.4,0.5)
\psline{-}(0,0)(2.4,0)
\psarc[linecolor=blue,linewidth=\elegant]{-}(0.8,0){0.2}{0}{180}
\psarc[linecolor=blue,linewidth=\elegant]{-}(2.0,0){0.2}{0}{180}
\psbezier[linecolor=blue,linewidth=\elegant]{-}(0.2,0)(0.2,0.7)(1.4,0.7)(1.4,0)
\rput(0.2,-0.25){\scriptsize $\big|$}
\rput(0.4,-0.25){\scriptsize $+$}
\rput(0.8,-0.25){\scriptsize $-$}
\rput(1.2,-0.25){\scriptsize $-$}
\rput(1.6,-0.25){\scriptsize $+$}
\rput(2.0,-0.25){\scriptsize $-$}
\rput(2.3,-0.25){\scriptsize $\big\rangle$}
\rput(2.6,0){.}
\end{pspicture}
\label{eq:depth}
\end{equation}
The integer $m_s^{i,j}$ evaluates to $1$ if the number of down spins under $(i,j)$ is equal to $\frac{j-i+1}2$
and increases (decreases) by one unit for every additional down (up) spin under $(i,j)$. In the example \eqref{eq:depth}, 
we find $m_s^{1,4} = m_s^{2,3} = m_s^{5,6} = 1$ and hence $c_s = 1$. For $|s\rangle=|+-+--\rangle$, on the other hand, we have
\begin{equation}\psset{unit=1.2}
\begin{pspicture}[shift=-0.325](-0.0,-0.45)(2.4,0.5)
\psline{-}(0,0)(2.4,0)
\psarc[linecolor=blue,linewidth=\elegant]{-}(0.8,0){0.2}{0}{180}
\psarc[linecolor=blue,linewidth=\elegant]{-}(2.0,0){0.2}{0}{180}
\psbezier[linecolor=blue,linewidth=\elegant]{-}(0.2,0)(0.2,0.7)(1.4,0.7)(1.4,0)
\rput(0.2,-0.25){\scriptsize $\big|$}
\rput(0.4,-0.25){\scriptsize $+$}
\rput(0.8,-0.25){\scriptsize $-$}
\rput(1.2,-0.25){\scriptsize $+$}
\rput(1.6,-0.25){\scriptsize $-$}
\rput(2.0,-0.25){\scriptsize $-$}
\rput(2.3,-0.25){\scriptsize $\big\rangle$}
\end{pspicture} \quad \rightarrow \quad m_s^{1,4} = 0, \quad m_s^{2,3} = m_s^{5,6} = 1  \quad \rightarrow \quad c_s = 0.
\label{eq:depth2}
\end{equation}
Because $\qn{0}= 0$, the expression \eqref{eq:Alexi} implies that many terms in the decomposition \eqref{eq:fnddecomp} are zero. This is detailed in the following lemma.

\begin{Lemme}\label{lem:nonzero}
For $q$ generic, the coefficient $c_s$ is nonzero if and only if $m_s^{i,j}>0$ for all $(i,j) \in \theArcs w$.
\end{Lemme}
{\scshape Proof } 
For $q$ generic, the integer $m_s^{i,j}$ is zero only if the arc $(i,j)$ overarches exactly 
$\frac{j-i-1}2$ down spins
and increases by one unit for each extra down spin. 
Arcs that join neighbouring sites have $m_s^{i,i+1}\in \{0,1\}$. Suppose that $w$ and $|s\rangle$ are such that $c_s \neq 0$ for $q$ 
generic and $w$ contains at least one arc $(i,j)$ with $m_s^{i,j}<0$ but none with $m_s^{i,j}=0$. Select an arc $(i,j)$ with $m_s^{i,j} < 0$ 
for which all arcs $(k, \ell)$ it overarches have $m_s^{k,\ell}\ge 1$. Let $\mathcal E$ be the subset of arcs immediately overarched by 
$(i,j)$, that is, arcs $(k,\ell)$ for which there is no $(k', \ell') \in \theArcs w$ satisfying $i<k'<k<\ell<\ell'<j$. The set $\mathcal E$ can be 
ordered so that it has the form $\mathcal E = \{(i+1, r_1),(r_1+1, r_2),\dots, (r_t+1, j-1)\}$ for some integer $t$. It then follows that
\begin{equation}
\big[\#\textrm{ of ``$-$'' under }(i,j)\big] \ge \sum_{(k,\ell) \,\in\, \mathcal E} \big[\#{\textrm{ of ``$-$'' under }(k,\ell)}\big] \ge 
 \sum_{(k,\ell)\, \in\, \mathcal E}\tfrac12(\ell-k+1) = \tfrac12(j-i-1),
\end{equation}
which implies $m_s^{i,j} \ge 0$ and thus contradicts the assumption. 
Note that $q$ is assumed generic to exclude cases where $c_s = 0$ because $\qn{m} = 0$ for some $m \neq 0$.\hfill $\square$ 

\medskip

\noindent Thus, if $w$ and $|s \rangle$ are such that $c_s \neq 0$, the number of down spins under the arc $(i,j)$ is greater than or equal to $\frac{j-i+1}2$. In addition, in order to reach the right magnetisation $s = \frac{d-1}{2}$, the outer arches $(i,j)$ must have exactly $\frac{j-i+1}2$ down spins, and all spins not overarched by any arc must be up. 

The examples \eqref{eq:depth} and \eqref{eq:depth2} above fail to reveal the full complexity of \eqref{eq:Alexi}. Indeed, 
higher $q$-numbers appear for larger system sizes, especially for links with imbricated patterns of arcs and only few defects. 
Here are some examples, first for $n=12$ and $d = 0$:
\begin{equation}
\psset{unit=1.1}
\begin{pspicture}[shift=-0.6](-0.0,-0.45)(4.8,1.6)
\psline{-}(0,0)(4.8,0)
\psarc[linecolor=blue,linewidth=\elegant]{-}(2.8,0){0.2}{0}{180}
\psarc[linecolor=blue,linewidth=\elegant]{-}(2.0,0){0.2}{0}{180}
\psbezier[linecolor=blue,linewidth=\elegant]{-}(1.4,0)(1.4,0.9)(3.4,0.9)(3.4,0)
\psbezier[linecolor=blue,linewidth=\elegant]{-}(1.0,0)(1.0,1.3)(3.8,1.3)(3.8,0)
\psbezier[linecolor=blue,linewidth=\elegant]{-}(0.6,0)(0.6,1.7)(4.2,1.7)(4.2,0)
\psbezier[linecolor=blue,linewidth=\elegant]{-}(0.2,0)(0.2,2.1)(4.6,2.1)(4.6,0)
\rput(0.2,-0.25){\scriptsize $\big|$}
\rput(0.4,-0.25){\scriptsize $+$}
\rput(0.8,-0.25){\scriptsize $+$}
\rput(1.2,-0.25){\scriptsize $-$}
\rput(1.6,-0.25){\scriptsize $-$}
\rput(2.0,-0.25){\scriptsize $-$}
\rput(2.4,-0.25){\scriptsize $-$}
\rput(2.8,-0.25){\scriptsize $-$}
\rput(3.2,-0.25){\scriptsize $+$}
\rput(3.6,-0.25){\scriptsize $-$}
\rput(4.0,-0.25){\scriptsize $+$}
\rput(4.4,-0.25){\scriptsize $+$}
\rput(4.7,-0.25){\scriptsize $\big\rangle$}
\end{pspicture} \hspace{0.3cm} \rightarrow \hspace{0.3cm}
\begin{array}{l}
m_s^{1,12} = m_s^{5,6} = m_s^{7,8}= 1, \\[0.2cm]
m_s^{2,11} = m_s^{4,9} = 2, \quad m_s^{3,10} = 3,
\end{array}
 \hspace{0.2cm} \rightarrow \hspace{0.3cm} c_s = \qn{2}^2\qn{3}, 
\end{equation}
and then for $n=6$ and $d=6,4,2,0$:
\begin{align}
&f_{6,6}(\psset{unit=0.6}\,
\begin{pspicture}[shift=-0.125](-0.0,0)(2.4,0.5)
\psline{-}(0,0)(2.4,0)
\psline[linecolor=blue,linewidth=\elegant]{-}(0.2,0)(0.2,0.5)
\psline[linecolor=blue,linewidth=\elegant]{-}(0.6,0)(0.6,0.5)
\psline[linecolor=blue,linewidth=\elegant]{-}(1.0,0)(1.0,0.5)
\psline[linecolor=blue,linewidth=\elegant]{-}(1.4,0)(1.4,0.5)
\psline[linecolor=blue,linewidth=\elegant]{-}(1.8,0)(1.8,0.5)
\psline[linecolor=blue,linewidth=\elegant]{-}(2.2,0)(2.2,0.5)
\end{pspicture}\,)=|++++\,+\rangle, \quad 
f_{6,4}(\,
\begin{pspicture}[shift=-0.125](-0.0,0)(2.4,0.5)
\psline{-}(0,0)(2.4,0)
\psline[linecolor=blue,linewidth=\elegant]{-}(0.2,0)(0.2,0.5)
\psline[linecolor=blue,linewidth=\elegant]{-}(0.6,0)(0.6,0.5)
\psline[linecolor=blue,linewidth=\elegant]{-}(1.0,0)(1.0,0.5)
\psline[linecolor=blue,linewidth=\elegant]{-}(1.4,0)(1.4,0.5)
\psarc[linecolor=blue,linewidth=\elegant]{-}(2.0,0){0.2}{0}{180}
\end{pspicture}\,)=|++++\,-\rangle,\quad
f_{6,4}(\,
\begin{pspicture}[shift=-0.125](-0.0,0)(2.4,0.5)
\psline{-}(0,0)(2.4,0)
\psline[linecolor=blue,linewidth=\elegant]{-}(0.2,0)(0.2,0.5)
\psline[linecolor=blue,linewidth=\elegant]{-}(0.6,0)(0.6,0.5)
\psline[linecolor=blue,linewidth=\elegant]{-}(1.0,0)(1.0,0.5)
\psarc[linecolor=blue,linewidth=\elegant]{-}(1.6,0){0.2}{0}{180}
\psline[linecolor=blue,linewidth=\elegant]{-}(2.2,0)(2.2,0.5)
\end{pspicture}\,)=|+++-\,+\rangle,
\nonumber
\\[0.2cm]
&
f_{6,2}(\psset{unit=0.6}\,
\begin{pspicture}[shift=-0.125](-0.0,0)(2.4,0.5)
\psline{-}(0,0)(2.4,0)
\psline[linecolor=blue,linewidth=\elegant]{-}(0.2,0)(0.2,0.5)
\psarc[linecolor=blue,linewidth=\elegant]{-}(0.8,0){0.2}{0}{180}
\psline[linecolor=blue,linewidth=\elegant]{-}(1.4,0)(1.4,0.5)
\psarc[linecolor=blue,linewidth=\elegant]{-}(2.0,0){0.2}{0}{180}
\end{pspicture}\,)=|+-++\,-\rangle,\quad
f_{6,2}(\,
\begin{pspicture}[shift=-0.125](-0.0,0)(2.4,0.5)
\psline{-}(0,0)(2.4,0)
\psline[linecolor=blue,linewidth=\elegant]{-}(0.2,0)(0.2,0.5)
\psline[linecolor=blue,linewidth=\elegant]{-}(0.6,0)(0.6,0.5)
\psarc[linecolor=blue,linewidth=\elegant]{-}(1.6,0){0.2}{0}{180}
\psbezier[linecolor=blue,linewidth=\elegant]{-}(1.0,0)(1.0,0.7)(2.2,0.7)(2.2,0)
\end{pspicture}\,)=|++--\,+\rangle+|+++-\,-\rangle, 
\nonumber
\\[0.2cm]
&
f_{6,0}(\psset{unit=0.6}\,
\begin{pspicture}[shift=-0.125](-0.0,0)(2.4,0.5)
\psline{-}(0,0)(2.4,0)
\psarc[linecolor=blue,linewidth=\elegant]{-}(0.4,0){0.2}{0}{180}
\psarc[linecolor=blue,linewidth=\elegant]{-}(1.2,0){0.2}{0}{180}
\psarc[linecolor=blue,linewidth=\elegant]{-}(2.0,0){0.2}{0}{180}
\end{pspicture}\,)=|-+-+\,-\rangle,\quad
f_{6,0}(\,
\begin{pspicture}[shift=-0.125](-0.0,0)(2.4,0.5)
\psline{-}(0,0)(2.4,0)
\psarc[linecolor=blue,linewidth=\elegant]{-}(0.4,0){0.2}{0}{180}
\psarc[linecolor=blue,linewidth=\elegant]{-}(1.6,0){0.2}{0}{180}
\psbezier[linecolor=blue,linewidth=\elegant]{-}(1.0,0)(1.0,0.7)(2.2,0.7)(2.2,0)
\end{pspicture}\,)=|-+--\,+\rangle+|-++-\,-\rangle, 
\\[0.2cm]
&
f_{6,0}(\psset{unit=0.6}\,
\begin{pspicture}[shift=-0.125](-0.0,0)(2.4,0.5)
\psline{-}(0,0)(2.4,0)
\psarc[linecolor=blue,linewidth=\elegant]{-}(0.8,0){0.2}{0}{180}
\psarc[linecolor=blue,linewidth=\elegant]{-}(1.6,0){0.2}{0}{180}
\psbezier[linecolor=blue,linewidth=\elegant]{-}(0.2,0)(0.2,1.1)(2.2,1.1)(2.2,0)
\end{pspicture}\,)=|--+-\,+\rangle+ |+---\,+\rangle + |+-+-\,-\rangle,
\nonumber
\\[0.2cm]
&
f_{6,0}(\psset{unit=0.6}\,
\begin{pspicture}[shift=-0.125](-0.0,0)(2.4,0.5)
\psline{-}(0,0)(2.4,0)
\psarc[linecolor=blue,linewidth=\elegant]{-}(1.2,0){0.2}{0}{180}
\psbezier[linecolor=blue,linewidth=\elegant]{-}(0.6,0)(0.6,0.7)(1.8,0.7)(1.8,0)
\psbezier[linecolor=blue,linewidth=\elegant]{-}(0.2,0)(0.2,1.1)(2.2,1.1)(2.2,0)
\end{pspicture}\,)=|---+\,+\rangle + |-+--\,+\rangle +
|+--+\,-\rangle 
+ |++--\,-\rangle + \qn 2 |+---\,+\rangle.
\nonumber
\end{align}
The matrix $f_{6,0}$ used in the construction (\ref{Sff}) of $S_{6,0}$ given in (\ref{eq:S60mat}) is readily computed using these expressions and left-right symmetry. 
Each column in the $10\times 5$ matrix $f_{6,0}$ corresponds to an element $w\in\mathcal B_{6,0}$ and lists the coefficients in 
$f_{6,0}(w)$. In the ordered spin basis 
\begin{align} \big \{\, &
| ++--\,- \rangle,
| +-+-\,- \rangle,
| +--+\,- \rangle,
| +---\,+ \rangle,
| -++-\,- \rangle,\notag\\ &
| -+-+\,- \rangle,
| -+--\,+ \rangle,
| --++\,- \rangle,
| --+-\,+ \rangle,
| ---+\,+ \rangle
\,\big \}
\label{eq:spinbasis}\end{align}
for $\eigenSz{5}{-1/2}$, the matrix is 
thus given by
\begin{equation}
f_{6,0} = \left(
\begin{array}{ccccc}
 0 & 0 & 0 & 0 & 1 \\
 0 & 0 & 0 & 1 & 0 \\
 0 & 0 & 1 & 0 & 1 \\
 0 & 0 & 0 & 1 & \qn 2  \\
 0 & 1 & 0 & 0 & 0 \\
 1 & 0 & 0 & 0 & 0 \\
 0 & 1 & 0 & 0 & 1 \\
 0 & 0 & 1 & 0 & 0 \\
 0 & 0 & 0 & 1 & 0 \\
 0 & 0 & 0 & 0 & 1 \\
\end{array}
\right).
\end{equation}
Recalling that $\beta=-\qn{2}$, it is readily verified that $f_{6,0}^\dagger f_{6,0}$ reproduces the matrix $S_{6,0}$ given in \eqref{eq:S60mat}, 
and that it 
intertwines
$\hloop_{6,0}$ and $\hspin_{5,0}$ as in \eqref{eq:entrelacement}. It is also easy to see from its matrix realisation 
that $f_{6,0}$ is injective. 
Indeed, its columns are linearly independent, as can be seen from the position of the lowest $1$ in each column. Its kernel is thus trivial. 
This observation is key to establishing the injectivity of $f_{n,d}$ in general.

The injectivity of $f_{n,d}$ is the content of Proposition~\ref{sec:f.injective} in Section~\ref{sec:injectivity}.
Its intertwining property is given in the next proposition whose technical proof is deferred until
Appendix~\ref{sec:proof}.
\begin{Proposition}\label{sec:int} The linear map 
$f_{n,d}$ satisfies the intertwining relation $f_{n,d}\,\hloop_{n,d}=\hspin_{n-1,d}\, f_{n,d}$.
\end{Proposition}

\subsection{Injectivity}
\label{sec:injectivity}

In establishing the injectivity of $f_{n,d}$ in Proposition~\ref{sec:f.injective} below, we use particular orderings of the bases of 
$\Stan{n,d}$ and $\eigenSz{n-1}{(d-1)/2}$. First, note that the set $\theArcsb{w}=\{i\,|\, (i,j)\in\theArcs w\}$ of {\em starting} (or leftmost) 
nodes of the arcs is sufficient to uniquely determine $w$. Indeed, given $\theArcsb{w}$, the $(n,d)$-link $w$ is reconstructed by first 
forming the pairs 
$(i, i+1)$ for which $i \in \theArcsb{w}$ but $i+1 \notin  \theArcsb w$. New pairs are formed with unpaired starting nodes 
$i' \in \theArcsb w$ for which the next unpaired node, to its right, is not in $\theArcsb w$. This is repeated until every $i \in \theArcsb{w}$ 
is paired. Defects are then inserted in the remaining unoccupied nodes. 
 
To define an ordering on links, we assign to each $w$ in $\mathcal B_{n,d}$ the dyadic fraction 
\be
 \mathsf d_{{\rm link}}(w) = \sum_{i \in \theArcsb w} 2^{-i}.
\ee
Clearly, two links share the same dyadic fraction if and only if they are identical. The link basis $\mathcal B_{n,d}$ is then ordered 
by increasing values of $\mathsf d_{{\rm link}}(w)$. The basis \eqref{eq:V60basis} of $\Stan{6,0}$ is in fact presented in this order, 
with the corresponding values of $\mathsf d_{{\rm link}}(w)$ given by  
$\left\{\frac{21}{32},\frac{22}{32},\frac{25}{32},\frac{26}{32},\frac{28}{32}\right\}$. 

Similarly, we associate the dyadic fraction 
\be
 \mathsf d_{{\rm spin}}(s) =\sum_{i=1}^{n-1}\,2^{-i}\delta_{s_i,-} 
\ee 
to $|s\rangle$ in the spin 
basis of $\eigenSz{n-1}{(d-1)/2}$, and order the basis with increasing dyadic fractions. For example, the basis \eqref{eq:spinbasis} is 
already ordered accordingly and the corresponding values of $\mathsf d_{{\rm spin}}(s)$ are
$\{\tfrac{7}{32},\tfrac{11}{32},\tfrac{13}{32},\tfrac{14}{32},\tfrac{19}{32},\tfrac{21}{32},\tfrac{22}{32},\tfrac{25}{32},\tfrac{26}{32},\tfrac{28}{32}\}.$ 

\begin{Proposition}The linear map $f_{n,d}:\Stan{n,d}\rightarrow
\eigenSz{n-1}{(d-1)/2}$ is injective.
\label{sec:f.injective}
\end{Proposition}
\noindent{\scshape Proof }
A direct computation shows that 
\be 
\dim\eigenSz{n-1}{(d-1)/2}-\dim\Stan{n,d} 
={n-1 \choose \frac{n-d-4}2} = \dim\eigenSz{n-1}{(d+3)/2},
\ee
which is nonnegative. This property is a necessary condition for the injectivity of $f_{n,d}$. 

Two observations about the image of a link $w\in\mathcal B_{n,d}$ under $f_{n,d}$ are crucial. 
Let $|s_w\rangle$ be the spin state with down spins at positions labeled by elements of $\theArcsb w$. The link $w$ and the corresponding spin state
$|s_w\rangle$ have equal dyadic fractions, $\mathsf d_{{\rm link}}(w) = \mathsf d_{{\rm spin}}(s_w)$. First, 
in $f_{n,d}(w)=\sum_s c_s|s\rangle$, the coefficient $c_{s_w}$ of $|s_w\rangle$ is $1$ because all arcs contain precisely the minimum number of down spins required for a 
nonzero coefficient, that is, $m_{s_w}^{i,j}=1$ for all arcs $(i,j)$ in $\theArcs{w}$. Second, all other nonzero coefficients $c_{s'}$ are 
associated to spin states $|s'\rangle$ that precede $|s_w\rangle$ in the ordered basis. Indeed, because each down spin in $|s_w\rangle$ is located
in the leftmost possible position $i$ under an arc $(i,j)$, 
any other $|s'\rangle$ contributing to $f_{n,d}(w)$ will have at least one of its down spins further to the right, compared to those of $|s_w\rangle$, and will thus correspond to a smaller dyadic fraction: 
$\mathsf d_{{\rm spin}}(s') < \mathsf d_{{\rm spin}}(s_w)$.

The map $f_{n,d}$ is injective if and only if the columns in its matrix representation are linearly independent. Suppose there exists a 
linear combination $\sum_{v}\alpha_vf_{n,d}(v)$ of these columns that is zero and let $f_{n,d}(w)$ be the leading column appearing in 
this linear combination, i.e.~the one for which the value of $\mathsf d_{{\rm link}}(w)$ is maximal. By the previous two observations, it is 
the only column of the sum that contains a nonzero matrix element at position $|s_w\rangle$. Its weight $\alpha_{w}$
must therefore be zero. The same argument shows that all terms in the linear combination are zero from which it follows 
that $f_{n,d}$ is injective.  \hfill $\square$

%
\section{Diagonalisability and reality of spectra}
\label{sec:diagandreal}
%

\subsection{Proof of Theorem~\ref{thm:thm0}}
\label{sub:proof}

{\scshape Proof } Section~\ref{sec:strategy} contains all the elements needed to prove Theorem~\ref{thm:thm0} presented in the introduction.
An inner product $S_{n,d}$ was thus constructed for each standard module $\Stan{n,d}$ over $\tl n(\beta)$, for $\beta \in \mathbb R$ and $n \ge 2$, and is built as $S_{n,d} = f_{n,d}^\dagger\,f_{n,d}$ using the intertwiner $f_{n,d}$ introduced in Section~\ref{sec:fnd}. The crucial intertwining property, $f_{n,d}\,\hloop_{n,d}=\hspin_{n-1,d}\, f_{n,d}$, between the loop Hamiltonian $\hloop_{n,d}$ and the spin-chain Hamiltonian $\hspin_{n-1,d}$ is the content of Proposition~\ref{sec:int}, while Proposition~\ref{sec:f.injective} in Section~\ref{sec:injectivity} asserts
that $f_{n,d}$ is injective. From the discussion in Section~\ref{sec:newH}, these properties
imply that $\hloop_{n,d}$ is self-adjoint with respect to the inner product $S_{n,d}$. 
This is the content of Theorem~\ref{thm:thm0} and thus completes its proof.\hfill$\square$

\medskip 

As an immediate consequence of Theorem~\ref{thm:thm0}, we see that $\hloop_{n,d}$ is diagonalisable and has a real spectrum for 
$0\le  d\le  n$, $d\equiv n\: \mathrm{mod}\: 2$, $n\ge 2$ and $\beta\in\mathbb{R}$.
As already mentioned in the introduction, this result was conjectured in~\cite{PRZ06}.

\subsection{Corollaries of Theorem~\ref{thm:thm0}}
\label{sub:consequences}

This subsection discusses some of the important corollaries of Theorem~\ref{thm:thm0},
in particular Theorem~\ref{thm:Theorem2} and Theorem~\ref{cor:cor1} presented in the introduction.

\begin{Proposition}\label{cor:Irre} 
The action of $h$ on any irreducible module $\Irre{n,d}$ is diagonalisable and has a real spectrum.
\end{Proposition}
\noindent{\scshape Proof } If $\tl n$ is semi-simple or $d$ critical, the statement is covered by Theorem~\ref{thm:thm0}. 
Let $\tl n$ be non-semi-simple and
$d$ be an element of a non-critical orbit.
Theorem \ref{thm:thm0} says that there exists an inner product $(\ |\ )_{n,d}$ on the standard 
module $\Stan{n,d}$ such that $(v|hw)=(hv|w)$ for all $v,w\in\Stan{n,d}$. This implies that the matrix $\hloop_{n,d}$ is diagonalisable 
with real eigenvalues. Since the restriction of an inner product to a subspace is likewise an inner product on the subspace, the statement 
follows for all irreducible modules $\Irre{n,d}$ that are submodules of a standard one, namely for all $\Irre{n,d}$ with $d\neq d_l$. 
Another argument is needed for $\Irre{n,d_l}$. The following one actually holds for all irreducible modules.

Since $\hloop_{n,d}$ is diagonalisable, its minimal polynomial is $p(x) = \prod_i (x - \lambda_i)$ where $\lambda_i$ runs over its (real) 
distinct eigenvalues (i.e.~each linear factor appears only once). Because of the reducible
module structure of $\Stan{n,d}$ given in \eqref{eq:Loewystandard}, one can form a basis of $\Stan{n,d}$ 
by starting with a basis of $\Irre{n,d_+}$ and completing it. In this basis, $\hloop_{n,d}$ is upper block triangular,
\be
\hloop_{n,d} = \begin{pmatrix} \hloop_{1} & X \\ 0 & \hloop_{2} \end{pmatrix},
\ee
where $\hloop_1$ and $\hloop_2$ are the actions of $h$ on $\Irre{n,d_+}$ and $\Irre{n,d} \simeq \Stan{n,d}/\Irre{n,d_+}$, respectively. The set of eigenvalues of $\hloop_{n,d}$ is then the union of those of $\hloop_1$ and $\hloop_2$, which are thus real. It also follows that
\be
p(\hloop_{n,d}) = \begin{pmatrix} p(\hloop_1) & Y \\ 0 & p(\hloop_2) \end{pmatrix} = 0,
\ee 
where $Y$ is a function of $\hloop_1$, $\hloop_2$ and $X$. This implies that $p(\hloop_1) = p(\hloop_2) = 0$ and that neither $\hloop_1$ nor $\hloop_2$ has nontrivial Jordan blocks. 
Because every irreducible module $\Irre {n,d}$ appears as the quotient of the
corresponding standard module $\Stan {n,d}$, and because the reasoning above applies to all standard modules, 
this completes the proof. \hfill $\square$

\medskip

\noindent{\scshape Proof of Theorem~\ref{thm:Theorem2} } 
By the Jordan-H\"older theorem, every (finite-dimensional) module over $\tl n$
has a composition series. It follows that there exists a basis in which all elements of 
$\tl n$ are represented by upper block triangular matrices with the diagonal blocks isomorphic to irreducible representations. 
The spectrum of $h$ in these representations is the union of the spectra of its diagonal blocks, that is, the union of the spectra on 
its irreducible composition factors. These spectra are real according to Proposition~\ref{cor:Irre}, 
so the spectrum of $h$ in any representation is real. \hfill$\square$

\medskip

Theorem~\ref{cor:cor1} is an immediate consequence of Theorem~\ref{thm:Theorem2}.
Indeed, the Hamiltonian $\hxxz$ is merely the matrix representative of $h$ in the representation $\chit$ (whose decomposition was discussed in Section~\ref{sub:decomposition}). 
Theorems~\ref{thm:Theorem2} and \ref{cor:cor1} do not preclude the possibility of nontrivial Jordan blocks in $\hxxz$. 
In fact, the Hamiltonian $\hxxz$ does have such Jordan blocks for $q$ a root of unity and $n$ large enough.
The simplest example is for $n=2$ at $q=\pm i$, in which case $\hxxz = - \chit(e_1)$ is a $4\times 4$ matrix. 
Indeed, because $e_1^2=\beta e_1$ and $\beta=q+q^{-1}=0$, $\hxxz$ is nilpotent but nonzero and must therefore 
have at least one nontrivial Jordan block. 
A simple direct computation shows that the number of nontrivial Jordan blocks is exactly one in this case, and that it is of rank $2$.

\subsection[Diagonalisability of $h$ on indecomposable modules]{Diagonalisability of $\boldsymbol h$ on indecomposable modules}
\label{sub:indec}

Unless $\beta=0$ and $n=2$ (see below), the set of indecomposable modules over $\tl n(\beta)$ can be organised in two disjoint 
families: the zigzag modules and the projective ones with more than two composition factors. 
The structure and Loewy diagrams of these projective modules are described in Section~\ref{sec:Reps}, while the zigzag modules
have Loewy diagrams of the form
\be
\begin{pspicture}[shift=-0.9](-0.2,-0.2)(7.1,1.4)
\rput(0,1.2){$\Irre{n,d}$}
\rput(1.2,0){$\Irre{n,d_+}$}
\rput(2.4,1.2){$\Irre{n,d_{++}}$}
\rput(3.6,0){$\dots$}
\psline[linewidth=\elegant]{->}(0.3,0.9)(0.9,0.3)
\psline[linewidth=\elegant]{->}(2.1,0.9)(1.5,0.3)
\psline[linewidth=\elegant]{->}(2.7,0.9)(3.3,0.3)
\psline[linewidth=\elegant]{->}(4.5,0.9)(3.9,0.3)
\rput(4.8,1.2){$\Irre{n,d'_{--}}$}
\rput(6.0,0){$\Irre{n,d'_-}$}
\rput(7.2,1.2){$\Irre{n,d'}$}
\psline[linewidth=\elegant]{->}(5.1,0.9)(5.7,0.3)
\psline[linewidth=\elegant]{->}(6.9,0.9)(6.3,0.3)
\end{pspicture}
\ee
for some $d, d'$ from the same non-critical orbit. The family of zigzag modules also includes the cases where one or both endpoints, $\Irre{n,d}$ and $\Irre{n,d'}$, are in the socle, i.e.~modules in the form of
$\psset{unit=0.28cm}
\begin{pspicture}[shift=0](-0.1,0)(4.4,1)
\psline[linewidth=\elegant]{-}(0,1)(0.8,0)(1.6,1)(2.4,0)(3.2,1)(4.0,0)
\end{pspicture}$, 
$\psset{unit=0.28cm}
\begin{pspicture}[shift=0](0.7,0)(5.2,1)
\psline[linewidth=\elegant]{-}(0.8,0)(1.6,1)(2.4,0)(3.2,1)(4.0,0)(4.8,1)
\end{pspicture}$
and 
$\psset{unit=0.28cm}
\begin{pspicture}[shift=0](-0.1,0)(5.2,1)
\psline[linewidth=\elegant]{-}(0,0)(0.8,1)(1.6,0)(2.4,1)(3.2,0)(4.0,1)(4.8,0)
\end{pspicture}$.
The restriction on the number of composition factors in the projective modules ensures that there is no 
overlap between the two families. In this classification, irreducible modules are zigzag modules with a single composition factor, 
whereas reducible standard modules have two. In the exceptional case $\tl {n=2}(\beta=0)$, the projective cover of $\Irre{2,2}$ is not 
classified as indicated. Its Loewy diagram is given to the right in (\ref{proj}).

As demonstrated above, for $\beta\in\mathbb{R}$, 
the matrix representative of $h$ on an irreducible or a standard module is diagonalisable with a real spectrum.
In fact, a simple argument shows that this property of $h$ extends to all zigzag modules.
We begin by considering the co-standard module $\Stan{n,d}^\star$, the module contragredient to $\Stan{n,d}$.
Because irreducible modules over $\tl n$ are self-contragredient, the Loewy diagram of $\Stan{n,d}^\star$ is given by
\be
\begin{pspicture}[shift=-0.7](-0.2,-0.2)(1.4,1.4)
\rput(0,0){$\Irre{n,d}$}
\rput(1.2,1.2){$\Irre{n,d_+}$}
\psline[linewidth=\elegant]{->}(0.9,0.9)(0.3,0.3)
\end{pspicture} \ .
\ee
The matrix representative of $a \in \tl n$ on $\Stan{n,d}^\star$ is the transpose of the matrix 
representative of $a^\dagger$ on $\Stan{n,d}$, where $a^\dagger$ is the
{\it dual} of $a$, obtained by reversing the order of composition in products of Temperley-Lieb generators,
\be
 a=\sum_k\sum_{i_1,i_2,\ldots,i_k}\alpha_{i_1,i_2,\dots, i_k}e_{i_1}e_{i_2}\ldots e_{i_k}\qquad \rightarrow\qquad 
  a^\dagger=\sum_k\sum_{i_1,i_2,\ldots,i_k}\alpha_{i_1,i_2,\dots, i_k}e_{i_k}\ldots e_{i_2}e_{i_1},
\ee
where $\alpha_{i_1,i_2,\dots, i_k}\in\mathbb C$. 
Because $e_j^\dagger=e_j$ and hence $h^\dagger=h$, the Hamiltonian on $\Stan{n,d}^\star$ is 
$\hloop_{n,d}^{\mathrm{T}}=\hloop_{n,d}^\dagger$, and is thus also diagonalisable with real eigenvalues. 

A zigzag module with three composition factors has a Loewy diagram of the form
\be
\begin{pspicture}[shift=-0.7](-0.2,-0.2)(2.6,1.4)
\rput(0,1.2){$\Irre{n,d}$}
\rput(1.2,0){$\Irre{n,d_+}$}
\rput(2.4,1.2){$\Irre{n,d_{++}}$}
\psline[linewidth=\elegant]{->}(0.3,0.9)(0.9,0.3)
\psline[linewidth=\elegant]{->}(2.1,0.9)(1.5,0.3)
\end{pspicture}
\qquad {\rm or}\qquad
\begin{pspicture}[shift=-0.7](-0.2,-0.2)(2.6,1.4)
\rput(0,0){$\Irre{n,d}$}
\rput(1.2,1.2){$\Irre{n,d_+}$}
\rput(2.4,0){$\Irre{n,d_{++}}$}
\psline[linewidth=\elegant]{->}(0.9,0.9)(0.3,0.3)
\psline[linewidth=\elegant]{->}(1.5,0.9)(2.1,0.3)
\end{pspicture}\ .
\ee
The module
$\mathsf M=\psset{unit=0.28cm}
\begin{pspicture}[shift=0](0,0)(2,1)
\psline[linewidth=\elegant]{-}(0,1)(0.8,0)(1.6,1)
\end{pspicture}$\!
has submodules isomorphic to $\Stan{n,d}$ and $\Stan{n,d+}^\star$, and eigenvectors of the action of $h$ on either are thus eigenvectors of the action of $h$ on $\mathsf M$. Since, as 
a vector space, $\mathsf M$ is the sum $\Stan{n,d}+\Stan{n,d+}^\star$, it is possible to extract a basis for $\mathsf M$ from bases of eigenvectors of $h$ on $\Stan{n,d}$ and $\Stan{n,d+}^\star$. 
Thus, $h$ is diagonalisable with real eigenvalues on $\mathsf M$ as well.
Because the module 
$\psset{unit=0.28cm}
\begin{pspicture}[shift=0](0,0)(2,1)
\psline[linewidth=\elegant]{-}(0,0)(0.8,1)(1.6,0)
\end{pspicture}$\!
is contragredient to 
$\psset{unit=0.28cm}
\begin{pspicture}[shift=0](0,0)(2,1)
\psline[linewidth=\elegant]{-}(0,1)(0.8,0)(1.6,1)
\end{pspicture}$\!,
the matrix representative of $h$ on
$\psset{unit=0.28cm}
\begin{pspicture}[shift=0](0,0)(2,1)
\psline[linewidth=\elegant]{-}(0,0)(0.8,1)(1.6,0)
\end{pspicture}$\!
is the transpose of the matrix representative of $h$ on
$\psset{unit=0.28cm}
\begin{pspicture}[shift=0](0,0)(2,1)
\psline[linewidth=\elegant]{-}(0,1)(0.8,0)(1.6,1)
\end{pspicture}$\!
and is therefore diagonalisable with real eigenvalues.
Similar arguments are used recursively for larger and larger zigzag modules to show that $h$ is diagonalisable with a real spectrum on each of them.

There nevertheless exist Temperley-Lieb modules on which $h$ is non-diagonalisable. This was observed in~\cite{PRZ06} and conjectured to hold in subsequent works on logarithmic minimal models. To see this, we merely have to consider $h$ acting on the exceptional projective cover of $\Irre{2,2}$ for $\tl 2(0)$, 
which is precisely the rank-2 Jordan block contained in $\hxxz$ for $n=2$ and $\beta=0$ (in the zero magnetisation sector), as discussed at the end of Section~\ref{sub:consequences}.

For general $\tl n(\beta)$, let us consider a projective module $\Proj{n,d}$ with more than two composition factors. According to the discussion in Section~\ref{sec:Reps}, it has three or four composition factors and its socle and head are both isomorphic to the irreducible module $\Irre{n,d}$.

If the action of $h$ on $\Proj{n,d}$ has a rank-$2$ Jordan block, then the eigenvector and its Jordan partner cannot be both in the maximal submodule (the bottom zigzag module
$\psset{unit=0.28cm}
\begin{pspicture}[shift=0](0,0)(1.8,1)
\psline[linewidth=\elegant]{-}(0,1)(0.8,0)(1.6,1)
\end{pspicture}$). The projection of the Jordan partner on the head of $\Proj{n,d}$ is therefore non-zero. Similarly, the projection of the eigenvector of this rank-$2$ Jordan block on $\Proj{n,d}/\Irre{n,d}$ must be zero, since this quotient is also a zigzag module,~$\psset{unit=0.28cm}
\begin{pspicture}[shift=0](0,0)(1.8,1)
\psline[linewidth=\elegant]{-}(0,0)(0.8,1)(1.6,0)
\end{pspicture}$. This eigenvector thus belongs to the socle. It follows that a rank-$2$ Jordan block in $\Proj{n,d}$, if any, ``ties'' the head and socle. The same argument also prevents the possibility of higher-rank Jordan blocks.
Explorations for small $n$ suggest that every eigenvector in the socle has a Jordan partner, 
an observation previously used in~\cite{RP07} to infer the structure of the limiting Virasoro modules. These explorations thus suggest that the number of rank-$2$ Jordan blocks of $h$ on $\Proj{n,d}$ is $\dim \Irre{n,d}$.

%
\section{Discussion} 
\label{sec:conclusion}
%

Let us review the main results. For each standard module $\Stan{n,d}$ over $\tl n(\beta)$, 
in Section~\ref{sec:strategy}, an inner product $(\ |\ )_{n,d}$ is constructed with respect to which the Hamiltonian $h$ is self-adjoint. This is the content of 
Theorem~\ref{thm:thm0}. From this construction, it immediately follows that the representative of $h$ on 
$\Stan{n,d}$, denoted by $\hloop_{n,d}$, is diagonalisable and has a real spectrum for all $n,d$ and $\beta\in\mathbb{R}$. 
Theorem~\ref{thm:Theorem2} then extends the reality of the spectra of $h$ to all Temperley-Lieb modules. 
Finally, Theorem~\ref{cor:cor1} is merely a specialisation of Theorem~\ref{thm:Theorem2} to the XXZ spin-chain representation $\chit$, 
thus establishing that the $U_q(sl_2)$-invariant Hamiltonian $\hxxz$ has real spectra for all $n \ge 2$ and $q+q^{-1} \in \mathbb R$. 

XXZ Hamiltonians with real boundary fields have been studied before, for instance by Yang and Fendley \cite{YF04} as well as by Nichols, Rittenberg and de Gier~\cite{NRdG05} who related them with the one-boundary Temperley-Lieb algebra. The connection between $\hspin_{n-1}$ and $\tl n$, however, seems new.
As our proof of the three theorems is based on 
this particular spin-chain Hamiltonian $\hspin_{n-1}$ and the nontrivial intertwiner $f_{n,d}$ (whose key properties were established using intricate technical manipulations), the reader may wonder how we came about this 
construction. This is briefly outlined in the next two paragraphs.

For $\beta = 0$, the map $f_{n,d}$ intertwines the standard loop hamiltonian $\hloop_{n,d}$ and $\hspin_{n-1}|_{\beta = 0}$, 
an XX Hamiltonian on the open spin chain of length $n-1$ with no boundary magnetic fields. In our recent study~\cite{MDRR14} 
of the dimer model, $\hspin_{n-1}|_{\beta = 0}$ was shown to belong to a representation of $\tl n(0)$ on $\ctwotimes {n-1}$. 
As one can show, in this case, $f_{n,d}$ is actually a Temperley-Lieb homomorphism between $\Stan{n,d}$ and the representation 
$\tau^\star$ on $\ctwotimes {n-1}$, 
where $\tau^\star$ is 
the contragredient of $\tau$, see (\ref{eq:tau}), defined by $\tau^\star(e_j)=\big(\tau(e_j)\big)^T$. The intertwining property 
$f_{n,d}\,\hloop_{n,d} = \hspin_{n-1,d}\,f_{n,d}$ can therefore be established in a straightforward way because the same property holds 
with the Hamiltonians replaced by any of the Temperley-Lieb generators $e_j$. 

The generalisation to $\beta\neq0$ was achieved in two steps. First, for $\beta \in \mathbb R$, we searched for a spin-chain Hamiltonian on $\ctwotimes{n-1}$ that satisfies the same Bethe ansatz equations as $\hxxz$,
and found $\hspin_{n-1}$. Second, the corresponding intertwiner $f_{n,d}$ was constructed as a solution to the intertwining relation $f_{n,d}\,\hloop_{n,d} = \hspin_{n-1,d}\,f_{n,d}$, initially for small system sizes, based on which we subsequently guessed, and ultimately proved, the general form described in Section~\ref{sec:fnd}. 
Physical interpretations of $f_{n,d}$ and $S_{n,d}$ are not clear at this point.

We note that the new inner product $(\ |\ )_{n,d}$ is in general not the only bilinear form with respect to which $h$ is self-adjoint. 
Indeed, the well-known Gram product $\langle\ |\ \rangle_{n,d}$ is a bilinear form on standard modules,
having the even stronger property $\langle v| e_jw\rangle_{n,d} =\langle e_j v| w\rangle_{n,d}$, $j=1,2,\ldots,n-1$. It plays a key role in the Temperley-Lieb representation theory~\cite{W95,RSA14}. In sharp contrast to the bilinear form constructed in this paper, however, the Gram product is not in general an inner product.
The determinant of its matrix realisation $\mathcal G_{n,d}$ evaluates to zero at a finite number of roots of unity 
(except for $d=n$, in which case $\det \mathcal G_{n,d}=1$). Normalised by the appropriate power of $\beta$, $\mathcal G_{n,d}$ 
tends to an identity matrix as $\beta \rightarrow \infty$, so all eigenvalues of $\mathcal G_{n,d}$ are positive for $\beta$ large enough. 
Let $\beta_c$ denote the largest $\beta$ for which $\det \mathcal G_{n,d}=0$. 
Then the eigenvalues of $\mathcal G_{n,d}$ are all positive for $\beta > \beta_c$. On this semi-infinite interval, 
all words $a \in \tl n$ that are invariant under vertical flips are self-adjoint with respect to the Gram product. 
This shows that, on the interval $\beta >\beta_c$, there are at least two positive-definite bilinear forms with respect to which $h$ is self-adjoint, 
namely $(\ |\ )_{n,d}$ and $\langle\ |\ \rangle_{n,d}$.

A significant part of the interest in many lattice models lies in their continuum scaling limits. For models built from diagrammatic algebras, 
features seen in this limit are often present in some form at the lattice level as well \cite{PS90,PRZ06,KS94,GRS13op1,GRS13op2}.
This typically includes the indecomposability of
representations, in particular the appearance of nontrivial Jordan blocks in the evolution operators. 
As Theorem~\ref{thm:thm0} proves a 2006 conjecture by Pearce, Rasmussen and Zuber~\cite{PRZ06} on the diagonalisability 
and reality of spectra of $h$ on standard modules, it also gives credence to the subsequent inference
that the limiting operator $L_0$ is diagonalisable on the corresponding Virasoro modules. The reality of spectra for the XXZ spin chains, given in Theorem~\ref{cor:cor1}, should also persist in the continuum scaling limit.
This theorem does not address the issue of diagonalisability, discussed instead in Section~\ref{sub:indec}.
In fact, the Virasoro mode $L_0$ is expected to exhibit nontrivial
Jordan blocks in the continuum scaling limit of the XXZ spin chains.

The role and properties of the new scalar product $(\ |\ )_{n,d}$ in the continuum scaling limit are unclear. 
In particular, the positive-definiteness of $(\ |\ )_{n,d}$ holds for all standard modules $\Stan{n,d}$, even when they are reducible yet indecomposable. It would be interesting to determine whether this property persists in the continuum scaling limit, and whether the limiting bilinear form is related to invariant forms like the Shapovalov form for Verma modules.

This paper opens several avenues for further investigation of the Temperley-Lieb algebras and related diagram algebras.
An interesting question regards the possibility of extending our analysis and results for the Temperley-Lieb Hamiltonians
to the double-row transfer matrices~\cite{PRZ06} which depend on a spectral parameter $u$ and in general also a set of
inhomogeneities $\xi_j, j=1, \dots, n$. If at all, these matrices may be self-adjoint with respect to $(\ |\ )_{n,d}$ for specific choices of the parameters only. 
Another interesting problem is to determine to what extent our results and approach can be applied to other algebras such as 
the blob algebras~\cite{MS93}, the dilute Temperley-Lieb algebras~\cite{BN89,BSA13}, 
the Birman-Wenzl-Murakami algebras~\cite{M87,BW89} and the Fuss-Catalan algebras~\cite{BJ97}. 
Considering the complexity of the proof of Theorem~\ref{thm:thm0} for the 
ordinary Temperley-Lieb algebras, an extension to any of these more complicated algebras is likely to be quite involved. 
Results in this direction would nevertheless shed important new light on these algebras and the statistical models they describe.

\subsection*{Acknowledgments}

AMD and YSA are supported by the National Sciences and Engineering Research Council of Canada, AMD by a Postdoctoral 
Fellowship and YSA by a Discovery Grant. JR is supported by the Australian Research Council under the Future Fellowship scheme, 
project number FT100100774. PR is Senior Research Associate of the Belgian Fonds National de la Recherche Scientifique (FNRS). 
PR and AMD acknowledge the support of the Belgian Interuniversity Attraction Poles Program P7/18 through the network DYGEST 
(Dynamical, Geometry and Statistical Physics). The authors thank Christian Korff and Robert Weston for helpful discussions about their 
paper~\cite{KW07}. AMD also thanks Christian Hagendorf for useful discussions, Jan de Gier for pointing out the references~\cite{YF04,NRdG05} and Vladimir Rittenberg for reviving his interest in this problem.

\bigskip

\bigskip

\appendix

%
\section{Proof of Proposition~\ref{sec:int}}
\label{sec:proof}
%

\subsection{Preliminaries}
\label{sec:prelims}

The objective of this appendix is to prove the intertwining relation
\be
\mathbb H_{n-1,d}f_{n,d}(w)-f_{n,d}(hw) =0, \qquad w \in \Stan{n,d},\qquad 0\le d \le n, \qquad d\equiv n \:{\rm mod}\: 2,
 \qquad n \ge 2,
\label{eq:intertwining}
\ee
stated in Proposition~\ref{sec:int}. 
Here $h w$ denotes the standard action of the Temperley-Lieb Hamiltonian $h$ on $w \in \Stan{n,d}$.
Because the labels $n-1$ and $d$ of $\hspin_{n-1,d}$
can be determined by the states it acts on, these labels are suppressed in the following and $\hspin_{n-1,d}$ is simply denoted by $\hspin$. 

The proof is by induction on $n$. The case $n=2$ for both $d=0$ and $2$ is easy as all matrices are $1\times 1$. Indeed, $\mathbb H_{1,0}=\hloop_{2,0}=(-\beta)$, $\mathbb H_{1,2}=\hloop_{2,2}=(0)$ and $f_{2,0}=f_{2,2}=(1)$, and \eqref{eq:intertwining} is verified.

Hereafter, we occasionally denote links $w \in \mathcal B_{n,d}$ by
$ 
\psset{unit=0.5}
\begin{pspicture}[shift=-0.05](-0.0,0)(1.5,0.5)
\psline{-}(0,0)(1.5,0)
\pspolygon[fillstyle=solid,fillcolor=white](0.2,0)(1.3,0)(1.3,0.6)(0.2,0.6)
\rput(0.75,0.3){$_w$}
\end{pspicture}\,$. 
These can be grouped into two families,
\be
\textrm{\bf Family I:} \qquad
w = \,\psset{unit=0.5}
\begin{pspicture}[shift=-1.00](-0.4,-1.0)(1.9,1.2)
\psline{-}(-0.4,0)(1.9,0)
\pspolygon[fillstyle=solid,fillcolor=white](0.2,0)(1.3,0)(1.3,0.6)(0.2,0.6)
\psbezier[linecolor=blue,linewidth=1.5pt](-0.2,0)(-0.2,1.5)(1.7,1.5)(1.7,0)
\rput(0.75,0.3){$_{w_1}$}
\rput(0.0,-0.4){$_\uparrow$}\rput(0.0,-1.0){$_1$}
\rput(1.5,-0.4){$_\uparrow$}\rput(1.5,-1.0){$_{n-1}$}
\end{pspicture}\ ,
\qquad \quad \textrm{\bf Family II:} \qquad
w = \,
\begin{pspicture}[shift=-1.00](-0.0,-1.0)(3.3,1.2)
\psline{-}(0,0)(3.3,0)
\pspolygon[fillstyle=solid,fillcolor=white](0.2,0)(1.3,0)(1.3,0.6)(0.2,0.6)
\rput(0.75,0.3){$_{w_1}$}
\pspolygon[fillstyle=solid,fillcolor=white](1.6,0)(3.1,0)(3.1,0.6)(1.6,0.6)
\rput(2.35,0.3){$_{w_2}$}
\rput(1.45,-0.4){$_\uparrow$}\rput(1.45,-1.0){$_i$}
\end{pspicture}\ ,
\label{eq:families}
\ee
which our analysis will distinguish. The integers appearing below $w$ indicate the labels $j=1,\dots, n-1$ of the 
spins when drawn under $w$, with $s_j$ appearing between the nodes $j$ and $j+1$ in the construction discussed in 
Section~\ref{sec:fnd}. For both families, the induction assumption is 
that \eqref{eq:intertwining} holds on $w_1$ and $w_2$ whose respective lengths
are less than $n$. It is convenient to let the notation cover extreme cases such as 
$w = \psset{unit=0.6}\begin{pspicture}[shift=0](0,0)(0.8,0.4)
\psline{-}(0,0)(0.8,0)
\psarc[linecolor=blue,linewidth=1.5pt](0.4,0){0.2}{0}{180}
\end{pspicture}$ in Family I, in which case we say that $w_1$ is of length zero.

For the recursive argument, the spin-chain Hamiltonian is split as 
\begin{alignat}{2}
&\textrm{\bf Family I:} \qquad &&\hspin = - \mathbb H_{(1)} 
+ \hspin_C - \mathbb H_{(n-1)},
\label{eq:Hspindec1}
\\ &\textrm{\bf Family II:} \qquad &&\hspin = \hspin_L - \mathbb H_{(i)} + \hspin_R,\label{eq:Hspindec2}
\end{alignat}
where $\mathbb H_{(i)}$ is minus the sum of the operators in \eqref{eq:hspin} affecting the $i$-th spin, that is
\be
\mathbb H_{(i)} = \sigma^-_{i-1}\sigma^+_i + \sigma^+_{i-1}\sigma^-_{i} + \sigma^-_{i}\sigma^+_{i+1} + \sigma^+_{i}\sigma^-_{i+1} + \{2\} \big( \sigma^-_{i-1}\sigma^+_{i-1}\sigma^-_i\sigma^+_i + \sigma^-_{i}\sigma^+_{i}\sigma^-_{i+1}\sigma^+_{i+1}
   -\sigma^-_i\sigma^+_i\big).
\label{hi}
\ee
Here, the convention $\sigma^\pm_0 = \sigma^\pm_n = 0$ is used and implies that $\mathbb H_{(1)}$ and $\mathbb H_{(n-1)}$ contain four terms each instead of seven as in the general case $2 \le i \le n-2$.
Crucially, the full Hamiltonian $\hspin$ {\it does not} equal $-\sum_i \mathbb H_{(i)}$. 
The remaining $\hspin_L$, $\hspin_C$ 
and $\hspin_R$ act as $\hspin$ on each their set of consecutive spins, and as the identity on the remaining spins.

The Temperley-Lieb Hamiltonian is likewise split,
\begin{alignat}{2}
&\textrm{\bf Family I:} \qquad &&h = - e_1 + h_C-e_{n-1},\label{eq:Hloopdec1}
\\ &\textrm{\bf Family II:} \qquad &&h = h_L-e_i + h_R,\label{eq:Hloopdec2}
\end{alignat}
where 
\be 
h_L = - \sum_{j = 1}^{i-1}e_j, \qquad 
h_C =  - \sum_{j = 2}^{n-2} e_j,\qquad 
h_R = -\! \sum_{j = i+1}^{n-1}e_j. 
\ee
Each of these Hamiltonian operators acts as the original Hamiltonian $h$ on a set of consecutive nodes 
and as the identity on the remaining ones. For instance,
\be
\psset{unit=0.5}
h_R \,
\begin{pspicture}[shift=-0.1](-0.0,0)(3.3,0.6)
\psline{-}(0,0)(3.3,0)
\pspolygon[fillstyle=solid,fillcolor=white](0.2,0)(1.3,0)(1.3,0.6)(0.2,0.6)
\rput(0.75,0.3){$_{w_1}$}
\pspolygon[fillstyle=solid,fillcolor=white](1.6,0)(3.1,0)(3.1,0.6)(1.6,0.6)
\rput(2.35,0.3){$_{w_2}$}
\end{pspicture}\, = 
\begin{pspicture}[shift=-0.1](-0.0,0)(3.3,0.6)
\psline{-}(0,0)(3.3,0)
\pspolygon[fillstyle=solid,fillcolor=white](0.2,0)(1.3,0)(1.3,0.6)(0.2,0.6)
\rput(0.75,0.3){$_{w_1}$}
\pspolygon[fillstyle=solid,fillcolor=white](1.6,0)(3.1,0)(3.1,0.6)(1.6,0.6)
\rput(2.35,0.3){$_{h w_2}$}
\end{pspicture}\, .
\ee

In the proof of (\ref{eq:intertwining}), we shall distinguish between the cases $d=0$ and $d>0$. 
The former is independent of the latter, but not the other way around. 
The analysis of the case $d=0$ is based on a family of 
maps $g^p$ labeled by an integer $p$ and constructed from $f_{n,d}$. 
They are defined in Section~\ref{sec:gnp} 
and shown in Section~\ref{sec:theproof} 
to satisfy a modified intertwining property, 
see Proposition~\ref{sec:toprovep}. The intertwining property 
\eqref{eq:intertwining} for $d = 0$ is then a specialisation of Proposition~\ref{sec:toprovep} to the case $p=0$.
Although the case $d>0$ does not require the new maps $g^p$, it uses the injective maps $f_{n,d}$ and induction 
on $n$ and thus relies on the validity of \eqref{eq:intertwining} for $d=0$ and $n'<n$. 
However, because its proof is simpler, we choose to present the case $d>0$ first (in Section~\ref{sec:d>0}), 
assuming that \eqref{eq:intertwining} holds for $d = 0$.

\subsection[The case $d>0$]{The case $\boldsymbol{d>0}$}
\label{sec:d>0}

By $(n,d)$, we mean a pair of integers as in (\ref{eq:intertwining}),
\be
 (n,d):\qquad 0\le d \le n, \qquad d\equiv n \:{\rm mod}\: 2,\qquad n \ge 2.
\ee
To establish the induction step in the proof of \eqref{eq:intertwining} for $d>0$, 
the next proposition shows that \eqref{eq:intertwining} holds for the pair $(n,d)$ with $d>0$, assuming that it holds for each pair 
$(n',d')$ with $n'<n$. 

\begin{Proposition} If the intertwining relation \eqref{eq:intertwining} holds for all pairs $(n',d')$ with $n'<n$, then it holds for the pair 
$(n,d)$ with $d>0$.
\end{Proposition}
{\scshape Proof } The case $d=n$ is trivial, as both $\hspin\hs f_{n,d}(w)$ and $f_{n,d}(h w)$ are zero. 
For $1\le d\le n-2$, all links $w\in\mathcal B_{n,d}$ can be written as
\be
w = \psset{unit=0.5}
 \,
\begin{pspicture}[shift=-1.05](0,-1.0)(3.3,1.2)
\psline{-}(0,0)(3.3,0)
\pspolygon[fillstyle=solid,fillcolor=white](0.2,0)(1.3,0)(1.3,0.6)(0.2,0.6)
\rput(0.75,0.3){$_{w_1}$}
\pspolygon[fillstyle=solid,fillcolor=white](1.6,0)(3.1,0)(3.1,0.6)(1.6,0.6)
\rput(2.35,0.3){$_{w_2}$}
\rput(1.45,-0.4){$_\uparrow$}\rput(1.45,-1.0){$_i$}
\end{pspicture}\,,
\qquad w_1 \in \Stan{i,d_1}, \qquad w_2 \in \Stan{n-i,d_2}
\label{eq:ww1w2}
\ee 
for some $i$, where $d_1$ and $d_2$ are the defect numbers of $w_1$ and $w_2$, respectively.
The link $w$ thus belongs to Family II, see \eqref{eq:families}, and we can express $f_{n,d}(w)$ as
\be
f_{n,d}(
\psset{unit=0.5}
 \,
\begin{pspicture}[shift=-0.10](0,0)(3.3,0.6)
\psline{-}(0,0)(3.3,0)
\pspolygon[fillstyle=solid,fillcolor=white](0.2,0)(1.3,0)(1.3,0.6)(0.2,0.6)
\rput(0.75,0.3){$_{w_1}$}
\pspolygon[fillstyle=solid,fillcolor=white](1.6,0)(3.1,0)(3.1,0.6)(1.6,0.6)
\rput(2.35,0.3){$_{w_2}$}
\end{pspicture}\,
) = \big| f_{i,d_1}(w_1), \,+\,, f_{n-i, d_2}(w_2)\big \rangle,
\ee
since any spin not under an arc must be up (see the discussion after the proof of Lemma~\ref{lem:nonzero}).
Using \eqref{eq:Hspindec1}, one finds
\begin{alignat}{2}
\hspin\hs f_{n,d}(w) &= -\mathbb H_{(i)} f_{n,d}(w) + \big(\hspin_L + \hspin_R\big)f_{n,d}(w)\label{eq:part1}\\[0.2cm]
&= -\mathbb H_{(i)} f_{n,d}(w) + \big| \hspin\hs f_{i,d_1}(w_1), \,+\,, f_{n-i, d_2}(w_2)\big \rangle + \big| f_{i,d_1}(w_1), \,+\,, \hspin\hs f_{n-i, d_2}(w_2)\big \rangle, \nonumber
\end{alignat}
and likewise using \eqref{eq:Hloopdec1},
\begin{alignat}{2}
 f_{n,d}(h w) &= -f_{n,d}\big(e_i w\big) + f_{n,d}\big((h_L + h_R)w\big) \label{eq:part2} \\[0.2cm]
&=-f_{n,d}\big(e_i w\big) +  \big| f_{i,d_1}(h w_1), \,+\,, f_{n-i, d_2}(w_2)\big \rangle + \big| f_{i,d_1}(w_1), \,+\,, f_{n-i, d_2}(h w_2)\big \rangle. \nonumber 
\end{alignat}
Subtracting \eqref{eq:part2} from \eqref{eq:part1} yields
\begin{alignat}{2}
\hspin\hs f_{n,d}(w)- f_{n,d}(h w) = -\mathbb H_{(i)} f_{n,d}(w) &+ f_{n,d}\big(e_iw\big)  + \big| \big(\hspin\hs f_{i,d_1}(w) - f_{i,d_1}(h w_1)\big), \,+\,, f_{n-i, d_2}(w_2)\big \rangle \nonumber \\[0.2cm]& 
+ \big| f_{i,d_1}(w_1), \,+\,, \big(\hspin\hs f_{n-i, d_2}(w)-f_{n-i, d_2}(h w_2)\big)\big \rangle.
\end{alignat}
From the induction assumption, the last two terms are zero. Proving the proposition then amounts to showing that
\be
f_{n,d}\big(e_iw\big) = \mathbb H_{(i)} f_{n,d}(w).
\label{eq:fe=hf}
\ee

We separate the verification of (\ref{eq:fe=hf}) into three cases. Because 
$w$ has at least one defect ($d \ge 1$), without loss of generality, we can choose 
the position $i$ such that a defect occupies either the node $i$ or the node $i+1$, or both nodes are occupied by defects. 
There are thus three possibilities
\be
\textrm{(i)} \quad w = \,
\psset{unit=0.5}
\begin{pspicture}[shift=-1.05](0,-1.0)(3.8,1.2)
\psline{-}(0,0)(3.8,0)
\pspolygon[fillstyle=solid,fillcolor=white](0.2,0)(1.3,0)(1.3,0.6)(0.2,0.6)
\rput(0.75,0.3){$_{w_3}$}
\pspolygon[fillstyle=solid,fillcolor=white](2.5,0)(3.6,0)(3.6,0.6)(2.5,0.6)
\rput(3.05,0.3){$_{w_4}$}
\rput(1.9,-0.5){$_\uparrow$}\rput(1.9,-1.1){$_i$}
\psline[linewidth=1.5pt,linecolor=blue]{-}(1.7,0)(1.7,1)
\psline[linewidth=1.5pt,linecolor=blue]{-}(2.1,0)(2.1,1)
\psline[linecolor=gray,linestyle=dashed,dash=2pt 1pt]{-}(1.9,-0.1)(1.9,1.4)
\end{pspicture}
\qquad\textrm{(ii)} \quad w = \,
\begin{pspicture}[shift=-1.05](0,-1.0)(5.7,1.2)
\psline{-}(0,0)(5.7,0)
\pspolygon[fillstyle=solid,fillcolor=white](0.2,0)(1.3,0)(1.3,0.6)(0.2,0.6)
\rput(0.75,0.3){$_{w_3}$}
\pspolygon[fillstyle=solid,fillcolor=white](2.5,0)(3.6,0)(3.6,0.6)(2.5,0.6)
\rput(3.05,0.3){$_{w_4}$}
\pspolygon[fillstyle=solid,fillcolor=white](4.4,0)(5.5,0)(5.5,0.6)(4.4,0.6)
\rput(4.95,0.3){$_{w_5}$}
\rput(1.9,-0.5){$_\uparrow$}\rput(1.9,-1.1){$_i$}
\rput(4.2,-0.5){$_\uparrow$}\rput(4.2,-1.1){$_j$}
\psline[linewidth=1.5pt,linecolor=blue]{-}(1.7,0)(1.7,1)
\psbezier[linewidth=1.5pt,linecolor=blue]{-}(2.1,0)(2.1,1.5)(4,1.5)(4,0)
\psline[linecolor=gray,linestyle=dashed,dash=2pt 1pt]{-}(1.9,-0.1)(1.9,1.4)
\end{pspicture}
\qquad\textrm{(iii)} \quad w = \,
\begin{pspicture}[shift=-1.05](-1.9,-1.0)(3.8,1.2)
\psline{-}(-1.9,0)(3.8,0)
\pspolygon[fillstyle=solid,fillcolor=white](-0.6,0)(-1.7,0)(-1.7,0.6)(-0.6,0.6)
\rput(-1.15,0.3){$_{w_3}$}
\pspolygon[fillstyle=solid,fillcolor=white](0.2,0)(1.3,0)(1.3,0.6)(0.2,0.6)
\rput(0.75,0.3){$_{w_4}$}
\pspolygon[fillstyle=solid,fillcolor=white](2.5,0)(3.6,0)(3.6,0.6)(2.5,0.6)
\rput(3.05,0.3){$_{w_5}$}
\rput(1.9,-0.5){$_\uparrow$}\rput(1.9,-1.1){$_i$}
\rput(-0.4,-0.5){$_\uparrow$}\rput(-0.4,-1.1){$_j$}
\psbezier[linewidth=1.5pt,linecolor=blue]{-}(1.7,0)(1.7,1.5)(-0.2,1.5)(-0.2,0)
\psline[linewidth=1.5pt,linecolor=blue]{-}(2.1,0)(2.1,1)
\psline[linecolor=gray,linestyle=dashed,dash=2pt 1pt]{-}(1.9,-0.1)(1.9,1.4)
\end{pspicture}
\ee
where $w_3, w_4$ and $w_5$ are generic links, some of which may have length zero. 
A dashed delimiter is included in these diagrams to indicate the separation between the original links $w_1$ and $w_2$ in \eqref{eq:ww1w2}. 

\paragraph{Case (i):} Under the standard action, connecting defects yields a zero result, so $f_{n,d}\big(e_iw\big)=0$. 
A straightforward computation also shows that 
\be
\mathbb H_{(i)} f_{n,d}(w) = \mathbb H_{(i)} \big| f_{i-1,d_1-1}(w_{3}), \,+\,, \,+\,, \,+\,, f_{n-i-1, d_2-1}(w_{4})\big \rangle = 0,
\ee
thereby completing the verification of \eqref{eq:fe=hf} in this case.

\paragraph{Case (ii):} For nontrivial $w_4$, the lefthand side of \eqref{eq:fe=hf} reads
\be f_{n,d}\big(e_iw\big) = f_{n,d}(\,
\psset{unit=0.5}
\begin{pspicture}[shift=-1.0](0,-1.0)(5.7,1.2)
\psline{-}(0,0)(5.7,0)
\pspolygon[fillstyle=solid,fillcolor=white](0.2,0)(1.3,0)(1.3,0.6)(0.2,0.6)
\rput(0.75,0.3){$_{w_3}$}
\pspolygon[fillstyle=solid,fillcolor=white](2.5,0)(3.6,0)(3.6,0.6)(2.5,0.6)
\rput(3.05,0.3){$_{w_4}$}
\pspolygon[fillstyle=solid,fillcolor=white](4.4,0)(5.5,0)(5.5,0.6)(4.4,0.6)
\rput(4.95,0.3){$_{w_5}$}
\rput(1.9,-0.5){$_\uparrow$}\rput(1.9,-1.1){$_i$}
\rput(4.2,-0.5){$_\uparrow$}\rput(4.2,-1.1){$_j$}
\psline[linewidth=1.5pt,linecolor=blue]{-}(4,0)(4,1)
\psarc[linewidth=1.5pt,linecolor=blue]{-}(1.9,0){0.2}{0}{180}
\psline[linecolor=gray,linestyle=dashed,dash=2pt 1pt]{-}(1.9,-0.1)(1.9,1.4)
\end{pspicture}\,
) = \big| f_{i-1,d_1-1}(w_3), \,+\,, \,-\,, \,+\,, f_{j-i-2, 0}(w_4),\,+\,, \,+\,, f_{n-j,d_2}(w_5)\big \rangle,
\ee
while the righthand side is expressed as
\begin{alignat}{2}
\mathbb H_{(i)} f_{n,d}(w) & = \mathbb H_{(i)} \big| f_{i-1,d_1-1}(w_3), \,+\,, \,+\,, f_{n-i, d_2}(w_2)\big \rangle 
 = \sigma^+_{i+1} \big| f_{i-1,d_1-1}(w_3), \,+\,, \,-\,, f_{n-i, d_2}(w_2)\big \rangle\nonumber\\[0.1cm]
& =  \big| f_{i-1,d_1-1}(w_3), \,+\,, \,-\,, \,+\,, f_{j-i-2, 0}(w_4),\,+\,, \,+\,, f_{n-j,d_2}(w_5)\big \rangle.
\end{alignat}
In the second expression, $s_{i-1}=s_i=+$, so only the term $\sigma_i^-\sigma_{i+1}^+$ in $\mathbb H_{(i)}$ acts non-trivially. This explains the second equality.
The last one relies on the fact that in any spin contribution to $f_{n,d}(w)$, the positions $i+1$ and $j-1$ cannot both be occupied 
by a down spin. Applying $\sigma^+_{i+1}$ imposes a down spin in position $i+1$, so
position $j-1$ is occupied by an up spin. 
Under these circumstances, the contribution of $w_4$ is simply $f_{j-i-2,0}(w_4)$. 

Case (ii) also includes the possibility that $w_4$ has length zero, that is if
$i+1 = j-1$ and a half-arc connects the nodes $i+1$ and $i+2$. Although this subcase is not covered by the computation above, it is straightforward to verify \eqref{eq:fe=hf} using similar arguments.

\paragraph{Case (iii):} As the link $w$ is the left-right mirror image of the corresponding one in case (ii), the proof is the same as in case (ii). 
\hfill $\square$

\subsection[The maps $g^p$]{The maps $\boldsymbol{g^p}$}
\label{sec:gnp}

The missing element in the inductive proof of \eqref{eq:intertwining} is the case $d=0$. As mentioned before, this case is independent of Section~\ref{sec:d>0}. 
Its proof is based on the new linear map
\be 
g^p\; : \; \Stan{n,d=0} \rightarrow \eigenSz{n-1}{-p-1/2}\,,
\qquad p = 0, \dots, \tfrac {n-2}2,
\ee
whose action on the link $w\in\mathcal B_{n,0}$ is defined as
\be
\big|\underbrace{\,+\,, \dots, \,+\,}_p, g^p(\,
\psset{unit=0.5}
\begin{pspicture}[shift=-0.05](-0.0,0)(1.5,0.5)
\psline{-}(0,0)(1.5,0)
\pspolygon[fillstyle=solid,fillcolor=white](0.2,0)(1.3,0)(1.3,0.6)(0.2,0.6)
\rput(0.75,0.3){$_w$}
\end{pspicture}
\,), \underbrace{\,+\,, \dots, \,+\,}_p\big\rangle= \frac1{\qn{p}!}f_{n+2p,d=0}\Big(
\,
\begin{pspicture}[shift=-0.85](-1.4,-0.8)(2.9,2.0)
\psline{-}(-1.4,0)(2.9,0)
\pspolygon[fillstyle=solid,fillcolor=white](0.2,0)(1.3,0)(1.3,0.6)(0.2,0.6)
\psbezier[linecolor=blue,linewidth=1.5pt](-0.2,0)(-0.2,1.5)(1.7,1.5)(1.7,0)
\psbezier[linecolor=blue,linewidth=1.5pt](-0.8,0)(-0.8,2.1)(2.3,2.1)(2.3,0)
\psbezier[linecolor=blue,linewidth=1.5pt](-1.2,0)(-1.2,2.5)(2.7,2.5)(2.7,0)
\rput(-0.47,0.3){$.\hspace{-0.03cm}.\hspace{-0.03cm}.$}
\rput(1.97,0.3){$.\hspace{-0.03cm}.\hspace{-0.03cm}.$}
\rput(0.75,0.3){$_w$}
\rput(-0.67,-0.75){$\underbrace{\ }_p$}
\end{pspicture}\, 
\Big)\Bigg|_{\substack{s_1 = s_2 = \dots = s_p=\,+ \\[0.1cm] s_{n+p}=s_{n+p+1} = \dots = s_{n+2p-1}=\,+}}
\ee
where 
\be
 \qn{m}! = \prod_{k=1}^m \,\qn{k}, \qquad \qn{0}! \equiv 1,\qquad \qn{m} = [m]_{-q}.
\ee
Because the value of $n$ follows from the argument $w$, we have not included it as a label for $g^p(w)$. 
Here are some examples for $n=6$:
\begin{alignat}{2}
&g^0(\psset{unit=0.6}\,
\begin{pspicture}[shift=-0.125](-0.0,0)(2.4,0.5)
\psline{-}(0,0)(2.4,0)
\psarc[linecolor=blue,linewidth=1.5pt]{-}(1.2,0){0.2}{0}{180}
\psbezier[linecolor=blue,linewidth=1.5pt]{-}(0.6,0)(0.6,0.7)(1.8,0.7)(1.8,0)
\psbezier[linecolor=blue,linewidth=1.5pt]{-}(0.2,0)(0.2,1.1)(2.2,1.1)(2.2,0)
\end{pspicture}\,)=|---+\,+\rangle+ |-+--\,+\rangle  +|+--+\,-\rangle + |++--\,-\rangle + \qn{2} |+---\,+\rangle, \nonumber \\[0.1cm]
&g^1(\psset{unit=0.6}\,
\begin{pspicture}[shift=-0.125](-0.0,0)(2.4,0.5)
\psline{-}(0,0)(2.4,0)
\psarc[linecolor=blue,linewidth=1.5pt]{-}(1.2,0){0.2}{0}{180}
\psbezier[linecolor=blue,linewidth=1.5pt]{-}(0.6,0)(0.6,0.7)(1.8,0.7)(1.8,0)
\psbezier[linecolor=blue,linewidth=1.5pt]{-}(0.2,0)(0.2,1.1)(2.2,1.1)(2.2,0)
\end{pspicture}\,)=\qn{2}^2\big(\,|----\,+\rangle+|+---\,-\rangle\big) +\qn{2} \big(\,|-+--\,-\rangle+ |---+\,-\rangle\big),\\[0.1cm]
&g^2(\psset{unit=0.6}\,
\begin{pspicture}[shift=-0.125](-0.0,0)(2.4,0.5)
\psline{-}(0,0)(2.4,0)
\psarc[linecolor=blue,linewidth=1.5pt]{-}(1.2,0){0.2}{0}{180}
\psbezier[linecolor=blue,linewidth=1.5pt]{-}(0.6,0)(0.6,0.7)(1.8,0.7)(1.8,0)
\psbezier[linecolor=blue,linewidth=1.5pt]{-}(0.2,0)(0.2,1.1)(2.2,1.1)(2.2,0)
\end{pspicture}\,)=\qn{2}\qn{3}\,|----\,-\rangle. \nonumber
\end{alignat}
For 
$w\in\mathcal B_{n,0}$, 
let $\hat w$ be the link where $w$ is overarched by $p$ arcs. Then $g^p(w)$ outputs the restriction of 
$f_{n+2p,d=0}(\hat w)$ to spin states of length $n+2p-1$ with only up spins under the $p$ outmost layers. The normalisation 
by $1/\qn{p}!$ removes the factors associated to the half-arcs in these outmost layers, as these are in fact independent of $w$. 

The integer $p$ counts the excess of down spins in $g^p(w)$ in comparison with $f_{n,0}(w)$. It is then natural and convenient for later computations to set 
\be
 g^p(w)=0,\qquad p<0\quad  {\rm or}  \quad p\ge \tfrac n2.
\label{g0}
\ee 
This extension of the definition of $g^p$ will simplify the proofs that follow.
The case $p = 0$ amounts to $g^{p=0}(w) = f_{n,d=0}(w)$. In fact, the 
intertwining property \eqref{eq:intertwining} at $d = 0$ is the $p=0$ specialisation of the following proposition.
\begin{Proposition}\label{sec:toprovep} 
Let $n \in \mathbb N$ and $p = 0, \dots, \tfrac{n-2}2$. Then
\be 
\hspin\hs g^p(w) - g^p(h w)  = (\sigma^-_1+\sigma^-_{n-1})g^{p-1}(w), \qquad  w \in \Stan{n,d=0}.
\label{eq:toprovep}
\ee
\end{Proposition}
Section~\ref{sec:theproof} is devoted to the proof of this proposition. 

Our interest lies in the case $p=0$, but the proof of Section~\ref{sec:theproof} is inductive on increasing values of $n$ and decreasing values of $p$. The case $p = 0$ thus relies on the maps $g^p$ with $p>0$. As a preliminary to the proof, it is useful to understand the recursive properties of the map $g^p$. For the two families of links \eqref{eq:families}, we find
\begin{alignat}{2}
&g^p\Big(\,
\psset{unit=0.5}
\begin{pspicture}[shift=-0.85](-0.4,-0.8)(1.9,1.2)
\psline{-}(-0.4,0)(1.9,0)
\pspolygon[fillstyle=solid,fillcolor=white](0.2,0)(1.3,0)(1.3,0.6)(0.2,0.6)
\psbezier[linecolor=blue,linewidth=1.5pt](-0.2,0)(-0.2,1.5)(1.7,1.5)(1.7,0)
\rput(0.75,0.3){$_{w_1}$}
\rput(0.0,-0.4){$_\uparrow$}\rput(0.0,-1.0){$_1$}
\rput(1.5,-0.4){$_\uparrow$}\rput(1.5,-1.0){$_{n-1}$}
\end{pspicture}
\,\Big) = \qn{p+1} \Big(
\big| \, - \,, \,  g^{p-1}(w_1)\, , \,-\, \big\rangle +
\big| \, + \,, \,  g^{p}(w_1)\, , \,-\, \big\rangle +
\big| \, - \,, \,  g^{p}(w_1)\, , \,+\, \big\rangle+
\big| \, + \,, \,  g^{p+1}(w_1)\, , \,+\, \big\rangle 
\Big), \label{eq:prop1} \\[-0.2cm]
&g^{p}(
\,
\psset{unit=0.5}
\begin{pspicture}[shift=-0.05](-0.0,0)(3.3,0.5)
\psline{-}(0,0)(3.3,0)
\pspolygon[fillstyle=solid,fillcolor=white](0.2,0)(1.3,0)(1.3,0.6)(0.2,0.6)
\rput(0.75,0.3){$_{w_1}$}
\rput(0.1,0){
\pspolygon[fillstyle=solid,fillcolor=white](1.5,0)(3.0,0)(3.0,0.6)(1.5,0.6)
\rput(2.25,0.3){$_{w_2}$}
}
\rput(1.45,-0.4){$_\uparrow$}\rput(1.45,-1.0){$_i$}
\end{pspicture}\,
) = 
\sum_{a = 0}^{p-1} 
\big|g^{a}(w_1), \, -\, ,
g^{p-1-a}(w_2)\big\rangle 
+
 \sum_{a = 0}^p 
\big|g^{a}(w_1), \,+\,,
g^{p-a}(w_2)\big\rangle.
\label{eq:prop2}
\end{alignat}
We note that because of the convention $g^{-1}(w)=0$, the first sum in \eqref{eq:prop2} can also be extended up to $p$. Equations \eqref{eq:prop1} and \eqref{eq:prop2} are then expressible in the following compact forms,
\begin{alignat}{2}
&g^p\Big(\,
\psset{unit=0.5}
\begin{pspicture}[shift=-0.85](-0.4,-0.8)(1.9,1.2)
\psline{-}(-0.4,0)(1.9,0)
\pspolygon[fillstyle=solid,fillcolor=white](0.2,0)(1.3,0)(1.3,0.6)(0.2,0.6)
\psbezier[linecolor=blue,linewidth=1.5pt](-0.2,0)(-0.2,1.5)(1.7,1.5)(1.7,0)
\rput(0.75,0.3){$_{w_1}$}
\rput(0.0,-0.4){$_\uparrow$}\rput(0.0,-1.0){$_1$}
\rput(1.5,-0.4){$_\uparrow$}\rput(1.5,-1.0){$_{n-1}$}
\end{pspicture}
\,\Big) = 
\qn{p+1}\sum_{\ell,r} \big| \, \ell , \,  g^{p+\frac{\ell+r}{2}}(w_1)\, , r \big\rangle,\label{eq:prop1v2}\\
&g^{p}(
\,
\psset{unit=0.5}
\begin{pspicture}[shift=-0.05](-0.0,0)(3.3,0.5)
\psline{-}(0,0)(3.3,0)
\pspolygon[fillstyle=solid,fillcolor=white](0.2,0)(1.3,0)(1.3,0.6)(0.2,0.6)
\rput(0.75,0.3){$_{w_1}$}
\rput(0.1,0){
\pspolygon[fillstyle=solid,fillcolor=white](1.5,0)(3.0,0)(3.0,0.6)(1.5,0.6)
\rput(2.25,0.3){$_{w_2}$}
}
\rput(1.45,-0.4){$_\uparrow$}\rput(1.45,-1.0){$_i$}
\end{pspicture}\,
) = 
\sum_a \sum_{s} \big|g^{a}(w_1), s, g^{p-a+\tfrac{s-1}2}(w_2)\big\rangle,
\label{eq:prop2v2}
\end{alignat}
where the sums over $\ell,r,s$ run over $\{+,-\}\simeq\{+1,-1\}$, while the sum over $a$ is on the set $\{0, \dots, p\}$. 

At any point, one may choose to re-express a state given as a sum over spins in terms of a state obtained by acting with Pauli matrices on 
a given distinguished state. Examples to play a role later are
\be
 \ket{\Omega_k}=\!\!\sum_{s_1,\dots,s_k}\!\!\ket{s_1,\dots,s_k}=\prod_{i=1}^k(I+\sigma_i^-)\,\ket{\underbrace{+,\dots,+}_k},\qquad
  k\in\mathbb{N}.
\label{eq:Omegak}
\ee
We also choose to keep the spins {\it free} whenever possible. Almost trivial such examples are
\be \label{-s}
 \ket{-}=\sigma^-\ket{+}=\sum_s\sigma^-\ket{s}
\ee
and
\be
 \sum_a\bket{g^a(
\psset{unit=0.6}
\begin{pspicture}[shift=-0.1](0,0)(0.8,0.4)
\psline{-}(0,0)(0.8,0)
\psarc[linecolor=blue,linewidth=1.5pt](0.4,0){0.2}{0}{180}
\end{pspicture}
),\ldots}=\sum_a\delta_{a,0}\bket{g^0(
\psset{unit=0.6}
\begin{pspicture}[shift=-0.1](0,0)(0.8,0.4)
\psline{-}(0,0)(0.8,0)
\psarc[linecolor=blue,linewidth=1.5pt](0.4,0){0.2}{0}{180}
\end{pspicture}
),\ldots}=\sum_a\delta_{a,0}\bket{-,\dots}=\sum_a\sum_s\delta_{a,0}\,\sigma_1^-\bket{s,\ldots},
\label{eq:ga0}
\ee
where the unspecified part may depend on $a$. 

Expressions for $g^p$ on more complicated states $w$ are obtained by applying the relations \eqref{eq:prop1v2}, 
\eqref{eq:prop2v2} and \eqref{eq:ga0} more than once. To illustrate, we evaluate
\begin{alignat}{2}
 g^p\Big(
\,
\psset{unit=0.5}
\begin{pspicture}[shift=-0.25](-0.4,0)(3.3,0.5)
\psline{-}(-0.4,0)(3.3,0)
\pspolygon[fillstyle=solid,fillcolor=white](0.2,0)(1.3,0)(1.3,0.6)(0.2,0.6)
\psbezier[linecolor=blue,linewidth=1.5pt](-0.2,0)(-0.2,1.8)(3.1,1.8)(3.1,0)
\rput(0.75,0.3){$_{w_1}$}
\pspolygon[fillstyle=solid,fillcolor=white](1.6,0)(2.7,0)(2.7,0.6)(1.6,0.6)
\rput(2.15,0.3){$_{w_2}$}
\end{pspicture}\,
\Big) &= \qn{p+1} \sum_{\ell, r} \bket{\ell, g^{p+\frac{\ell+r}2} \big(
\,
\psset{unit=0.5}
\begin{pspicture}[shift=-0.25](0,0)(2.9,0.5)
\psline{-}(0.0,0)(2.9,0)
\pspolygon[fillstyle=solid,fillcolor=white](0.2,0)(1.3,0)(1.3,0.6)(0.2,0.6)
\rput(0.75,0.3){$_{w_1}$}
\pspolygon[fillstyle=solid,fillcolor=white](1.6,0)(2.7,0)(2.7,0.6)(1.6,0.6)
\rput(2.15,0.3){$_{w_2}$}
\end{pspicture}\,
\big), r} \nonumber \\
& = \qn{p+1} \sum_a \sum_{\ell, r, s} \bket{\ell, g^a(w_1), s, g^{p-a+\frac{\ell + r + s -1}2}(w_2), r},
\label{eq:ex1}
\end{alignat}
where the rewriting follows by first applying \eqref{eq:prop1v2} and then \eqref{eq:prop2v2} on the term
$
g^{p+\frac{\ell+r}2} \big(
\,
\psset{unit=0.5}
\begin{pspicture}[shift=-0.15](0,0)(2.9,0.5)
\psline{-}(0.0,0)(2.9,0)
\pspolygon[fillstyle=solid,fillcolor=white](0.2,0)(1.3,0)(1.3,0.6)(0.2,0.6)
\rput(0.75,0.3){$_{w_1}$}
\pspolygon[fillstyle=solid,fillcolor=white](1.6,0)(2.7,0)(2.7,0.6)(1.6,0.6)
\rput(2.15,0.3){$_{w_2}$}
\end{pspicture}\,
\big)
$. 
Another example is
\begin{alignat}{2}
 g^p\Big(
\,
\psset{unit=0.5}
\begin{pspicture}[shift=-0.85](-0.4,-0.6)(3.4,0.5)
\psline{-}(-0.4,0)(3.4,0)
\psbezier[linecolor=blue,linewidth=1.5pt](-0.2,0)(-0.2,1.6)(1.7,1.6)(1.7,0)
\pspolygon[fillstyle=solid,fillcolor=white](0.2,0)(1.3,0)(1.3,0.6)(0.2,0.6)
\rput(0.75,0.3){$_{w_1}$}
\pspolygon[fillstyle=solid,fillcolor=white](2.1,0)(3.2,0)(3.2,0.6)(2.1,0.6)
\rput(2.65,0.3){$_{w_2}$}
\rput(1.9,-0.4){$_\uparrow$}\rput(1.9,-1.0){$_i$}
\end{pspicture}\,
\Big) &= \sum_a \sum_s \bket{g^a\Big(
\,
\psset{unit=0.5}
\begin{pspicture}[shift=-0.85](-0.4,-0.6)(1.9,0.5)
\psline{-}(-0.4,0)(1.9,0)
\psbezier[linecolor=blue,linewidth=1.5pt](-0.2,0)(-0.2,1.6)(1.7,1.6)(1.7,0)
\pspolygon[fillstyle=solid,fillcolor=white](0.2,0)(1.3,0)(1.3,0.6)(0.2,0.6)
\rput(0.75,0.3){$_{w_1}$}
\end{pspicture}\,
\Big), s, g^{p-a+\frac{s-1}2}(w_2)} \nonumber\\ 
& = \sum_a \sum_{\ell, r, s} \qn{a+1} \bket{\ell, g^{a+\frac{\ell+r}2}(w_1), r, s, g^{p-a+\frac{s-1}2}(w_2)} \nonumber\\
& = \sum_b \sum_{\ell, r, s} \qqn{b+1 - \tfrac{\ell + r}{2}}\bket{\ell, g^{b}(w_1), r, s, g^{p-b+\frac{\ell+r+s-1}2}(w_2)} \nonumber\\
& = \sum_b  \qqn{b+1 - \tfrac{\sz_1 + \sz_{i-1}}{2}} \sum_{\ell, r, s}\bket{\ell, g^{b}(w_1), r, s, g^{p-b+\frac{\ell+r+s-1}2}(w_2)},
\label{a35}
\end{alignat}
where \eqref{eq:prop1v2} and \eqref{eq:prop2v2} were both used but in the opposite order compared to \eqref{eq:ex1}. At the third equality, the change of summation index  $b = a + \frac{\ell + r}2$ was performed. Although it is not explicitly indicated, 
the change of index modifies the bounds of the sum over $b$. 
However, the sum over $b$ can be set to run over the set $\{0, 1, \dots, p\}$, 
whatever the value of $\tfrac{\ell + r}{2}$ is. Indeed, the terms added or removed vanish identically, either because $\{0\}=0$ or because $g^k(w)=0$ for negative $k$ in virtue of the convention \eqref{g0}. In this way, the summations over $b$ and over $\ell,r,s$ are independent and
their order can be interchanged. Finally, at the last equality, we have written the integers $\ell$ and $r$ inside the modified $q$-number brackets $\qn \cdot$ using 
$\sz$ matrices, thus allowing us to pull the brackets out of the sum over $\ell, r$ and $s$. 
Whenever possible, we apply this procedure to pull out brackets.

The computations in Section~\ref{sec:theproof} 
below rely on explicit expressions for $g^p(w)$ for the six special links
\be
\begin{array}{c}
\psset{unit=0.5}
\begin{pspicture}[shift=-1.25](-0.4,-1.2)(1.9,1.5)
\psline{-}(-0.8,0)(2.3,0)
\pspolygon[fillstyle=solid,fillcolor=white](0.2,0)(1.3,0)(1.3,0.6)(0.2,0.6)
\psbezier[linecolor=blue,linewidth=1.5pt](-0.2,0)(-0.2,1.5)(1.7,1.5)(1.7,0)
\psbezier[linecolor=blue,linewidth=1.5pt](-0.6,0)(-0.6,2.0)(2.1,2.0)(2.1,0)
\rput(0.75,0.3){$_v$}
\rput(-0.4,-0.4){$_\uparrow$}\rput(-0.4,-1.0){$_1$}
\rput(1.9,-0.4){$_\uparrow$}\rput(1.9,-1.0){$_{n-1}$}
\end{pspicture} \\[0.25cm]
\psset{unit=0.5}
\begin{pspicture}[shift=-1.25](-0.4,-1.2)(1.9,2.0)
\psline{-}(-0.8,0)(2.3,0)
\pspolygon[fillstyle=solid,fillcolor=white](0.2,0)(1.3,0)(1.3,0.6)(0.2,0.6)
\psarc[linecolor=blue,linewidth=1.5pt](-0.4,0){0.2}{0}{180}
\psarc[linecolor=blue,linewidth=1.5pt](1.9,0){0.2}{0}{180}
\rput(0.75,0.3){$_v$}
\rput(-0.4,-0.4){$_\uparrow$}\rput(-0.4,-1.0){$_1$}
\rput(1.9,-0.4){$_\uparrow$}\rput(1.9,-1.0){$_{n-1}$}
\end{pspicture}
\end{array}
\qquad \qquad 
\begin{array}{c}
\psset{unit=0.5}
\begin{pspicture}[shift=-1.25](-0.8,-1.2)(3.8,1.5)
\psline{-}(-0.8,0)(3.8,0)
\psbezier[linecolor=blue,linewidth=1.5pt](-0.2,0)(-0.2,1.6)(1.7,1.6)(1.7,0)
\psbezier[linecolor=blue,linewidth=1.5pt](-0.6,0)(-0.6,2.5)(3.6,2.5)(3.6,0)
\pspolygon[fillstyle=solid,fillcolor=white](0.2,0)(1.3,0)(1.3,0.6)(0.2,0.6)
\rput(0.75,0.3){$_{v_1}$}
\pspolygon[fillstyle=solid,fillcolor=white](2.1,0)(3.2,0)(3.2,0.6)(2.1,0.6)
\rput(2.65,0.3){$_{v_2}$}
\rput(-0.4,-0.4){$_\uparrow$}\rput(-0.4,-1.0){$_1$}
\rput(1.9,-0.4){$_\uparrow$}\rput(1.9,-1.0){$_i$}
\rput(3.4,-0.4){$_\uparrow$}\rput(3.4,-1.0){$_{n-1}$}
\end{pspicture} \\[0.25cm]
\psset{unit=0.5}
\begin{pspicture}[shift=-1.25](-0.8,-1.2)(3.8,2.0)
\psline{-}(-0.8,0)(3.8,0)
\psbezier[linecolor=blue,linewidth=1.5pt](1.7,0)(1.7,1.6)(3.6,1.6)(3.6,0)
\psarc[linecolor=blue,linewidth=1.5pt](-0.4,0){0.2}{0}{180}
\pspolygon[fillstyle=solid,fillcolor=white](0.2,0)(1.3,0)(1.3,0.6)(0.2,0.6)
\rput(0.75,0.3){$_{v_1}$}
\pspolygon[fillstyle=solid,fillcolor=white](2.1,0)(3.2,0)(3.2,0.6)(2.1,0.6)
\rput(2.65,0.3){$_{v_2}$}
\rput(-0.4,-0.4){$_\uparrow$}\rput(-0.4,-1.0){$_1$}
\rput(1.9,-0.4){$_\uparrow$}\rput(1.9,-1.0){$_i$}
\rput(3.4,-0.4){$_\uparrow$}\rput(3.4,-1.0){$_{n-1}$}
\end{pspicture}
\end{array}
\qquad \qquad 
\begin{array}{c}
\psset{unit=0.5}
\begin{pspicture}[shift=-1.25](-0.4,-1.2)(5.7,1.5)
\psline{-}(-0.4,0)(5.7,0)
\psbezier[linecolor=blue,linewidth=1.5pt](-0.2,0)(-0.2,1.6)(1.7,1.6)(1.7,0)
\psbezier[linecolor=blue,linewidth=1.5pt](2.1,0)(2.1,1.6)(4,1.6)(4,0)
\pspolygon[fillstyle=solid,fillcolor=white](0.2,0)(1.3,0)(1.3,0.6)(0.2,0.6)
\rput(0.75,0.3){$_{v_1}$}
\pspolygon[fillstyle=solid,fillcolor=white](2.5,0)(3.6,0)(3.6,0.6)(2.5,0.6)
\rput(3.05,0.3){$_{v_2}$}
\pspolygon[fillstyle=solid,fillcolor=white](4.4,0)(5.5,0)(5.5,0.6)(4.4,0.6)
\rput(4.95,0.3){$_{v_3}$}
\rput(0.0,-0.4){$_\uparrow$}\rput(0.0,-1.0){$_1$}
\rput(1.9,-0.4){$_\uparrow$}\rput(1.9,-1.0){$_i$}
\rput(4.2,-0.4){$_\uparrow$}\rput(4.2,-1.0){$_j$}
\end{pspicture} \\[0.25cm]
\psset{unit=0.5}
\begin{pspicture}[shift=-1.25](-0.4,-1.2)(5.7,2.0)
\psline{-}(-0.4,0)(5.7,0)
\psbezier[linecolor=blue,linewidth=1.5pt](-0.2,0)(-0.2,2.1)(4,2.1)(4,0)
\psarc[linecolor=blue,linewidth=1.5pt](1.9,0){0.2}{0}{180}
\pspolygon[fillstyle=solid,fillcolor=white](0.2,0)(1.3,0)(1.3,0.6)(0.2,0.6)
\rput(0.75,0.3){$_{v_1}$}
\pspolygon[fillstyle=solid,fillcolor=white](2.5,0)(3.6,0)(3.6,0.6)(2.5,0.6)
\rput(3.05,0.3){$_{v_2}$}
\pspolygon[fillstyle=solid,fillcolor=white](4.4,0)(5.5,0)(5.5,0.6)(4.4,0.6)
\rput(4.95,0.3){$_{v_3}$}
\rput(0.0,-0.4){$_\uparrow$}\rput(0.0,-1.0){$_1$}
\rput(1.9,-0.4){$_\uparrow$}\rput(1.9,-1.0){$_i$}
\rput(4.2,-0.4){$_\uparrow$}\rput(4.2,-1.0){$_j$}
\end{pspicture}
\end{array}
\label{eq:sixstates}
\ee
These are computed using the rules \eqref{eq:prop1v2}, \eqref{eq:prop2v2} and \eqref{eq:ga0}, with the final expressions 
compiled in Section~\ref{app:fantastic6}. 

\subsection[The case $d=0$]{The case $\boldsymbol{d=0}$}\label{sec:theproof}

This subsection provides the proof of Proposition~\ref{sec:toprovep}. As already mentioned, the case $p=0$ of this 
statement is the case $d=0$ of Proposition~\ref{sec:int}. The proof is by induction and is specific to each family of 
links \eqref{eq:families}. 

\paragraph{Family I:} Let 
$w = \,\psset{unit=0.5}
\begin{pspicture}[shift=-0.15](-0.4,-0.0)(1.9,1.0)
\psline{-}(-0.4,0)(1.9,0)
\pspolygon[fillstyle=solid,fillcolor=white](0.2,0)(1.3,0)(1.3,0.6)(0.2,0.6)
\psbezier[linecolor=blue,linewidth=1.5pt](-0.2,0)(-0.2,1.5)(1.7,1.5)(1.7,0)
\rput(0.75,0.3){$_{w_1}$}
\end{pspicture} $. Simplified using \eqref{eq:Hspindec1} and \eqref{eq:prop1v2}, the action of $\hspin$ on $g^p(w)$ reads
\begin{alignat}{2}
\hspin\hs g^p(w) &= -\mathbb H_{(1)} g^p(w) - \mathbb H_{(n-1)}g^p(w) + \hspin_C\, g^p(w) \nonumber\\&
=  -\mathbb H_{(1)} g^p(w) - \mathbb H_{(n-1)}g^p(w) + \qn{p+1} \sum_{\ell, r} \bket{\ell, \hspin\hs g^{p+\tfrac{\ell+r}2}(w_1),r}.
\end{alignat}
From \eqref{eq:Hloopdec1} and \eqref{eq:prop1v2}, the similar decomposition of
$g^p(hw)$ is given by
\begin{alignat}{2}
g^p(h w) &= -g^p\big(e_1w\big) -g^p\big(e_{n-1}w\big) + 
g^p\big(h_Cw\big),
\nonumber\\[0.1cm]
& = -g^p\big(e_1w\big) -g^p\big(e_{n-1}w\big) + \qn{p+1} \sum_{\ell,r} \bket{\ell, g^{p+\frac{\ell+r}2}(h w_1),r}.
\end{alignat}
Subtracting these, we find
\begin{alignat}{2}
\hspin\hs g^p(w) - g^p(hw) &= \qn{p+1} \sum_{\ell, r} \bket{\ell, \Big(\hspin\hs g^{p+\frac{\ell + r}2}(w_1) - g^{p+\frac{\ell + r}2}(hw_1)\Big),r} \nonumber\\[-0.2cm]
& \hspace{3.5cm} 
- (\mathbb H_{(1)}+ \mathbb H_{(n-1)})g^p(w) 
+ g^p\big(e_1w\big) + g^p\big(e_{n-1}w\big)\nonumber \\
& \hspace{-2.5cm}= \qn{p+1}(\sm_2 + \sm_{n-2}) \sum_{\ell,r}\bket{\ell, 
g^{p-1+\frac{\ell + r}2}(w_1),r} 
- (\mathbb H_{(1)}+ \mathbb H_{(n-1)}) g^p(w)
 +g^p\big(e_1w\big) + g^p\big(e_{n-1}w\big)\nonumber \\
& \hspace{-2.5cm}= \frac{\qn{p+1}}{\qn{p}}(\sm_2 + \sm_{n-2})\, g^{p-1}(w) -(\mathbb H_{(1)}+ \mathbb H_{(n-1)}) g^p(w) + g^p\big(e_1w\big) + g^p\big(e_{n-1}w\big).
\label{a39}
\end{alignat}
Here, the induction assumption was used at the second equality and the first term was recognised as a multiple of $g^{p-1}(w)$ at the last equality. Although \eqref{a39} looks singular for $p=0$, upon expanding $g^{p-1}(w)$ using \eqref{eq:prop1v2} (with $p$ changed to $p-1$), one readily sees that it is not and that the rest of the calculation below handles this particular case correctly.

The intertwining relation in this case is thus equivalent to
\be 
\Big(\frac{\qn{p+1}}{\qn p } (\sm_2 + \sm_{n-2}) - (\sm_1 + \sm_{n-1}) \Big) g^{p-1}(w)  -(\mathbb H_{(1)}+ \mathbb H_{(n-1)}) g^p(w)
 + g^p\big(e_1w\big) + g^p\big(e_{n-1}w\big) =0.
\ee
A sufficient condition is that the following two equations hold separately,
\begin{alignat}{2}
&\Big(\frac{\qn{p+1}}{\qn p} \sm_2 - \sm_1 
\Big) g^{p-1}(w) - \mathbb H_{(1)} g^p(w) + g^p\big(e_1w\big)  = 0,\label{eq:zero1}\\
&\Big(\frac{\qn{p+1}}{\qn p} \sm_{n-2} - \sm_{n-1} \Big) g^{p-1}(w) - \mathbb H_{(n-1)} g^p(w)+ g^p\big(e_{n-1}w\big)  = 0.\label{eq:zero2}
\end{alignat}
As they are mirror images under left-right reflection, it suffices to establish the first. 

Family I splits into two subfamilies,
\be
\textrm{\bf I (a)\,:} \qquad
w = \, 
\psset{unit=0.5}
\begin{pspicture}[shift=-1.25](-0.6,-1.2)(2.1,1.8)
\psline{-}(-0.8,0)(2.3,0)
\pspolygon[fillstyle=solid,fillcolor=white](0.2,0)(1.3,0)(1.3,0.6)(0.2,0.6)
\psbezier[linecolor=blue,linewidth=1.5pt](-0.2,0)(-0.2,1.5)(1.7,1.5)(1.7,0)
\psbezier[linecolor=blue,linewidth=1.5pt](-0.6,0)(-0.6,2.0)(2.1,2.0)(2.1,0)
\rput(0.75,0.3){$_{w_2}$}
\rput(-0.4,-0.4){$_\uparrow$}\rput(-0.4,-1.0){$_1$}
\rput(1.9,-0.4){$_\uparrow$}\rput(1.9,-1.0){$_{n-1}$}
\end{pspicture} \ \ ,
\qquad \qquad \textrm{\bf I (b)\,:} \qquad
w = \,
\begin{pspicture}[shift=-1.25](-0.8,-1.2)(3.8,2.4)
\psline{-}(-0.8,0)(3.8,0)
\psbezier[linecolor=blue,linewidth=1.5pt](-0.2,0)(-0.2,1.6)(1.7,1.6)(1.7,0)
\psbezier[linecolor=blue,linewidth=1.5pt](-0.6,0)(-0.6,2.5)(3.6,2.5)(3.6,0)
\pspolygon[fillstyle=solid,fillcolor=white](0.2,0)(1.3,0)(1.3,0.6)(0.2,0.6)
\rput(0.75,0.3){$_{w_3}$}
\pspolygon[fillstyle=solid,fillcolor=white](2.1,0)(3.2,0)(3.2,0.6)(2.1,0.6)
\rput(2.65,0.3){$_{w_4}$}
\rput(-0.4,-0.4){$_\uparrow$}\rput(-0.4,-1.0){$_1$}
\rput(1.9,-0.4){$_\uparrow$}\rput(1.9,-1.0){$_i$}
\rput(3.4,-0.4){$_\uparrow$}\rput(3.4,-1.0){$_{n-1}$}
\end{pspicture} \ \, .
\label{eq:familiesIab}
\ee
We will assume that $w_2$, $w_3$ and $w_4$ are proper links (of nonzero length). If one or more of them is not, the analysis is similar but considerably simpler. The next steps use the expressions of $g^p(w)$, $g^{p-1}(w)$ and $g^{p}(e_{1}w)$ given in Section~\ref{app:fantastic6}. 

For Family I (a), the requirement \eqref{eq:zero1} translates into 
\begin{alignat}{2}
&\big(\qn{p+1}\sm_2 - \qn p \sm_1\big)\qqn{p+\tfrac{\sz_1 + \sz_{n-1}}{2}} \sum_{r,s,t,u} \bket{r,s,g^{\lambda_1-1}(w_2),t,u} \label{eq:zerofamilyIa}\\
& -\mathbb H_{(1)} \qn{p+1}
\qqn{p+1+\tfrac{\sz_1 + \sz_{n-1}}{2}}
\hspace{-0.1cm}  \sum_{r,s,t,u} \bket{r,s,g^{\lambda_1}(w_2),t,u} 
 + \sm_1 \sm_{n-1}\hspace{-0.1cm} \sum_{r,s,t,u} \bket{r,s,g^{\lambda_1-2}(w_2),t,u} =0 \nonumber
\end{alignat}
where 
\be
\lambda_1 = p+\frac{r+s+t+u}2.
\ee 
As Section~\ref{app:g} will show,
a sufficient condition for the vanishing of 
\eqref{eq:zerofamilyIa} is the simpler relation
\be
\Big[\big(\qn{p+1}\sm_2 - \qn p \sm_1\big)\qqn{p+\tfrac{\sz_1+\sz_4}{2}}-\mathbb H_{(1)}\qn{p+1}\qqn{p+1+\tfrac{\sz_1+\sz_4}{2}}+\sm_1\sm_4\Big] | \Omega_4 \rangle = 0,
\label{eq:suf1}
\ee
which must hold for all $p$. Establishing it is a straightforward exercise with Pauli matrices acting on $\ctwotimes 4$ and is easily done either by hand or with a computer.

For links $w$ in Family I (b), the states $g^p(w)$, $g^{p-1}(w)$ and $g^{p}(e_{1}w)$ contain an extra sum over an integer $a$, and the requirement \eqref{eq:zero1} translates into a family of equations labeled by this integer,
\begin{alignat}{2}
&\big(\qn{p+1}\sm_2 - \qn p \sm_1\big)\qqn{a+1 - \tfrac{\sz_2 + \sz_{i-1}}2} \hspace{-0.1cm}\sum_{\ell,r,s,t,u} \hspace{-0.1cm} \bket{\ell, r, g^a(w_3), s, t, g^{\lambda_2-a-1}(w_4),u} \nonumber\\
& \hspace{1.00cm} - \mathbb H_{(1)} \qn{p+1} \qqn{a+1 - \tfrac{\sz_2 + \sz_{i-1}}2} \hspace{-0.1cm} \sum_{\ell,r,s,t,u} \hspace{-0.1cm} \bket{\ell, r, g^a(w_3), s, t, g^{\lambda_2-a}(w_4),u} \nonumber\\
& \hspace{1.00cm} + \sm_1 \qqn{p-a+\tfrac{\sz_2 + \sz_{i-1}}2} \hspace{-0.1cm}\sum_{\ell,r,s,t,u} \hspace{-0.1cm} \bket{\ell, r, g^a(w_3), s, t, g^{\lambda_2-a-1}(w_4),u} =0,
\end{alignat}
with 
\be 
 \lambda_2 = p+\frac{\ell+r+s+t+u-1}2.
\label{la2}
\ee 
From the discussion in Section~\ref{app:g}, it is sufficient to check that
\be
\Big[\big(\qn{p+1}(\sm_2-\mathbb H_{(1)}) - \qn p \sm_1\big) \qqn{a+1 - \tfrac{\sz_2 + \sz_{3}}2} + \sm_1 \qqn{p-a + \tfrac{\sz_2 + \sz_3}2 }\Big]
 |\Omega_5 \rangle = 0\label{eq:suf2}
\ee
holds for all $p$ and $a$. 
This is a computation in $\ctwotimes 5$ and is readily verified, thus concluding the proof of \eqref{eq:toprovep} for links in Family I.

\paragraph{Family II:} Let 
$\psset{unit=0.5}
w =
\begin{pspicture}[shift=-0.85](-0.1,-0.8)(3.3,1.2)
\psline{-}(0,0)(3.3,0)
\pspolygon[fillstyle=solid,fillcolor=white](0.2,0)(1.3,0)(1.3,0.6)(0.2,0.6)
\rput(0.75,0.3){$_{w_1}$}
\pspolygon[fillstyle=solid,fillcolor=white](1.6,0)(3.1,0)(3.1,0.6)(1.6,0.6)
\rput(2.35,0.3){$_{w_2}$}
\rput(1.45,-0.4){$_\uparrow$}\rput(1.45,-1.0){$_i$}
\end{pspicture}\,
$. Using \eqref{eq:Hspindec2} and \eqref{eq:prop2v2}, $\hspin\hs g^p(w)$ is expressed as
\begin{alignat}{2}
\hspin\hs g^p(w) &= -\mathbb H_{(i)} g^p(w) + \hspin_L\, g^p(w) + \hspin_R \,g^p(w)  \label{eq:spinpart}\\
& = -\mathbb H_{(i)} g^p(w) + \sum_a \sum_s \Big(\bket{\hspin\hs g^a(w_1), s, g^{p-a+\frac{s-1}2}(w_2)} + \bket{g^a(w_1), s, \hspin\hs g^{p-a+\frac{s-1}2}(w_2)}\Big). \nonumber
\end{alignat}
Using \eqref{eq:Hloopdec1} and \eqref{eq:prop2v2}, we likewise find
\begin{alignat}{2}
g^p(h w) &= -g^p\big(e_iw\big) + g^p(h_Lw) + g^p(h_Rw), \label{eq:looppart}\\
& = -g^p\big(e_iw\big) + \sum_a \sum_s \Big(\bket{ g^a(hw_1), s, g^{p-a+\frac{s-1}2}(w_2)} + \bket{g^a(w_1), s, g^{p-a+\frac{s-1}2}(hw_2)}\Big).\nonumber
\end{alignat}
Subtracting \eqref{eq:looppart} from \eqref{eq:spinpart} yields
\begin{alignat}{2}
\hspin\hs g^p(w)-g^p(hw) &=  - \mathbb H_{(i)} g^p(w) + g^p\big(e_iw\big)  +\sum_a \sum_s \bket{\big(\hspin\hs g^a(w_1)-g^a(hw_1)\big), s, g^{p-a+\frac{s-1}2}(w_2)} \nonumber\\ 
&\hspace{1cm}+ \sum_a \sum_s \bket{g^a(w_1), s, \big(\hspin\hs g^{p-a+\frac{s-1}2}(w_2)-g^{p-a+\frac{s-1}2}(hw_2)\big)}\nonumber\\
& = - \mathbb H_{(i)} g^p(w) + g^p\big(e_iw\big) + \big(\sm_1+\sm_{i-1}\big)\sum_a \sum_s \bket{ g^{a-1}(w_1), s, g^{p-a+\frac{s-1}2}(w_2)} \nonumber\\
&\hspace{1cm} + \big(\sm_{i+1}+\sm_{n-1}\big)\sum_a \sum_s \bket{ g^{a}(w_1), s, g^{p-1-a+\frac{s-1}2}(w_2)}  \nonumber\\
& = - \mathbb H_{(i)} g^p(w) + g^p\big(e_iw\big) +\big(\sm_1+\sm_{i-1} + \sm_{i+1} + \sm_{n-1}\big) g^{p-1}(w),
\end{alignat}
where the second equality is a consequence of the induction assumption, while the last equality follows by recombining 
$g^{p-1}(w)$ using \eqref{eq:prop2v2}. The intertwining relation is now equivalent to
\be
 \big(\sm_{i-1}+ \sm_{i+1}\big)g^{p-1}(w) - \mathbb H_{(i)} g^p(w)+ g^p\big(e_iw\big)  =0. \label{eq:a-b2}
\ee
Without loss of generality, the position $i$ is chosen to be between the two leftmost outer arches. The link $w$ then has the form
\be
w = \psset{unit=0.5}
\begin{pspicture}[shift=-1.25](-0.4,-1.2)(5.7,1.8)
\psline{-}(-0.4,0)(5.7,0)
\psbezier[linecolor=blue,linewidth=1.5pt](-0.2,0)(-0.2,1.6)(1.7,1.6)(1.7,0)
\psbezier[linecolor=blue,linewidth=1.5pt](2.1,0)(2.1,1.6)(4,1.6)(4,0)
\pspolygon[fillstyle=solid,fillcolor=white](0.2,0)(1.3,0)(1.3,0.6)(0.2,0.6)
\rput(0.75,0.3){$_{w_3}$}
\pspolygon[fillstyle=solid,fillcolor=white](2.5,0)(3.6,0)(3.6,0.6)(2.5,0.6)
\rput(3.05,0.3){$_{w_4}$}
\pspolygon[fillstyle=solid,fillcolor=white](4.4,0)(5.5,0)(5.5,0.6)(4.4,0.6)
\rput(4.95,0.3){$_{w_5}$}
\rput(0.0,-0.5){$_\uparrow$}\rput(0.0,-1.1){$_1$}
\rput(1.9,-0.5){$_\uparrow$}\rput(1.9,-1.1){$_i$}
\rput(4.2,-0.5){$_\uparrow$}\rput(4.2,-1.1){$_j$}
\psline[linecolor=gray,linestyle=dashed,dash=2pt 1pt]{-}(1.9,-0.1)(1.9,1.6)
\end{pspicture} 
\qquad \quad \Big(\, \textrm{i.e.}\,\quad 
w_1 = 
\begin{pspicture}[shift=-1.25](-0.4,-1.2)(1.8,1.8)
\psline{-}(-0.4,0)(1.8,0)
\psbezier[linecolor=blue,linewidth=1.5pt](-0.2,0)(-0.2,1.6)(1.7,1.6)(1.7,0)
\pspolygon[fillstyle=solid,fillcolor=white](0.2,0)(1.3,0)(1.3,0.6)(0.2,0.6)
\rput(0.75,0.3){$_{w_3}$}
\rput(0.0,-0.4){$_\uparrow$}\rput(0.0,-1.0){$_1$}
\rput(1.5,-0.4){$_\uparrow$}\rput(1.5,-1.0){$_{i-1}$}
\end{pspicture}
\qquad 
w_2 = \begin{pspicture}[shift=-1.25](1.9,-1.2)(5.7,1.8)
\psline{-}(1.9,0)(5.7,0)
\psbezier[linecolor=blue,linewidth=1.5pt](2.1,0)(2.1,1.6)(4,1.6)(4,0)
\pspolygon[fillstyle=solid,fillcolor=white](2.5,0)(3.6,0)(3.6,0.6)(2.5,0.6)
\rput(3.05,0.3){$_{w_4}$}
\pspolygon[fillstyle=solid,fillcolor=white](4.4,0)(5.5,0)(5.5,0.6)(4.4,0.6)
\rput(4.95,0.3){$_{w_5}$}
\rput(2.3,-0.4){$_\uparrow$}\rput(2.3,-1.0){$_{i+1}$}
\rput(4.2,-0.4){$_\uparrow$}\rput(4.2,-1.0){$_j$}
\end{pspicture} \Big).
\ee 
Henceforth, we assume that $w_3$, $w_4$ and $w_5$ are generic links of nonzero lengths, the other cases being simpler. The final step is to use the expressions for $g^{p-1}(w)$, $g^p(e_iw)$ and $g^p(w)$ given in Section~\ref{app:fantastic6}. 
These involve sums over two integers, $a$ and $b$, and \eqref{eq:a-b2} then translates into a family of equations labeled 
by these integers,
\begin{alignat}{2}
&\big(\sm_{i-1} + \sm_{i+1}\big) \qqn{a+1-\tfrac{\sz_1 + \sz_{i-1}}2}\qqn{b+1-\tfrac{\sz_{i+1} + \sz_{j-1}}2}\hspace{-0.1cm} \sum_{k,\ell, r,s,t,u}\hspace{-0.1cm} \bket{k,g^a(w_3),\ell, r,s, g^b(w_4), t, u, g^{\lambda_3-a-b-1}(w_5)} \nonumber\\
& - \mathbb H_{(i)} \qqn{a+1-\tfrac{\sz_1 + \sz_{i-1}}2}\qqn{b+1-\tfrac{\sz_{i+1} + \sz_{j-1}}2}\hspace{-0.1cm} \sum_{k,\ell, r,s,t,u}\hspace{-0.1cm} \bket{k,g^a(w_3),\ell, r,s, g^b(w_4), t, u, g^{\lambda_3-a-b}(w_5)} \nonumber \\
& + \sm_i \qqn{a+b - \tfrac{\sz_1+\sz_{i-1}+\sz_{i+1}+\sz_{j-1}-4}2}\hspace{-0.1cm} \sum_{k,\ell, r,s,t,u}\hspace{-0.1cm} \bket{k,g^a(w_3),\ell, r,s, g^b(w_4), t, u, g^{\lambda_3-a-b-1}(w_5)} = 0, 
\end{alignat}
with 
\be 
 \lambda_3 = p+\frac{k+\ell+r+s+t+u-2}2.
\label{la3}
\ee 
As discussed in Section~\ref{app:g}, 
a sufficient condition is that
\be
\Big(\big(\sm_2 + \sm_4 - \mathbb H_{(3)}\big) \qqn{a+1-\tfrac{\sz_1 + \sz_{2}}2}\qqn{b+1-\tfrac{\sz_4 + \sz_5}2} 
  + \sm_3 \qqn{a+b - 
  \tfrac{\sz_1 + \sz_2 + \sz_4 +  \sz_5-4}2}
   \Big) |\Omega_6 \rangle = 0
\label{eq:suf3}
\ee
holds for all $p$, $a$ and $b$. It is a straightforward exercise on $(\mathbb C^2)^{\otimes 6}$ to verify this, thereby concluding 
the proof for Family II and thus of Proposition~\ref{sec:toprovep}.\hfill $\square$ 

\subsection{Sufficient conditions}
\label{app:g}

In this appendix, we discuss the sufficient conditions that were used in Section~\ref{sec:theproof}. 
Consider the relation (\ref{eq:zerofamilyIa}), which has the general form
\be
M_0 \sum_{r,s,t,u}\bket{r,s,g^{\lambda_1}(w),t,u} + M_{-1} \sum_{r,s,t,u}\bket{r,s,g^{\lambda_1-1}(w),t,u} + M_{-2} \sum_{r,s,t,u}\bket{r,s,g^{\lambda_1-2}(w),t,u} = 0,
\label{MM}
\ee
where $M_0,M_{-1}$ and $M_{-2}$ are operators that only act on the first two and last two spins (namely $r,s,t,u$), in such a way that $M_j$ increases the total magnetisation by $j$. 
As $\lambda_1=p+\sigma$ with $\sigma \equiv (r+s+t+u)/2$, the upper index on $g$ takes on seven different values, so that the spin state appearing in the middle has the form $g^{p+k}(w)$, where $k$ depends on the value of $\sigma$ and varies from $-4$ to $+2$. Because $M_0,M_{-1}$ and $M_{-2}$ act on the first and last spins only, the terms pertaining to different values of $k$ are linearly independent (they contain a different number of down spins) and must vanish separately. So the relation (\ref{MM}) reduces to distinct equations labelled by $k$ (seven in the present case), 
\be 
\sum_{j=-2}^0 \; M_j \: \sum_{r,s,t,u} \, \delta_{\sigma+j,k}\, \bket{r,s,g^{p+k}(w),t,u} = 0.
\ee
The key feature is that in any of these relations, the same $g^{p+k}(w)$ appears in all terms. When $g^{p+k}(w)$ is not identically zero, that is if $0 \le p+k \le \frac{n}2-3$, it plays a spectator role and can be ignored, replacing each relation by a simpler condition in the reduced space $\ctwotimes 4$,
\be
\sum_{j=-2}^0 \; \bar M_j \: \sum_{r,s,t,u} \, \delta_{\sigma+j,k}\, \bket{r,s,t,u} = 
\sum_{j=-2}^0 \; \bar M_j \: \mathbb P_{k-j} |\Omega_4\rangle = 0,
\ee
with $|\Omega_k \rangle$ defined in \eqref{eq:Omegak}. Here the actions of $\bar M_0,\bar M_{-1}$ and $\bar M_{-2}$ are identical to those of $M_0,M_{-1}$ and $M_{-2}$, except they are applied to the spins in positions $1,2,3,4$ instead of $1,2,n-2$ and $n-1$, with the other positions absent. We have also used the projector $\mathbb P_k \,:\, \ctwotimes 4 \to \eigenSz{4}{k}$ on the subspace of magnetisation $k$ (evidently $\mathbb P_k=0$ if $k \neq 0, \pm 1,\pm 2$). Finally, because $M_j$ increases the magnetisation by $j$, the previous equation can be rewritten as
\be
\mathbb P_{k} \: \sum_{j=-2}^0 \; \bar M_j \, |\Omega_4\rangle = 0, \qquad 0 \le p+k \le \frac{n}2-3.
\label{pk}
\ee
Whether or not all values of $k=0,\pm 1,\pm 2$ are included, the following single relation provides a sufficient condition to (\ref{pk}),
\be
(\bar M_0 + \bar M_{-1} + \bar M_{-2})\, |\Omega_4\rangle = 0.
\ee
This was precisely the first sufficient condition used in Section~\ref{sec:theproof}. 
For a generic value of $p$, namely not too close to either 0 or $\frac n 2$ so that all five values of $k$ are included in (\ref{pk}), it is in fact equivalent to the original condition (\ref{MM}).

This property readily extends to situations where the operators $M_j$ act on more than four specified spin positions or where there are more than one $g$ term in a given ket. In all of these scenarios, the conclusion is that we may replace the corresponding relation by simpler sufficient conditions provided the upper indices of the $g$-parts match the action of the operators $M_j$. This property is used in Section~\ref{sec:theproof} 
to produce the two sufficient conditions, \eqref{eq:suf2} and \eqref{eq:suf3}. They respectively take the form
\begin{alignat}{3}
& \hspace{-3mm} \sum_a\hspace{0.05cm} \sum_{j = -1}^0\hspace{0.1cm} \sum_{\ell, r,s,t,u} M_j \bket{\ell, r, g^a(w_1), s, t, g^{\lambda_2+j-a}(w_2),u} = 0&\rightarrow\ \Big(\sum_{j = -1}^0 \bar M_j\Big) | \Omega_5 \rangle= 0, \\
& \hspace{-3mm} \sum_{a,b}\hspace{0.05cm}\sum_{j = -1}^0 \hspace{0.1cm}\sum_{k,\ell, r,s,t,u} M_j \bket{k, g^a(w_1), \ell, r, s, g^b(w_2), t, u, g^{\lambda_3+j-a-b}(w_3)}= 0\ \ &\rightarrow\ \Big(\sum_{j = -1}^0 \bar M_j\Big) | \Omega_6 \rangle = 0, 
\end{alignat}
where it is recalled that $\lambda_2$ and $\lambda_3$ given in (\ref{la2}) and (\ref{la3}) depend on the partial magnetisation related to the spins explicitly shown. Here the operators $\bar M_j$ are obtained from $M_j$ by changing the spin positions $\ell, r,s, \dots$ where the action is performed to 
positions $1, 2, 3, \dots$, and by removing all spectators. Each of these sufficient conditions is established following the above discussion.

\subsection[Expressions for $g^p(w)$]{Expressions for $\boldsymbol{g^p(w)}$}\label{app:fantastic6}

Explicit expressions for $g^p(w)$ for the six links in \eqref{eq:sixstates} are given by
\vspace{0.2cm}
\begin{alignat}{2}
 g^p\Big(
\,
\psset{unit=0.5}
\begin{pspicture}[shift=-1.45](-0.8,-1.2)(2.3,1.8)
\psline{-}(-0.8,0)(2.3,0)
\pspolygon[fillstyle=solid,fillcolor=white](0.2,0)(1.3,0)(1.3,0.6)(0.2,0.6)
\psbezier[linecolor=blue,linewidth=1.5pt](-0.2,0)(-0.2,1.5)(1.7,1.5)(1.7,0)
\psbezier[linecolor=blue,linewidth=1.5pt](-0.6,0)(-0.6,2.0)(2.1,2.0)(2.1,0)
\rput(-0.4,-0.4){$_\uparrow$}\rput(-0.4,-1.0){$_1$}
\rput(1.9,-0.4){$_\uparrow$}\rput(1.9,-1.0){$_{n-1}$}
\rput(0.75,0.3){$_v$}
\end{pspicture}\,
\Big) &= \qn{p+1} \qqn{p+1+\tfrac{\sz_1 + \sz_{n-1}}2} \sum_{r,s,t,u} \bket{r,s,g^{\lambda_1}(v),t,u},\\\displaybreak[0]
 g^p\Big(
\,
\psset{unit=0.45}
\begin{pspicture}[shift=-1.45](-0.8,-1.2)(2.3,1.8)
\psline{-}(-0.8,0)(2.3,0)
\pspolygon[fillstyle=solid,fillcolor=white](0.2,0)(1.3,0)(1.3,0.6)(0.2,0.6)
\psarc[linecolor=blue,linewidth=1.5pt](-0.4,0){0.2}{0}{180}
\psarc[linecolor=blue,linewidth=1.5pt](1.9,0){0.2}{0}{180}
\rput(-0.4,-0.4){$_\uparrow$}\rput(-0.4,-1.0){$_1$}
\rput(1.9,-0.4){$_\uparrow$}\rput(1.9,-1.0){$_{n-1}$}
\rput(0.75,0.3){$_v$}
\end{pspicture}\,
\Big) &= \sm_1 \sm_{n-1} \sum_{r,s,t,u} \bket{r,s,g^{\lambda_1-2}(v),t,u},\\\displaybreak[0]
 g^p\Big(
\,
\psset{unit=0.45}
\begin{pspicture}[shift=-1.45](-0.8,-1.2)(3.8,2.4)
\psline{-}(-0.8,0)(3.8,0)
\psbezier[linecolor=blue,linewidth=1.5pt](-0.2,0)(-0.2,1.6)(1.7,1.6)(1.7,0)
\psbezier[linecolor=blue,linewidth=1.5pt](-0.6,0)(-0.6,2.5)(3.6,2.5)(3.6,0)
\pspolygon[fillstyle=solid,fillcolor=white](0.2,0)(1.3,0)(1.3,0.6)(0.2,0.6)
\rput(0.75,0.3){$_{v_1}$}
\pspolygon[fillstyle=solid,fillcolor=white](2.1,0)(3.2,0)(3.2,0.6)(2.1,0.6)
\rput(2.65,0.3){$_{v_2}$}
\rput(-0.4,-0.4){$_\uparrow$}\rput(-0.4,-1.0){$_1$}
\rput(1.9,-0.4){$_\uparrow$}\rput(1.9,-1.0){$_i$}
\rput(3.4,-0.4){$_\uparrow$}\rput(3.4,-1.0){$_{n-1}$}
\end{pspicture}\,
\Big) &= \qn{p+1} \sum_a \qqn{a+1-\tfrac{\sz_2 + \sz_{i-1}}2} \sum_{\ell, r,s,t,u} \bket{\ell, r, g^a(v_1), s, t, g^{\lambda_2-a}(v_2),u},\\\displaybreak[0]
 g^p\Big(
\,
\psset{unit=0.45}
\begin{pspicture}[shift=-1.25](-0.8,-1.2)(3.8,2.4)
\psline{-}(-0.8,0)(3.8,0)
\psbezier[linecolor=blue,linewidth=1.5pt](1.7,0)(1.7,1.6)(3.6,1.6)(3.6,0)
\psarc[linecolor=blue,linewidth=1.5pt](-0.4,0){0.2}{0}{180}
\pspolygon[fillstyle=solid,fillcolor=white](0.2,0)(1.3,0)(1.3,0.6)(0.2,0.6)
\rput(0.75,0.3){$_{v_1}$}
\pspolygon[fillstyle=solid,fillcolor=white](2.1,0)(3.2,0)(3.2,0.6)(2.1,0.6)
\rput(2.65,0.3){$_{v_2}$}
\rput(-0.4,-0.4){$_\uparrow$}\rput(-0.4,-1.0){$_1$}
\rput(1.9,-0.4){$_\uparrow$}\rput(1.9,-1.0){$_i$}
\rput(3.4,-0.4){$_\uparrow$}\rput(3.4,-1.0){$_{n-1}$}
\end{pspicture}\,
\Big) &= \sm_1 \sum_a 
\qqn{p-a+\tfrac{\sz_2+\sz_{i-1}}2} \sum_{\ell, r,s,t,u} \bket{\ell, r, g^a(v_1), s, t, g^{\lambda_2-a-1}(v_2),u}, \\\displaybreak[0]
 g^p\Big(
\,
\psset{unit=0.45}
\begin{pspicture}[shift=-1.45](-0.4,-1.2)(5.7,2.4)
\psline{-}(-0.4,0)(5.7,0)
\psbezier[linecolor=blue,linewidth=1.5pt](-0.2,0)(-0.2,1.6)(1.7,1.6)(1.7,0)
\psbezier[linecolor=blue,linewidth=1.5pt](2.1,0)(2.1,1.6)(4,1.6)(4,0)
\pspolygon[fillstyle=solid,fillcolor=white](0.2,0)(1.3,0)(1.3,0.6)(0.2,0.6)
\rput(0.75,0.3){$_{v_1}$}
\pspolygon[fillstyle=solid,fillcolor=white](2.5,0)(3.6,0)(3.6,0.6)(2.5,0.6)
\rput(3.05,0.3){$_{v_2}$}
\pspolygon[fillstyle=solid,fillcolor=white](4.4,0)(5.5,0)(5.5,0.6)(4.4,0.6)
\rput(4.95,0.3){$_{v_3}$}
\rput(0.0,-0.4){$_\uparrow$}\rput(0.0,-1.0){$_1$}
\rput(1.9,-0.4){$_\uparrow$}\rput(1.9,-1.0){$_i$}
\rput(4.2,-0.4){$_\uparrow$}\rput(4.2,-1.0){$_j$}
\end{pspicture}\,
\Big) &= \sum_{a,b} \qqn{a-\tfrac{\sz_1+\sz_{i-1}-2}2}\qqn{b-\tfrac{\sz_{i+1}+\sz_{j-1}-2}2}\nonumber\\
&\hspace{3cm} \times \sum_{k,\ell, r,s,t,u}\hspace{-0.2cm} 
\bket{k, g^a(v_1), \ell, r,s,g^b(v_2),t,u,g^{\lambda_3-a-b}(v_3)},
\\\displaybreak[0]
 g^p\Big(
\,
\psset{unit=0.45}
\begin{pspicture}[shift=-1.45](-0.4,-1.2)(5.7,2.4)
\psline{-}(-0.4,0)(5.7,0)
\psbezier[linecolor=blue,linewidth=1.5pt](-0.2,0)(-0.2,2.1)(4,2.1)(4,0)
\psarc[linecolor=blue,linewidth=1.5pt](1.9,0){0.2}{0}{180}
\pspolygon[fillstyle=solid,fillcolor=white](0.2,0)(1.3,0)(1.3,0.6)(0.2,0.6)
\rput(0.75,0.3){$_{v_1}$}
\pspolygon[fillstyle=solid,fillcolor=white](2.5,0)(3.6,0)(3.6,0.6)(2.5,0.6)
\rput(3.05,0.3){$_{v_2}$}
\pspolygon[fillstyle=solid,fillcolor=white](4.4,0)(5.5,0)(5.5,0.6)(4.4,0.6)
\rput(4.95,0.3){$_{v_3}$}
\rput(0.0,-0.4){$_\uparrow$}\rput(0.0,-1.0){$_1$}
\rput(1.9,-0.4){$_\uparrow$}\rput(1.9,-1.0){$_i$}
\rput(4.2,-0.4){$_\uparrow$}\rput(4.2,-1.0){$_j$}
\end{pspicture}\,
\Big) &= \sm_i \sum_{a,b}
\qqn{a+b-\tfrac{\sz_1+\sz_{i-1}+\sz_{i+1}+\sz_{j-1}-4}2}\hspace{-0.15cm} \nonumber\\
& \hspace{3cm} \times \sum_{k,\ell, r,s,t,u}\hspace{-0.2cm} 
\bket{k, g^a(v_1), \ell, r,s,g^b(v_2),t,u,g^{\lambda_3-a-b-1}(v_3)}.
\end{alignat}
As in the calculation (\ref{a35}), the summations over $a,b$ run over the set $\{0, 1, \dots, p\}$. The three $\lambda_i$ are 
\be
 \lambda_1= p+\tfrac{r+s+t+u}2,\qquad \lambda_2=p+\tfrac{\ell+r+s+t+u-1}2,\qquad \lambda_3=p+\tfrac{k+\ell+r+s+t+u-2}2.
\label{eq:lambdas}
\ee

%
 
%

\end{document}